\documentstyle[12pt]{article}

\catcode`\@=11
\long\def\@makefntext#1{
\protect\noindent \hbox to 3.2pt {\hskip-.9pt
$^{{\ninerm\@thefnmark}}$\hfil}#1\hfill}		

\def\@makefnmark{\hbox to 0pt{$^{\@thefnmark}$\hss}}  

\def\ps@myheadings{\let\@mkboth\@gobbletwo
\def\@oddhead{\hbox{}
\rightmark\hfil\ninerm\thepage}
\def\@oddfoot{}\def\@evenhead{\ninerm\thepage\hfil
\leftmark\hbox{}}\def\@evenfoot{}
\def\sectionmark##1{}\def\subsectionmark##1{}}

\renewcommand{\thefootnote}{\fnsymbol{footnote}}

\newcounter{sectionc}\newcounter{subsectionc}\newcounter{subsubsectionc}
\renewcommand{\section}[1] {\vspace*{0.6cm}\addtocounter{sectionc}{1}
\setcounter{subsectionc}{0}\setcounter{subsubsectionc}{0}\noindent
	{\normalsize\bf\thesectionc. #1}\par\vspace*{0.4cm}}
\renewcommand{\subsection}[1] {\vspace*{0.6cm}\addtocounter{subsectionc}{1}
	\setcounter{subsubsectionc}{0}\noindent
	{\normalsize\it\thesectionc.\thesubsectionc. #1}\par\vspace*{0.4cm}}
\renewcommand{\subsubsection}[1]
{\vspace*{0.6cm}\addtocounter{subsubsectionc}{1}
	\noindent {\normalsize\rm\thesectionc.\thesubsectionc.\thesubsubsectionc.
	#1}\par\vspace*{0.4cm}}

\newcounter{appendixc}
\newcounter{subappendixc}[appendixc]
\newcounter{subsubappendixc}[subappendixc]

\renewcommand{\appendix}[1] {\vspace*{0.6cm}
        \refstepcounter{appendixc}
        \setcounter{figure}{0}
        \setcounter{table}{0}
        \setcounter{equation}{0}
        \renewcommand{\thefigure}{\Alph{appendixc}.\arabic{figure}}
        \renewcommand{\thetable}{\Alph{appendixc}.\arabic{table}}
        \renewcommand{\theappendixc}{\Alph{appendixc}}
        \renewcommand{\theequation}{\Alph{appendixc}.\arabic{equation}}
        \noindent{\bf Appendix \theappendixc #1}\par\vspace*{0.4cm}}

\def\abstracts#1{{

\centering{\begin{minipage}{12.2truecm}\footnotesize\baselineskip=12pt\noindent
	\centerline{\footnotesize ABSTRACT}\vspace*{0.3cm}
	\parindent=0pt #1
	\end{minipage}}\par}}

\renewenvironment{thebibliography}[1]
	{\begin{list}{\arabic{enumi}.}
	{\usecounter{enumi}\setlength{\parsep}{0pt}
\setlength{\leftmargin 1.25cm}{\rightmargin 0pt}
	 \setlength{\itemsep}{0pt} \settowidth
	{\labelwidth}{#1.}\sloppy}}{\end{list}}

\newcounter{itemlistc}
\newcounter{romanlistc}
\newcounter{alphlistc}
\newcounter{arabiclistc}

\newcommand{\fcaption}[1]{
        \refstepcounter{figure}
        \setbox\@tempboxa = \hbox{\footnotesize Fig.~\thefigure. #1}
        \ifdim \wd\@tempboxa > 6in
           {\begin{center}
        \parbox{6in}{\footnotesize\baselineskip=12pt Fig.~\thefigure. #1}
            \end{center}}
        \else
             {\begin{center}
             {\footnotesize Fig.~\thefigure. #1}
              \end{center}}
        \fi}

\newcommand{\tcaption}[1]{
        \refstepcounter{table}
        \setbox\@tempboxa = \hbox{\footnotesize Table~\thetable. #1}
        \ifdim \wd\@tempboxa > 6in
           {\begin{center}
        \parbox{6in}{\footnotesize\baselineskip=12pt Table~\thetable. #1}
            \end{center}}
        \else
             {\begin{center}
             {\footnotesize Table~\thetable. #1}
              \end{center}}
        \fi}

\def\@citex[#1]#2{\if@filesw\immediate\write\@auxout
	{\string\citation{#2}}\fi
\def\@citea{}\@cite{\@for\@citeb:=#2\do
	{\@citea\def\@citea{,}\@ifundefined
	{b@\@citeb}{{\bf ?}\@warning
	{Citation `\@citeb' on page \thepage \space undefined}}
	{\csname b@\@citeb\endcsname}}}{#1}}

\newif\if@cghi
\def\cite{\@cghitrue\@ifnextchar [{\@tempswatrue
	\@citex}{\@tempswafalse\@citex[]}}
\def\citelow{\@cghifalse\@ifnextchar [{\@tempswatrue
	\@citex}{\@tempswafalse\@citex[]}}
\def\@cite#1#2{{$\null^{#1}$\if@tempswa\typeout
	{IJCGA warning: optional citation argument
	ignored: `#2'} \fi}}

 1
 1
 1

\font\ninerm=cmr9

\newsavebox{\rgluonarrow}
\savebox{\rgluonarrow}(0,0)[l]{
\thicklines
\put(11.5,2){\line(-1,0){4}}
\put(11.5,2){\line(0,-1){4}}
\thinlines
}

\newsavebox{\lgluonarrow}
\savebox{\lgluonarrow}(0,0)[r]{
\thicklines
\put(-11.5,-2){\line(1,0){4}}
\put(-11.5,-2){\line(0,1){4}}
\thinlines
}

\newsavebox{\ugluonarrow}
\savebox{\ugluonarrow}(0,0)[b]{
\thicklines
\put(-2,11.5){\line(1,0){4}}
\put(-2,11.5){\line(0,-1){4}}
\thinlines
}

\newsavebox{\trgluonarrow}
\savebox{\trgluonarrow}(0,0)[bl]{
\multiput( 18, 10)(-0.1,-0.1){28}{\circle*{0.2}}
\multiput( 18, 10)( 0.1,-0.1){28}{\circle*{0.2}}
}

\newsavebox{\brgluonarrow}
\savebox{\brgluonarrow}(0,0)[tl]{
\multiput( 17.7,-7.8)(-0.1,-0.1){28}{\circle*{0.2}}
\multiput( 17.7,-7.8)( 0.1,-0.1){28}{\circle*{0.2}}
}

\newsavebox{\hblob}
\savebox{\hblob}(0,0)[c]{
\put(0,0){\oval(40,20)}
\put(-10,-10){\line(1,1){20}}
\put(-5,-10){\line(1,1){19.1}}
\put( 0,-10){\line(1,1){17.1}}
\put( 5,-10){\line(1,1){14.1}}
\put(10,-10){\line(1,1){10}}
\put( 5, 10){\line(-1,-1){19.1}}
\put( 0, 10){\line(-1,-1){17.1}}
\put(-5, 10){\line(-1,-1){14.1}}
\put(-10, 10){\line(-1,-1){10}}
}

\newsavebox{\roundblob}
\savebox{\roundblob}(0,0)[c]{
\put(  0,0){\circle{40}}
\thicklines
\put(-10,10){\line(1,1){10}}
\put(-10,10){\line(-1,-1){10}}
\put(-5, 5){\line(1,1){13.2}}
\put(-5, 5){\line(-1,-1){13.2}}
\put( 0,0){\line(1,1){14.1}}
\put( 0,0){\line(-1,-1){14.1}}
\put( 5, -5){\line(1,1){13.2}}
\put( 5, -5){\line(-1,-1){13.2}}
\put(10,-10){\line(1,1){10}}
\put(10,-10){\line(-1,-1){10}}
\thinlines
}

\newsavebox{\trgluonloop}
\savebox{\trgluonloop}(0,0)[bl]{
\put(  0.00, 20.00){\circle*{0.2}}
\put(  0.40, 19.92){\circle*{0.2}}
\put(  0.79, 19.70){\circle*{0.2}}
\put(  1.16, 19.34){\circle*{0.2}}
\put(  1.51, 18.87){\circle*{0.2}}
\put(  1.84, 18.31){\circle*{0.2}}
\put(  2.13, 17.70){\circle*{0.2}}
\put(  2.40, 17.06){\circle*{0.2}}
\put(  2.65, 16.43){\circle*{0.2}}
\put(  2.88, 15.85){\circle*{0.2}}
\put(  3.11, 15.34){\circle*{0.2}}
\put(  3.34, 14.94){\circle*{0.2}}
\put(  3.59, 14.65){\circle*{0.2}}
\put(  3.86, 14.50){\circle*{0.2}}
\put(  4.16, 14.47){\circle*{0.2}}
\put(  4.51, 14.58){\circle*{0.2}}
\put(  4.90, 14.79){\circle*{0.2}}
\put(  5.34, 15.11){\circle*{0.2}}
\put(  5.83, 15.48){\circle*{0.2}}
\put(  6.35, 15.90){\circle*{0.2}}
\put(  6.90, 16.32){\circle*{0.2}}
\put(  7.46, 16.71){\circle*{0.2}}
\put(  8.03, 17.05){\circle*{0.2}}
\put(  8.57, 17.30){\circle*{0.2}}
\put(  9.08, 17.44){\circle*{0.2}}
\put(  9.54, 17.46){\circle*{0.2}}
\put(  9.94, 17.35){\circle*{0.2}}
\put( 10.26, 17.11){\circle*{0.2}}
\put( 10.50, 16.75){\circle*{0.2}}
\put( 10.66, 16.27){\circle*{0.2}}
\put( 10.74, 15.70){\circle*{0.2}}
\put( 10.75, 15.06){\circle*{0.2}}
\put( 10.71, 14.38){\circle*{0.2}}
\put( 10.63, 13.69){\circle*{0.2}}
\put( 10.53, 13.02){\circle*{0.2}}
\put( 10.44, 12.39){\circle*{0.2}}
\put( 10.37, 11.83){\circle*{0.2}}
\put( 10.36, 11.34){\circle*{0.2}}
\put( 10.41, 10.96){\circle*{0.2}}
\put( 10.55, 10.67){\circle*{0.2}}
\put( 10.79, 10.48){\circle*{0.2}}
\put( 11.12, 10.38){\circle*{0.2}}
\put( 11.56, 10.36){\circle*{0.2}}
\put( 12.08, 10.40){\circle*{0.2}}
\put( 12.67, 10.48){\circle*{0.2}}
\put( 13.33, 10.58){\circle*{0.2}}
\put( 14.01, 10.67){\circle*{0.2}}
\put( 14.70, 10.74){\circle*{0.2}}
\put( 15.36, 10.75){\circle*{0.2}}
\put( 15.97, 10.71){\circle*{0.2}}
\put( 16.50, 10.60){\circle*{0.2}}
\put( 16.93, 10.40){\circle*{0.2}}
\put( 17.24, 10.12){\circle*{0.2}}
\put( 17.42,  9.76){\circle*{0.2}}
\put( 17.47,  9.34){\circle*{0.2}}
\put( 17.39,  8.85){\circle*{0.2}}
\put( 17.20,  8.32){\circle*{0.2}}
\put( 16.90,  7.77){\circle*{0.2}}
\put( 16.54,  7.20){\circle*{0.2}}
\put( 16.13,  6.64){\circle*{0.2}}
\put( 15.71,  6.11){\circle*{0.2}}
\put( 15.30,  5.60){\circle*{0.2}}
\put( 14.95,  5.13){\circle*{0.2}}
\put( 14.68,  4.72){\circle*{0.2}}
\put( 14.51,  4.34){\circle*{0.2}}
\put( 14.47,  4.02){\circle*{0.2}}
\put( 14.55,  3.73){\circle*{0.2}}
\put( 14.77,  3.47){\circle*{0.2}}
\put( 15.11,  3.23){\circle*{0.2}}
\put( 15.57,  3.01){\circle*{0.2}}
\put( 16.11,  2.78){\circle*{0.2}}
\put( 16.72,  2.54){\circle*{0.2}}
\put( 17.35,  2.28){\circle*{0.2}}
\put( 17.99,  2.00){\circle*{0.2}}
\put( 18.58,  1.69){\circle*{0.2}}
\put( 19.10,  1.35){\circle*{0.2}}
\put( 19.52,  0.99){\circle*{0.2}}
\put( 19.82,  0.61){\circle*{0.2}}
\put( 19.98,  0.22){\circle*{0.2}}
}

\newsavebox{\brgluonloop}
\savebox{\brgluonloop}(0,0)[tl]{
\put( 19.98, -0.18){\circle*{0.2}}
\put( 19.84, -0.58){\circle*{0.2}}
\put( 19.55, -0.96){\circle*{0.2}}
\put( 19.14, -1.33){\circle*{0.2}}
\put( 18.62, -1.67){\circle*{0.2}}
\put( 18.04, -1.98){\circle*{0.2}}
\put( 17.40, -2.26){\circle*{0.2}}
\put( 16.77, -2.52){\circle*{0.2}}
\put( 16.16, -2.76){\circle*{0.2}}
\put( 15.61, -2.99){\circle*{0.2}}
\put( 15.14, -3.22){\circle*{0.2}}
\put( 14.79, -3.45){\circle*{0.2}}
\put( 14.56, -3.71){\circle*{0.2}}
\put( 14.47, -3.99){\circle*{0.2}}
\put( 14.51, -4.32){\circle*{0.2}}
\put( 14.66, -4.68){\circle*{0.2}}
\put( 14.93, -5.10){\circle*{0.2}}
\put( 15.27, -5.56){\circle*{0.2}}
\put( 15.67, -6.06){\circle*{0.2}}
\put( 16.09, -6.60){\circle*{0.2}}
\put( 16.51, -7.16){\circle*{0.2}}
\put( 16.88, -7.72){\circle*{0.2}}
\put( 17.18, -8.28){\circle*{0.2}}
\put( 17.38, -8.81){\circle*{0.2}}
\put( 17.47, -9.30){\circle*{0.2}}
\put( 17.43, -9.73){\circle*{0.2}}
\put( 17.26,-10.09){\circle*{0.2}}
\put( 16.96,-10.38){\circle*{0.2}}
\put( 16.54,-10.58){\circle*{0.2}}
\put( 16.02,-10.71){\circle*{0.2}}
\put( 15.41,-10.75){\circle*{0.2}}
\put( 14.75,-10.74){\circle*{0.2}}
\put( 14.07,-10.68){\circle*{0.2}}
\put( 13.38,-10.58){\circle*{0.2}}
\put( 12.73,-10.49){\circle*{0.2}}
\put( 12.12,-10.40){\circle*{0.2}}
\put( 11.59,-10.36){\circle*{0.2}}
\put( 11.15,-10.37){\circle*{0.2}}
\put( 10.81,-10.47){\circle*{0.2}}
\put( 10.57,-10.65){\circle*{0.2}}
\put( 10.42,-10.93){\circle*{0.2}}
\put( 10.36,-11.31){\circle*{0.2}}
\put( 10.37,-11.79){\circle*{0.2}}
\put( 10.43,-12.34){\circle*{0.2}}
\put( 10.52,-12.97){\circle*{0.2}}
\put( 10.62,-13.64){\circle*{0.2}}
\put( 10.70,-14.33){\circle*{0.2}}
\put( 10.75,-15.01){\circle*{0.2}}
\put( 10.74,-15.65){\circle*{0.2}}
\put( 10.67,-16.23){\circle*{0.2}}
\put( 10.51,-16.71){\circle*{0.2}}
\put( 10.28,-17.09){\circle*{0.2}}
\put(  9.96,-17.34){\circle*{0.2}}
\put(  9.57,-17.46){\circle*{0.2}}
\put(  9.12,-17.45){\circle*{0.2}}
\put(  8.61,-17.32){\circle*{0.2}}
\put(  8.07,-17.07){\circle*{0.2}}
\put(  7.51,-16.74){\circle*{0.2}}
\put(  6.94,-16.35){\circle*{0.2}}
\put(  6.39,-15.93){\circle*{0.2}}
\put(  5.87,-15.52){\circle*{0.2}}
\put(  5.38,-15.13){\circle*{0.2}}
\put(  4.94,-14.82){\circle*{0.2}}
\put(  4.54,-14.59){\circle*{0.2}}
\put(  4.19,-14.48){\circle*{0.2}}
\put(  3.88,-14.49){\circle*{0.2}}
\put(  3.61,-14.64){\circle*{0.2}}
\put(  3.36,-14.91){\circle*{0.2}}
\put(  3.13,-15.31){\circle*{0.2}}
\put(  2.90,-15.81){\circle*{0.2}}
\put(  2.67,-16.38){\circle*{0.2}}
\put(  2.42,-17.01){\circle*{0.2}}
\put(  2.16,-17.65){\circle*{0.2}}
\put(  1.86,-18.27){\circle*{0.2}}
\put(  1.54,-18.83){\circle*{0.2}}
\put(  1.19,-19.31){\circle*{0.2}}
\put(  0.82,-19.68){\circle*{0.2}}
\put(  0.43,-19.91){\circle*{0.2}}
}

\newsavebox{\blgluonloop}
\savebox{\blgluonloop}(0,0)[tr]{
\put(  0.03,-20.00){\circle*{0.2}}
\put( -0.37,-19.94){\circle*{0.2}}
\put( -0.76,-19.72){\circle*{0.2}}
\put( -1.13,-19.38){\circle*{0.2}}
\put( -1.49,-18.91){\circle*{0.2}}
\put( -1.81,-18.36){\circle*{0.2}}
\put( -2.11,-17.75){\circle*{0.2}}
\put( -2.38,-17.11){\circle*{0.2}}
\put( -2.63,-16.48){\circle*{0.2}}
\put( -2.87,-15.89){\circle*{0.2}}
\put( -3.09,-15.38){\circle*{0.2}}
\put( -3.32,-14.97){\circle*{0.2}}
\put( -3.57,-14.67){\circle*{0.2}}
\put( -3.83,-14.50){\circle*{0.2}}
\put( -4.14,-14.47){\circle*{0.2}}
\put( -4.48,-14.56){\circle*{0.2}}
\put( -4.87,-14.77){\circle*{0.2}}
\put( -5.31,-15.08){\circle*{0.2}}
\put( -5.79,-15.45){\circle*{0.2}}
\put( -6.31,-15.87){\circle*{0.2}}
\put( -6.86,-16.29){\circle*{0.2}}
\put( -7.42,-16.68){\circle*{0.2}}
\put( -7.98,-17.03){\circle*{0.2}}
\put( -8.53,-17.28){\circle*{0.2}}
\put( -9.04,-17.44){\circle*{0.2}}
\put( -9.51,-17.47){\circle*{0.2}}
\put( -9.91,-17.37){\circle*{0.2}}
\put(-10.24,-17.14){\circle*{0.2}}
\put(-10.48,-16.78){\circle*{0.2}}
\put(-10.65,-16.31){\circle*{0.2}}
\put(-10.74,-15.75){\circle*{0.2}}
\put(-10.75,-15.11){\circle*{0.2}}
\put(-10.71,-14.44){\circle*{0.2}}
\put(-10.64,-13.75){\circle*{0.2}}
\put(-10.54,-13.07){\circle*{0.2}}
\put(-10.44,-12.44){\circle*{0.2}}
\put(-10.38,-11.87){\circle*{0.2}}
\put(-10.36,-11.38){\circle*{0.2}}
\put(-10.41,-10.98){\circle*{0.2}}
\put(-10.54,-10.69){\circle*{0.2}}
\put(-10.77,-10.49){\circle*{0.2}}
\put(-11.09,-10.38){\circle*{0.2}}
\put(-11.52,-10.36){\circle*{0.2}}
\put(-12.03,-10.39){\circle*{0.2}}
\put(-12.63,-10.47){\circle*{0.2}}
\put(-13.27,-10.57){\circle*{0.2}}
\put(-13.96,-10.66){\circle*{0.2}}
\put(-14.64,-10.73){\circle*{0.2}}
\put(-15.31,-10.76){\circle*{0.2}}
\put(-15.92,-10.72){\circle*{0.2}}
\put(-16.46,-10.61){\circle*{0.2}}
\put(-16.90,-10.42){\circle*{0.2}}
\put(-17.22,-10.14){\circle*{0.2}}
\put(-17.41, -9.79){\circle*{0.2}}
\put(-17.47, -9.37){\circle*{0.2}}
\put(-17.40, -8.89){\circle*{0.2}}
\put(-17.22, -8.37){\circle*{0.2}}
\put(-16.93, -7.81){\circle*{0.2}}
\put(-16.57, -7.25){\circle*{0.2}}
\put(-16.16, -6.69){\circle*{0.2}}
\put(-15.74, -6.15){\circle*{0.2}}
\put(-15.33, -5.64){\circle*{0.2}}
\put(-14.98, -5.17){\circle*{0.2}}
\put(-14.70, -4.75){\circle*{0.2}}
\put(-14.52, -4.37){\circle*{0.2}}
\put(-14.47, -4.04){\circle*{0.2}}
\put(-14.54, -3.75){\circle*{0.2}}
\put(-14.75, -3.49){\circle*{0.2}}
\put(-15.08, -3.25){\circle*{0.2}}
\put(-15.53, -3.02){\circle*{0.2}}
\put(-16.06, -2.80){\circle*{0.2}}
\put(-16.67, -2.56){\circle*{0.2}}
\put(-17.30, -2.30){\circle*{0.2}}
\put(-17.94, -2.02){\circle*{0.2}}
\put(-18.54, -1.72){\circle*{0.2}}
\put(-19.06, -1.38){\circle*{0.2}}
\put(-19.50, -1.02){\circle*{0.2}}
\put(-19.80, -0.64){\circle*{0.2}}
\put(-19.97, -0.25){\circle*{0.2}}
\put(-19.99,  0.15){\circle*{0.2}}
}

\newsavebox{\tlgluonloop}
\savebox{\tlgluonloop}(0,0)[br]{
\put(-19.86,  0.55){\circle*{0.2}}
\put(-19.58,  0.93){\circle*{0.2}}
\put(-19.18,  1.30){\circle*{0.2}}
\put(-18.67,  1.64){\circle*{0.2}}
\put(-18.08,  1.95){\circle*{0.2}}
\put(-17.46,  2.24){\circle*{0.2}}
\put(-16.82,  2.50){\circle*{0.2}}
\put(-16.20,  2.74){\circle*{0.2}}
\put(-15.65,  2.97){\circle*{0.2}}
\put(-15.18,  3.20){\circle*{0.2}}
\put(-14.82,  3.43){\circle*{0.2}}
\put(-14.58,  3.69){\circle*{0.2}}
\put(-14.47,  3.97){\circle*{0.2}}
\put(-14.50,  4.29){\circle*{0.2}}
\put(-14.65,  4.65){\circle*{0.2}}
\put(-14.90,  5.06){\circle*{0.2}}
\put(-15.24,  5.52){\circle*{0.2}}
\put(-15.64,  6.02){\circle*{0.2}}
\put(-16.06,  6.56){\circle*{0.2}}
\put(-16.47,  7.11){\circle*{0.2}}
\put(-16.85,  7.68){\circle*{0.2}}
\put(-17.16,  8.24){\circle*{0.2}}
\put(-17.37,  8.77){\circle*{0.2}}
\put(-17.47,  9.26){\circle*{0.2}}
\put(-17.44,  9.70){\circle*{0.2}}
\put(-17.28, 10.07){\circle*{0.2}}
\put(-16.99, 10.36){\circle*{0.2}}
\put(-16.58, 10.57){\circle*{0.2}}
\put(-16.06, 10.70){\circle*{0.2}}
\put(-15.46, 10.75){\circle*{0.2}}
\put(-14.81, 10.74){\circle*{0.2}}
\put(-14.12, 10.68){\circle*{0.2}}
\put(-13.44, 10.59){\circle*{0.2}}
\put(-12.78, 10.49){\circle*{0.2}}
\put(-12.17, 10.41){\circle*{0.2}}
\put(-11.63, 10.36){\circle*{0.2}}
\put(-11.18, 10.37){\circle*{0.2}}
\put(-10.83, 10.46){\circle*{0.2}}
\put(-10.58, 10.63){\circle*{0.2}}
\put(-10.43, 10.90){\circle*{0.2}}
\put(-10.36, 11.28){\circle*{0.2}}
\put(-10.37, 11.74){\circle*{0.2}}
\put(-10.42, 12.30){\circle*{0.2}}
\put(-10.51, 12.92){\circle*{0.2}}
\put(-10.61, 13.59){\circle*{0.2}}
\put(-10.70, 14.27){\circle*{0.2}}
\put(-10.75, 14.96){\circle*{0.2}}
\put(-10.75, 15.60){\circle*{0.2}}
\put(-10.68, 16.18){\circle*{0.2}}
\put(-10.53, 16.68){\circle*{0.2}}
\put(-10.30, 17.06){\circle*{0.2}}
\put( -9.99, 17.32){\circle*{0.2}}
\put( -9.61, 17.46){\circle*{0.2}}
\put( -9.16, 17.46){\circle*{0.2}}
\put( -8.66, 17.33){\circle*{0.2}}
\put( -8.11, 17.10){\circle*{0.2}}
\put( -7.55, 16.77){\circle*{0.2}}
\put( -6.99, 16.39){\circle*{0.2}}
\put( -6.44, 15.97){\circle*{0.2}}
\put( -5.91, 15.55){\circle*{0.2}}
\put( -5.42, 15.16){\circle*{0.2}}
\put( -4.97, 14.84){\circle*{0.2}}
\put( -4.57, 14.60){\circle*{0.2}}
\put( -4.21, 14.48){\circle*{0.2}}
\put( -3.90, 14.48){\circle*{0.2}}
\put( -3.63, 14.62){\circle*{0.2}}
\put( -3.38, 14.89){\circle*{0.2}}
\put( -3.15, 15.27){\circle*{0.2}}
\put( -2.92, 15.76){\circle*{0.2}}
\put( -2.69, 16.34){\circle*{0.2}}
\put( -2.44, 16.96){\circle*{0.2}}
\put( -2.18, 17.60){\circle*{0.2}}
\put( -1.89, 18.22){\circle*{0.2}}
\put( -1.57, 18.79){\circle*{0.2}}
\put( -1.22, 19.28){\circle*{0.2}}
\put( -0.85, 19.65){\circle*{0.2}}
\put( -0.46, 19.90){\circle*{0.2}}
\put( -0.06, 20.00){\circle*{0.2}}
}

\newsavebox{\gluon}
\savebox{\gluon}(0,0)[c]{
\put(  0.00,  0.00){\circle*{0.2}}
\put(  0.16,  0.25){\circle*{0.2}}
\put(  0.32,  0.50){\circle*{0.2}}
\put(  0.48,  0.74){\circle*{0.2}}
\put(  0.64,  0.97){\circle*{0.2}}
\put(  0.80,  1.20){\circle*{0.2}}
\put(  0.96,  1.41){\circle*{0.2}}
\put(  1.11,  1.61){\circle*{0.2}}
\put(  1.27,  1.79){\circle*{0.2}}
\put(  1.43,  1.96){\circle*{0.2}}
\put(  1.59,  2.10){\circle*{0.2}}
\put(  1.75,  2.23){\circle*{0.2}}
\put(  1.91,  2.33){\circle*{0.2}}
\put(  2.07,  2.41){\circle*{0.2}}
\put(  2.23,  2.46){\circle*{0.2}}
\put(  2.39,  2.49){\circle*{0.2}}
\put(  2.55,  2.50){\circle*{0.2}}
\put(  2.71,  2.48){\circle*{0.2}}
\put(  2.87,  2.43){\circle*{0.2}}
\put(  3.03,  2.37){\circle*{0.2}}
\put(  3.18,  2.27){\circle*{0.2}}
\put(  3.34,  2.16){\circle*{0.2}}
\put(  3.50,  2.02){\circle*{0.2}}
\put(  3.66,  1.86){\circle*{0.2}}
\put(  3.82,  1.69){\circle*{0.2}}
\put(  3.98,  1.50){\circle*{0.2}}
\put(  4.14,  1.29){\circle*{0.2}}
\put(  4.30,  1.07){\circle*{0.2}}
\put(  4.46,  0.84){\circle*{0.2}}
\put(  4.62,  0.60){\circle*{0.2}}
\put(  4.78,  0.35){\circle*{0.2}}
\put(  4.94,  0.10){\circle*{0.2}}
\put(  5.10, -0.15){\circle*{0.2}}
\put(  5.25, -0.39){\circle*{0.2}}
\put(  5.41, -0.64){\circle*{0.2}}
\put(  5.57, -0.88){\circle*{0.2}}
\put(  5.73, -1.11){\circle*{0.2}}
\put(  5.89, -1.32){\circle*{0.2}}
\put(  6.05, -1.53){\circle*{0.2}}
\put(  6.21, -1.72){\circle*{0.2}}
\put(  6.37, -1.89){\circle*{0.2}}
\put(  6.53, -2.05){\circle*{0.2}}
\put(  6.69, -2.18){\circle*{0.2}}
\put(  6.85, -2.29){\circle*{0.2}}
\put(  7.01, -2.38){\circle*{0.2}}
\put(  7.17, -2.44){\circle*{0.2}}
\put(  7.32, -2.48){\circle*{0.2}}
\put(  7.48, -2.50){\circle*{0.2}}
\put(  7.64, -2.49){\circle*{0.2}}
\put(  7.80, -2.46){\circle*{0.2}}
\put(  7.96, -2.40){\circle*{0.2}}
\put(  8.12, -2.31){\circle*{0.2}}
\put(  8.28, -2.21){\circle*{0.2}}
\put(  8.44, -2.08){\circle*{0.2}}
\put(  8.60, -1.93){\circle*{0.2}}
\put(  8.76, -1.76){\circle*{0.2}}
\put(  8.92, -1.58){\circle*{0.2}}
\put(  9.08, -1.38){\circle*{0.2}}
\put(  9.24, -1.16){\circle*{0.2}}
\put(  9.39, -0.93){\circle*{0.2}}
\put(  9.55, -0.70){\circle*{0.2}}
\put(  9.71, -0.46){\circle*{0.2}}
\put(  9.87, -0.21){\circle*{0.2}}
\put( 10.03,  0.04){\circle*{0.2}}
\put( 10.19,  0.29){\circle*{0.2}}
\put( 10.35,  0.54){\circle*{0.2}}
\put( 10.51,  0.78){\circle*{0.2}}
\put( 10.67,  1.01){\circle*{0.2}}
\put( 10.83,  1.24){\circle*{0.2}}
\put( 10.99,  1.45){\circle*{0.2}}
\put( 11.15,  1.64){\circle*{0.2}}
\put( 11.31,  1.82){\circle*{0.2}}
\put( 11.46,  1.98){\circle*{0.2}}
\put( 11.62,  2.13){\circle*{0.2}}
\put( 11.78,  2.25){\circle*{0.2}}
\put( 11.94,  2.34){\circle*{0.2}}
\put( 12.10,  2.42){\circle*{0.2}}
\put( 12.26,  2.47){\circle*{0.2}}
\put( 12.42,  2.50){\circle*{0.2}}
\put( 12.58,  2.50){\circle*{0.2}}
\put( 12.74,  2.47){\circle*{0.2}}
\put( 12.90,  2.42){\circle*{0.2}}
\put( 13.06,  2.35){\circle*{0.2}}
\put( 13.22,  2.26){\circle*{0.2}}
\put( 13.38,  2.14){\circle*{0.2}}
\put( 13.54,  2.00){\circle*{0.2}}
\put( 13.69,  1.84){\circle*{0.2}}
\put( 13.85,  1.66){\circle*{0.2}}
\put( 14.01,  1.46){\circle*{0.2}}
\put( 14.17,  1.25){\circle*{0.2}}
\put( 14.33,  1.03){\circle*{0.2}}
\put( 14.49,  0.80){\circle*{0.2}}
\put( 14.65,  0.56){\circle*{0.2}}
\put( 14.81,  0.31){\circle*{0.2}}
\put( 14.97,  0.06){\circle*{0.2}}
\put( 15.13, -0.19){\circle*{0.2}}
\put( 15.29, -0.44){\circle*{0.2}}
\put( 15.45, -0.68){\circle*{0.2}}
\put( 15.61, -0.92){\circle*{0.2}}
\put( 15.76, -1.14){\circle*{0.2}}
\put( 15.92, -1.36){\circle*{0.2}}
\put( 16.08, -1.56){\circle*{0.2}}
\put( 16.24, -1.75){\circle*{0.2}}
\put( 16.40, -1.92){\circle*{0.2}}
\put( 16.56, -2.07){\circle*{0.2}}
\put( 16.72, -2.20){\circle*{0.2}}
\put( 16.88, -2.31){\circle*{0.2}}
\put( 17.04, -2.39){\circle*{0.2}}
\put( 17.20, -2.45){\circle*{0.2}}
\put( 17.36, -2.49){\circle*{0.2}}
\put( 17.52, -2.50){\circle*{0.2}}
\put( 17.68, -2.49){\circle*{0.2}}
\put( 17.83, -2.45){\circle*{0.2}}
\put( 17.99, -2.39){\circle*{0.2}}
\put( 18.15, -2.30){\circle*{0.2}}
\put( 18.31, -2.19){\circle*{0.2}}
\put( 18.47, -2.06){\circle*{0.2}}
\put( 18.63, -1.90){\circle*{0.2}}
\put( 18.79, -1.73){\circle*{0.2}}
\put( 18.95, -1.55){\circle*{0.2}}
\put( 19.11, -1.34){\circle*{0.2}}
\put( 19.27, -1.12){\circle*{0.2}}
\put( 19.43, -0.90){\circle*{0.2}}
\put( 19.59, -0.66){\circle*{0.2}}
\put( 19.75, -0.41){\circle*{0.2}}
\put( 19.90, -0.17){\circle*{0.2}}
}

\newsavebox{\vgluon}
\savebox{\vgluon}(0,0)[b]{
\put(  0.00,  0.00){\circle*{0.2}}
\put( -0.20,  0.13){\circle*{0.2}}
\put( -0.40,  0.25){\circle*{0.2}}
\put( -0.59,  0.38){\circle*{0.2}}
\put( -0.79,  0.51){\circle*{0.2}}
\put( -0.97,  0.64){\circle*{0.2}}
\put( -1.15,  0.76){\circle*{0.2}}
\put( -1.33,  0.89){\circle*{0.2}}
\put( -1.49,  1.02){\circle*{0.2}}
\put( -1.65,  1.15){\circle*{0.2}}
\put( -1.79,  1.27){\circle*{0.2}}
\put( -1.93,  1.40){\circle*{0.2}}
\put( -2.05,  1.53){\circle*{0.2}}
\put( -2.16,  1.66){\circle*{0.2}}
\put( -2.25,  1.78){\circle*{0.2}}
\put( -2.33,  1.91){\circle*{0.2}}
\put( -2.40,  2.04){\circle*{0.2}}
\put( -2.44,  2.17){\circle*{0.2}}
\put( -2.48,  2.29){\circle*{0.2}}
\put( -2.50,  2.42){\circle*{0.2}}
\put( -2.50,  2.55){\circle*{0.2}}
\put( -2.49,  2.68){\circle*{0.2}}
\put( -2.46,  2.80){\circle*{0.2}}
\put( -2.41,  2.93){\circle*{0.2}}
\put( -2.35,  3.06){\circle*{0.2}}
\put( -2.27,  3.18){\circle*{0.2}}
\put( -2.18,  3.31){\circle*{0.2}}
\put( -2.08,  3.44){\circle*{0.2}}
\put( -1.96,  3.57){\circle*{0.2}}
\put( -1.83,  3.69){\circle*{0.2}}
\put( -1.69,  3.82){\circle*{0.2}}
\put( -1.54,  3.95){\circle*{0.2}}
\put( -1.37,  4.08){\circle*{0.2}}
\put( -1.20,  4.20){\circle*{0.2}}
\put( -1.02,  4.33){\circle*{0.2}}
\put( -0.84,  4.46){\circle*{0.2}}
\put( -0.65,  4.59){\circle*{0.2}}
\put( -0.45,  4.71){\circle*{0.2}}
\put( -0.25,  4.84){\circle*{0.2}}
\put( -0.05,  4.97){\circle*{0.2}}
\put(  0.15,  5.10){\circle*{0.2}}
\put(  0.34,  5.22){\circle*{0.2}}
\put(  0.54,  5.35){\circle*{0.2}}
\put(  0.73,  5.48){\circle*{0.2}}
\put(  0.92,  5.61){\circle*{0.2}}
\put(  1.11,  5.73){\circle*{0.2}}
\put(  1.28,  5.86){\circle*{0.2}}
\put(  1.45,  5.99){\circle*{0.2}}
\put(  1.61,  6.11){\circle*{0.2}}
\put(  1.76,  6.24){\circle*{0.2}}
\put(  1.89,  6.37){\circle*{0.2}}
\put(  2.02,  6.50){\circle*{0.2}}
\put(  2.13,  6.62){\circle*{0.2}}
\put(  2.23,  6.75){\circle*{0.2}}
\put(  2.31,  6.88){\circle*{0.2}}
\put(  2.38,  7.01){\circle*{0.2}}
\put(  2.43,  7.13){\circle*{0.2}}
\put(  2.47,  7.26){\circle*{0.2}}
\put(  2.49,  7.39){\circle*{0.2}}
\put(  2.50,  7.52){\circle*{0.2}}
\put(  2.49,  7.64){\circle*{0.2}}
\put(  2.46,  7.77){\circle*{0.2}}
\put(  2.42,  7.90){\circle*{0.2}}
\put(  2.37,  8.03){\circle*{0.2}}
\put(  2.30,  8.15){\circle*{0.2}}
\put(  2.21,  8.28){\circle*{0.2}}
\put(  2.11,  8.41){\circle*{0.2}}
\put(  1.99,  8.54){\circle*{0.2}}
\put(  1.87,  8.66){\circle*{0.2}}
\put(  1.73,  8.79){\circle*{0.2}}
\put(  1.58,  8.92){\circle*{0.2}}
\put(  1.42,  9.04){\circle*{0.2}}
\put(  1.25,  9.17){\circle*{0.2}}
\put(  1.07,  9.30){\circle*{0.2}}
\put(  0.89,  9.43){\circle*{0.2}}
\put(  0.70,  9.55){\circle*{0.2}}
\put(  0.50,  9.68){\circle*{0.2}}
\put(  0.31,  9.81){\circle*{0.2}}
\put(  0.11,  9.94){\circle*{0.2}}
\put( -0.09, 10.06){\circle*{0.2}}
\put( -0.29, 10.19){\circle*{0.2}}
\put( -0.49, 10.32){\circle*{0.2}}
\put( -0.68, 10.45){\circle*{0.2}}
\put( -0.87, 10.57){\circle*{0.2}}
\put( -1.06, 10.70){\circle*{0.2}}
\put( -1.24, 10.83){\circle*{0.2}}
\put( -1.41, 10.96){\circle*{0.2}}
\put( -1.57, 11.08){\circle*{0.2}}
\put( -1.72, 11.21){\circle*{0.2}}
\put( -1.86, 11.34){\circle*{0.2}}
\put( -1.98, 11.46){\circle*{0.2}}
\put( -2.10, 11.59){\circle*{0.2}}
\put( -2.20, 11.72){\circle*{0.2}}
\put( -2.29, 11.85){\circle*{0.2}}
\put( -2.36, 11.97){\circle*{0.2}}
\put( -2.42, 12.10){\circle*{0.2}}
\put( -2.46, 12.23){\circle*{0.2}}
\put( -2.49, 12.36){\circle*{0.2}}
\put( -2.50, 12.48){\circle*{0.2}}
\put( -2.49, 12.61){\circle*{0.2}}
\put( -2.47, 12.74){\circle*{0.2}}
\put( -2.44, 12.87){\circle*{0.2}}
\put( -2.38, 12.99){\circle*{0.2}}
\put( -2.32, 13.12){\circle*{0.2}}
\put( -2.23, 13.25){\circle*{0.2}}
\put( -2.14, 13.38){\circle*{0.2}}
\put( -2.03, 13.50){\circle*{0.2}}
\put( -1.90, 13.63){\circle*{0.2}}
\put( -1.77, 13.76){\circle*{0.2}}
\put( -1.62, 13.89){\circle*{0.2}}
\put( -1.46, 14.01){\circle*{0.2}}
\put( -1.30, 14.14){\circle*{0.2}}
\put( -1.12, 14.27){\circle*{0.2}}
\put( -0.94, 14.39){\circle*{0.2}}
\put( -0.75, 14.52){\circle*{0.2}}
\put( -0.56, 14.65){\circle*{0.2}}
\put( -0.36, 14.78){\circle*{0.2}}
\put( -0.16, 14.90){\circle*{0.2}}
\put(  0.04, 15.03){\circle*{0.2}}
\put(  0.24, 15.16){\circle*{0.2}}
\put(  0.44, 15.29){\circle*{0.2}}
\put(  0.63, 15.41){\circle*{0.2}}
\put(  0.82, 15.54){\circle*{0.2}}
\put(  1.01, 15.67){\circle*{0.2}}
\put(  1.19, 15.80){\circle*{0.2}}
\put(  1.36, 15.92){\circle*{0.2}}
\put(  1.52, 16.05){\circle*{0.2}}
\put(  1.68, 16.18){\circle*{0.2}}
\put(  1.82, 16.31){\circle*{0.2}}
\put(  1.95, 16.43){\circle*{0.2}}
\put(  2.07, 16.56){\circle*{0.2}}
\put(  2.18, 16.69){\circle*{0.2}}
\put(  2.27, 16.82){\circle*{0.2}}
\put(  2.34, 16.94){\circle*{0.2}}
\put(  2.41, 17.07){\circle*{0.2}}
\put(  2.45, 17.20){\circle*{0.2}}
\put(  2.48, 17.32){\circle*{0.2}}
\put(  2.50, 17.45){\circle*{0.2}}
\put(  2.50, 17.58){\circle*{0.2}}
\put(  2.48, 17.71){\circle*{0.2}}
\put(  2.45, 17.83){\circle*{0.2}}
\put(  2.40, 17.96){\circle*{0.2}}
\put(  2.34, 18.09){\circle*{0.2}}
\put(  2.26, 18.22){\circle*{0.2}}
\put(  2.16, 18.34){\circle*{0.2}}
\put(  2.06, 18.47){\circle*{0.2}}
\put(  1.94, 18.60){\circle*{0.2}}
\put(  1.80, 18.73){\circle*{0.2}}
\put(  1.66, 18.85){\circle*{0.2}}
\put(  1.51, 18.98){\circle*{0.2}}
\put(  1.34, 19.11){\circle*{0.2}}
\put(  1.17, 19.24){\circle*{0.2}}
\put(  0.99, 19.36){\circle*{0.2}}
\put(  0.80, 19.49){\circle*{0.2}}
\put(  0.61, 19.62){\circle*{0.2}}
\put(  0.41, 19.75){\circle*{0.2}}
\put(  0.22, 19.87){\circle*{0.2}}
}

\newsavebox{\trgluonone}
\savebox{\trgluonone}(0,0)[bl]{
\put(  0.00,  0.00){\circle*{0.2}}
\put(  0.03,  0.26){\circle*{0.2}}
\put(  0.06,  0.53){\circle*{0.2}}
\put(  0.08,  0.79){\circle*{0.2}}
\put(  0.12,  1.04){\circle*{0.2}}
\put(  0.15,  1.29){\circle*{0.2}}
\put(  0.19,  1.53){\circle*{0.2}}
\put(  0.24,  1.76){\circle*{0.2}}
\put(  0.28,  1.98){\circle*{0.2}}
\put(  0.34,  2.19){\circle*{0.2}}
\put(  0.40,  2.39){\circle*{0.2}}
\put(  0.47,  2.57){\circle*{0.2}}
\put(  0.55,  2.74){\circle*{0.2}}
\put(  0.63,  2.89){\circle*{0.2}}
\put(  0.72,  3.02){\circle*{0.2}}
\put(  0.82,  3.14){\circle*{0.2}}
\put(  0.93,  3.24){\circle*{0.2}}
\put(  1.05,  3.32){\circle*{0.2}}
\put(  1.18,  3.38){\circle*{0.2}}
\put(  1.32,  3.43){\circle*{0.2}}
\put(  1.47,  3.45){\circle*{0.2}}
\put(  1.62,  3.46){\circle*{0.2}}
\put(  1.78,  3.45){\circle*{0.2}}
\put(  1.96,  3.43){\circle*{0.2}}
\put(  2.14,  3.39){\circle*{0.2}}
\put(  2.32,  3.33){\circle*{0.2}}
\put(  2.52,  3.27){\circle*{0.2}}
\put(  2.72,  3.18){\circle*{0.2}}
\put(  2.92,  3.09){\circle*{0.2}}
\put(  3.14,  2.98){\circle*{0.2}}
\put(  3.35,  2.87){\circle*{0.2}}
\put(  3.57,  2.75){\circle*{0.2}}
\put(  3.80,  2.62){\circle*{0.2}}
\put(  4.02,  2.49){\circle*{0.2}}
\put(  4.25,  2.35){\circle*{0.2}}
\put(  4.48,  2.22){\circle*{0.2}}
\put(  4.71,  2.08){\circle*{0.2}}
\put(  4.93,  1.94){\circle*{0.2}}
}

\newsavebox{\trgluontwo}
\savebox{\trgluontwo}(0,0)[bl]{
\put(  5.16,  1.81){\circle*{0.2}}
\put(  5.38,  1.68){\circle*{0.2}}
\put(  5.60,  1.56){\circle*{0.2}}
\put(  5.82,  1.45){\circle*{0.2}}
\put(  6.03,  1.35){\circle*{0.2}}
\put(  6.23,  1.26){\circle*{0.2}}
\put(  6.43,  1.18){\circle*{0.2}}
\put(  6.63,  1.11){\circle*{0.2}}
\put(  6.81,  1.06){\circle*{0.2}}
\put(  6.99,  1.02){\circle*{0.2}}
\put(  7.16,  1.00){\circle*{0.2}}
\put(  7.32,  0.99){\circle*{0.2}}
\put(  7.48,  1.01){\circle*{0.2}}
\put(  7.62,  1.04){\circle*{0.2}}
\put(  7.76,  1.09){\circle*{0.2}}
\put(  7.88,  1.15){\circle*{0.2}}
\put(  8.00,  1.24){\circle*{0.2}}
\put(  8.11,  1.34){\circle*{0.2}}
\put(  8.21,  1.46){\circle*{0.2}}
\put(  8.30,  1.60){\circle*{0.2}}
\put(  8.38,  1.75){\circle*{0.2}}
\put(  8.46,  1.92){\circle*{0.2}}
\put(  8.52,  2.10){\circle*{0.2}}
\put(  8.58,  2.30){\circle*{0.2}}
\put(  8.64,  2.51){\circle*{0.2}}
\put(  8.69,  2.74){\circle*{0.2}}
\put(  8.73,  2.97){\circle*{0.2}}
\put(  8.77,  3.21){\circle*{0.2}}
\put(  8.80,  3.46){\circle*{0.2}}
\put(  8.83,  3.72){\circle*{0.2}}
\put(  8.86,  3.98){\circle*{0.2}}
\put(  8.89,  4.24){\circle*{0.2}}
\put(  8.92,  4.51){\circle*{0.2}}
\put(  8.95,  4.77){\circle*{0.2}}
\put(  8.97,  5.03){\circle*{0.2}}
\put(  9.00,  5.29){\circle*{0.2}}
\put(  9.04,  5.55){\circle*{0.2}}
\put(  9.07,  5.79){\circle*{0.2}}
\put(  9.11,  6.03){\circle*{0.2}}
\put(  9.16,  6.26){\circle*{0.2}}
\put(  9.21,  6.48){\circle*{0.2}}
\put(  9.26,  6.69){\circle*{0.2}}
\put(  9.33,  6.88){\circle*{0.2}}
\put(  9.40,  7.06){\circle*{0.2}}
\put(  9.47,  7.22){\circle*{0.2}}
\put(  9.56,  7.37){\circle*{0.2}}
\put(  9.66,  7.50){\circle*{0.2}}
\put(  9.76,  7.61){\circle*{0.2}}
\put(  9.87,  7.71){\circle*{0.2}}
\put(  9.99,  7.79){\circle*{0.2}}
\put( 10.12,  7.85){\circle*{0.2}}
\put( 10.26,  7.89){\circle*{0.2}}
\put( 10.41,  7.91){\circle*{0.2}}
\put( 10.56,  7.92){\circle*{0.2}}
\put( 10.73,  7.91){\circle*{0.2}}
\put( 10.90,  7.88){\circle*{0.2}}
\put( 11.08,  7.84){\circle*{0.2}}
\put( 11.27,  7.78){\circle*{0.2}}
\put( 11.47,  7.71){\circle*{0.2}}
\put( 11.67,  7.62){\circle*{0.2}}
\put( 11.88,  7.53){\circle*{0.2}}
\put( 12.09,  7.42){\circle*{0.2}}
\put( 12.31,  7.30){\circle*{0.2}}
\put( 12.53,  7.18){\circle*{0.2}}
\put( 12.75,  7.05){\circle*{0.2}}
\put( 12.98,  6.92){\circle*{0.2}}
\put( 13.21,  6.78){\circle*{0.2}}
\put( 13.43,  6.65){\circle*{0.2}}
\put( 13.66,  6.51){\circle*{0.2}}
\put( 13.89,  6.38){\circle*{0.2}}
\put( 14.11,  6.24){\circle*{0.2}}
\put( 14.34,  6.12){\circle*{0.2}}
\put( 14.56,  6.00){\circle*{0.2}}
\put( 14.77,  5.89){\circle*{0.2}}
\put( 14.98,  5.79){\circle*{0.2}}
\put( 15.18,  5.70){\circle*{0.2}}
\put( 15.38,  5.62){\circle*{0.2}}
\put( 15.58,  5.56){\circle*{0.2}}
\put( 15.76,  5.51){\circle*{0.2}}
\put( 15.94,  5.47){\circle*{0.2}}
\put( 16.11,  5.45){\circle*{0.2}}
\put( 16.27,  5.45){\circle*{0.2}}
\put( 16.42,  5.47){\circle*{0.2}}
\put( 16.56,  5.50){\circle*{0.2}}
\put( 16.69,  5.55){\circle*{0.2}}
\put( 16.82,  5.62){\circle*{0.2}}
\put( 16.93,  5.71){\circle*{0.2}}
\put( 17.04,  5.82){\circle*{0.2}}
\put( 17.14,  5.94){\circle*{0.2}}
\put( 17.23,  6.08){\circle*{0.2}}
\put( 17.31,  6.24){\circle*{0.2}}
\put( 17.38,  6.41){\circle*{0.2}}
\put( 17.45,  6.60){\circle*{0.2}}
\put( 17.51,  6.80){\circle*{0.2}}
\put( 17.56,  7.01){\circle*{0.2}}
\put( 17.61,  7.24){\circle*{0.2}}
\put( 17.65,  7.47){\circle*{0.2}}
\put( 17.69,  7.71){\circle*{0.2}}
\put( 17.72,  7.97){\circle*{0.2}}
\put( 17.75,  8.22){\circle*{0.2}}
\put( 17.78,  8.48){\circle*{0.2}}
\put( 17.81,  8.75){\circle*{0.2}}
\put( 17.84,  9.01){\circle*{0.2}}
\put( 17.86,  9.28){\circle*{0.2}}
\put( 17.89,  9.54){\circle*{0.2}}
\put( 17.92,  9.80){\circle*{0.2}}
\put( 17.96, 10.05){\circle*{0.2}}
\put( 17.99, 10.30){\circle*{0.2}}
\put( 18.03, 10.53){\circle*{0.2}}
\put( 18.08, 10.76){\circle*{0.2}}
\put( 18.13, 10.98){\circle*{0.2}}
\put( 18.19, 11.18){\circle*{0.2}}
\put( 18.25, 11.37){\circle*{0.2}}
\put( 18.32, 11.55){\circle*{0.2}}
\put( 18.40, 11.71){\circle*{0.2}}
\put( 18.49, 11.85){\circle*{0.2}}
\put( 18.59, 11.98){\circle*{0.2}}
\put( 18.69, 12.09){\circle*{0.2}}
\put( 18.80, 12.18){\circle*{0.2}}
\put( 18.93, 12.26){\circle*{0.2}}
\put( 19.06, 12.31){\circle*{0.2}}
\put( 19.20, 12.35){\circle*{0.2}}
\put( 19.35, 12.37){\circle*{0.2}}
\put( 19.51, 12.38){\circle*{0.2}}
\put( 19.67, 12.36){\circle*{0.2}}
\put( 19.85, 12.33){\circle*{0.2}}
\put( 20.03, 12.28){\circle*{0.2}}
\put( 20.22, 12.22){\circle*{0.2}}
\put( 20.42, 12.15){\circle*{0.2}}
\put( 20.62, 12.06){\circle*{0.2}}
\put( 20.83, 11.96){\circle*{0.2}}
\put( 21.04, 11.86){\circle*{0.2}}
\put( 21.26, 11.74){\circle*{0.2}}
\put( 21.48, 11.61){\circle*{0.2}}
\put( 21.71, 11.49){\circle*{0.2}}
\put( 21.93, 11.35){\circle*{0.2}}
\put( 22.16, 11.21){\circle*{0.2}}
\put( 22.39, 11.08){\circle*{0.2}}
\put( 22.62, 10.94){\circle*{0.2}}
\put( 22.84, 10.81){\circle*{0.2}}
\put( 23.07, 10.68){\circle*{0.2}}
\put( 23.29, 10.55){\circle*{0.2}}
\put( 23.51, 10.43){\circle*{0.2}}
\put( 23.72, 10.33){\circle*{0.2}}
\put( 23.93, 10.23){\circle*{0.2}}
\put( 24.14, 10.14){\circle*{0.2}}
\put( 24.33, 10.06){\circle*{0.2}}
\put( 24.52, 10.00){\circle*{0.2}}
\put( 24.71,  9.96){\circle*{0.2}}
\put( 24.88,  9.92){\circle*{0.2}}
\put( 25.05,  9.91){\circle*{0.2}}
\put( 25.21,  9.91){\circle*{0.2}}
\put( 25.36,  9.93){\circle*{0.2}}
\put( 25.50,  9.97){\circle*{0.2}}
\put( 25.63, 10.02){\circle*{0.2}}
\put( 25.75, 10.09){\circle*{0.2}}
\put( 25.87, 10.19){\circle*{0.2}}
\put( 25.97, 10.29){\circle*{0.2}}
\put( 26.07, 10.42){\circle*{0.2}}
\put( 26.16, 10.56){\circle*{0.2}}
\put( 26.24, 10.72){\circle*{0.2}}
\put( 26.31, 10.90){\circle*{0.2}}
\put( 26.37, 11.09){\circle*{0.2}}
\put( 26.43, 11.29){\circle*{0.2}}
\put( 26.48, 11.51){\circle*{0.2}}
\put( 26.53, 11.73){\circle*{0.2}}
\put( 26.57, 11.97){\circle*{0.2}}
\put( 26.61, 12.22){\circle*{0.2}}
\put( 26.64, 12.47){\circle*{0.2}}
\put( 26.67, 12.73){\circle*{0.2}}
\put( 26.70, 12.99){\circle*{0.2}}
\put( 26.73, 13.25){\circle*{0.2}}
\put( 26.76, 13.52){\circle*{0.2}}
\put( 26.78, 13.78){\circle*{0.2}}
\put( 26.81, 14.04){\circle*{0.2}}
\put( 26.84, 14.30){\circle*{0.2}}
\put( 26.88, 14.55){\circle*{0.2}}
\put( 26.91, 14.80){\circle*{0.2}}
\put( 26.96, 15.03){\circle*{0.2}}
\put( 27.00, 15.26){\circle*{0.2}}
\put( 27.05, 15.47){\circle*{0.2}}
\put( 27.11, 15.67){\circle*{0.2}}
\put( 27.18, 15.86){\circle*{0.2}}
\put( 27.25, 16.04){\circle*{0.2}}
\put( 27.33, 16.19){\circle*{0.2}}
\put( 27.42, 16.33){\circle*{0.2}}
\put( 27.52, 16.46){\circle*{0.2}}
\put( 27.63, 16.57){\circle*{0.2}}
\put( 27.74, 16.65){\circle*{0.2}}
\put( 27.86, 16.73){\circle*{0.2}}
\put( 28.00, 16.78){\circle*{0.2}}
\put( 28.14, 16.81){\circle*{0.2}}
\put( 28.29, 16.83){\circle*{0.2}}
\put( 28.45, 16.83){\circle*{0.2}}
\put( 28.62, 16.81){\circle*{0.2}}
\put( 28.80, 16.78){\circle*{0.2}}
\put( 28.98, 16.73){\circle*{0.2}}
\put( 29.17, 16.67){\circle*{0.2}}
\put( 29.37, 16.59){\circle*{0.2}}
\put( 29.57, 16.50){\circle*{0.2}}
\put( 29.78, 16.40){\circle*{0.2}}
\put( 30.00, 16.29){\circle*{0.2}}
\put( 30.22, 16.17){\circle*{0.2}}
\put( 30.44, 16.05){\circle*{0.2}}
\put( 30.66, 15.92){\circle*{0.2}}
\put( 30.89, 15.78){\circle*{0.2}}
\put( 31.12, 15.65){\circle*{0.2}}
\put( 31.35, 15.51){\circle*{0.2}}
\put( 31.57, 15.37){\circle*{0.2}}
\put( 31.80, 15.24){\circle*{0.2}}
\put( 32.02, 15.11){\circle*{0.2}}
\put( 32.25, 14.99){\circle*{0.2}}
\put( 32.46, 14.87){\circle*{0.2}}
\put( 32.68, 14.76){\circle*{0.2}}
\put( 32.88, 14.67){\circle*{0.2}}
\put( 33.09, 14.58){\circle*{0.2}}
\put( 33.28, 14.51){\circle*{0.2}}
\put( 33.47, 14.45){\circle*{0.2}}
\put( 33.65, 14.40){\circle*{0.2}}
\put( 33.83, 14.38){\circle*{0.2}}
\put( 33.99, 14.36){\circle*{0.2}}
\put( 34.15, 14.37){\circle*{0.2}}
\put( 34.30, 14.39){\circle*{0.2}}
\put( 34.44, 14.43){\circle*{0.2}}
\put( 34.57, 14.49){\circle*{0.2}}
\put( 34.69, 14.57){\circle*{0.2}}
\put( 34.80, 14.66){\circle*{0.2}}
\put( 34.90, 14.77){\circle*{0.2}}
\put( 35.00, 14.90){\circle*{0.2}}
\put( 35.09, 15.05){\circle*{0.2}}
\put( 35.16, 15.21){\circle*{0.2}}
\put( 35.24, 15.39){\circle*{0.2}}
\put( 35.30, 15.58){\circle*{0.2}}
\put( 35.36, 15.79){\circle*{0.2}}
\put( 35.41, 16.01){\circle*{0.2}}
\put( 35.45, 16.24){\circle*{0.2}}
\put( 35.49, 16.47){\circle*{0.2}}
\put( 35.53, 16.72){\circle*{0.2}}
\put( 35.56, 16.97){\circle*{0.2}}
\put( 35.59, 17.23){\circle*{0.2}}
\put( 35.62, 17.50){\circle*{0.2}}
\put( 35.65, 17.76){\circle*{0.2}}
\put( 35.67, 18.02){\circle*{0.2}}
\put( 35.70, 18.29){\circle*{0.2}}
\put( 35.73, 18.55){\circle*{0.2}}
\put( 35.76, 18.81){\circle*{0.2}}
\put( 35.80, 19.06){\circle*{0.2}}
\put( 35.83, 19.30){\circle*{0.2}}
\put( 35.88, 19.53){\circle*{0.2}}
\put( 35.92, 19.76){\circle*{0.2}}
\put( 35.98, 19.97){\circle*{0.2}}
\put( 36.04, 20.17){\circle*{0.2}}
\put( 36.11, 20.35){\circle*{0.2}}
\put( 36.18, 20.52){\circle*{0.2}}
\put( 36.26, 20.68){\circle*{0.2}}
\put( 36.35, 20.82){\circle*{0.2}}
\put( 36.45, 20.94){\circle*{0.2}}
\put( 36.56, 21.04){\circle*{0.2}}
\put( 36.68, 21.13){\circle*{0.2}}
\put( 36.80, 21.19){\circle*{0.2}}
\put( 36.94, 21.24){\circle*{0.2}}
\put( 37.08, 21.27){\circle*{0.2}}
\put( 37.23, 21.29){\circle*{0.2}}
\put( 37.39, 21.29){\circle*{0.2}}
\put( 37.56, 21.26){\circle*{0.2}}
\put( 37.74, 21.23){\circle*{0.2}}
\put( 37.93, 21.18){\circle*{0.2}}
\put( 38.12, 21.11){\circle*{0.2}}
\put( 38.32, 21.03){\circle*{0.2}}
\put( 38.52, 20.94){\circle*{0.2}}
\put( 38.74, 20.84){\circle*{0.2}}
\put( 38.95, 20.73){\circle*{0.2}}
\put( 39.17, 20.61){\circle*{0.2}}
\put( 39.39, 20.48){\circle*{0.2}}
\put( 39.62, 20.35){\circle*{0.2}}
\put( 39.85, 20.21){\circle*{0.2}}
\put( 40.07, 20.08){\circle*{0.2}}
}

\newsavebox{\blgluonone}
\savebox{\blgluonone}(0,0)[tr]{
\put(   0.00,  -0.00){\circle*{0.2}}
\put( - 0.03,  -0.26){\circle*{0.2}}
\put( - 0.06,  -0.53){\circle*{0.2}}
\put( - 0.08,  -0.79){\circle*{0.2}}
\put( - 0.12,  -1.04){\circle*{0.2}}
\put( - 0.15,  -1.29){\circle*{0.2}}
\put( - 0.19,  -1.53){\circle*{0.2}}
\put( - 0.24,  -1.76){\circle*{0.2}}
\put( - 0.28,  -1.98){\circle*{0.2}}
\put( - 0.34,  -2.19){\circle*{0.2}}
\put( - 0.40,  -2.39){\circle*{0.2}}
\put( - 0.47,  -2.57){\circle*{0.2}}
\put( - 0.55,  -2.74){\circle*{0.2}}
\put( - 0.63,  -2.89){\circle*{0.2}}
\put( - 0.72,  -3.02){\circle*{0.2}}
\put( - 0.82,  -3.14){\circle*{0.2}}
\put( - 0.93,  -3.24){\circle*{0.2}}
\put( - 1.05,  -3.32){\circle*{0.2}}
\put( - 1.18,  -3.38){\circle*{0.2}}
\put( - 1.32,  -3.43){\circle*{0.2}}
\put( - 1.47,  -3.45){\circle*{0.2}}
\put( - 1.62,  -3.46){\circle*{0.2}}
\put( - 1.78,  -3.45){\circle*{0.2}}
\put( - 1.96,  -3.43){\circle*{0.2}}
\put( - 2.14,  -3.39){\circle*{0.2}}
\put( - 2.32,  -3.33){\circle*{0.2}}
\put( - 2.52,  -3.27){\circle*{0.2}}
\put( - 2.72,  -3.18){\circle*{0.2}}
\put( - 2.92,  -3.09){\circle*{0.2}}
\put( - 3.14,  -2.98){\circle*{0.2}}
\put( - 3.35,  -2.87){\circle*{0.2}}
\put( - 3.57,  -2.75){\circle*{0.2}}
\put( - 3.80,  -2.62){\circle*{0.2}}
\put( - 4.02,  -2.49){\circle*{0.2}}
\put( - 4.25,  -2.35){\circle*{0.2}}
\put( - 4.48,  -2.22){\circle*{0.2}}
\put( - 4.71,  -2.08){\circle*{0.2}}
\put( - 4.93,  -1.94){\circle*{0.2}}
}

\newsavebox{\blgluontwo}
\savebox{\blgluontwo}(0,0)[tr]{
\put( - 5.16,  -1.81){\circle*{0.2}}
\put( - 5.38,  -1.68){\circle*{0.2}}
\put( - 5.60,  -1.56){\circle*{0.2}}
\put( - 5.82,  -1.45){\circle*{0.2}}
\put( - 6.03,  -1.35){\circle*{0.2}}
\put( - 6.23,  -1.26){\circle*{0.2}}
\put( - 6.43,  -1.18){\circle*{0.2}}
\put( - 6.63,  -1.11){\circle*{0.2}}
\put( - 6.81,  -1.06){\circle*{0.2}}
\put( - 6.99,  -1.02){\circle*{0.2}}
\put( - 7.16,  -1.00){\circle*{0.2}}
\put( - 7.32,  -0.99){\circle*{0.2}}
\put( - 7.48,  -1.01){\circle*{0.2}}
\put( - 7.62,  -1.04){\circle*{0.2}}
\put( - 7.76,  -1.09){\circle*{0.2}}
\put( - 7.88,  -1.15){\circle*{0.2}}
\put( - 8.00,  -1.24){\circle*{0.2}}
\put( - 8.11,  -1.34){\circle*{0.2}}
\put( - 8.21,  -1.46){\circle*{0.2}}
\put( - 8.30,  -1.60){\circle*{0.2}}
\put( - 8.38,  -1.75){\circle*{0.2}}
\put( - 8.46,  -1.92){\circle*{0.2}}
\put( - 8.52,  -2.10){\circle*{0.2}}
\put( - 8.58,  -2.30){\circle*{0.2}}
\put( - 8.64,  -2.51){\circle*{0.2}}
\put( - 8.69,  -2.74){\circle*{0.2}}
\put( - 8.73,  -2.97){\circle*{0.2}}
\put( - 8.77,  -3.21){\circle*{0.2}}
\put( - 8.80,  -3.46){\circle*{0.2}}
\put( - 8.83,  -3.72){\circle*{0.2}}
\put( - 8.86,  -3.98){\circle*{0.2}}
\put( - 8.89,  -4.24){\circle*{0.2}}
\put( - 8.92,  -4.51){\circle*{0.2}}
\put( - 8.95,  -4.77){\circle*{0.2}}
\put( - 8.97,  -5.03){\circle*{0.2}}
\put( - 9.00,  -5.29){\circle*{0.2}}
\put( - 9.04,  -5.55){\circle*{0.2}}
\put( - 9.07,  -5.79){\circle*{0.2}}
\put( - 9.11,  -6.03){\circle*{0.2}}
\put( - 9.16,  -6.26){\circle*{0.2}}
\put( - 9.21,  -6.48){\circle*{0.2}}
\put( - 9.26,  -6.69){\circle*{0.2}}
\put( - 9.33,  -6.88){\circle*{0.2}}
\put( - 9.40,  -7.06){\circle*{0.2}}
\put( - 9.47,  -7.22){\circle*{0.2}}
\put( - 9.56,  -7.37){\circle*{0.2}}
\put( - 9.66,  -7.50){\circle*{0.2}}
\put( - 9.76,  -7.61){\circle*{0.2}}
\put( - 9.87,  -7.71){\circle*{0.2}}
\put( - 9.99,  -7.79){\circle*{0.2}}
\put( -10.12,  -7.85){\circle*{0.2}}
\put( -10.26,  -7.89){\circle*{0.2}}
\put( -10.41,  -7.91){\circle*{0.2}}
\put( -10.56,  -7.92){\circle*{0.2}}
\put( -10.73,  -7.91){\circle*{0.2}}
\put( -10.90,  -7.88){\circle*{0.2}}
\put( -11.08,  -7.84){\circle*{0.2}}
\put( -11.27,  -7.78){\circle*{0.2}}
\put( -11.47,  -7.71){\circle*{0.2}}
\put( -11.67,  -7.62){\circle*{0.2}}
\put( -11.88,  -7.53){\circle*{0.2}}
\put( -12.09,  -7.42){\circle*{0.2}}
\put( -12.31,  -7.30){\circle*{0.2}}
\put( -12.53,  -7.18){\circle*{0.2}}
\put( -12.75,  -7.05){\circle*{0.2}}
\put( -12.98,  -6.92){\circle*{0.2}}
\put( -13.21,  -6.78){\circle*{0.2}}
\put( -13.43,  -6.65){\circle*{0.2}}
\put( -13.66,  -6.51){\circle*{0.2}}
\put( -13.89,  -6.38){\circle*{0.2}}
\put( -14.11,  -6.24){\circle*{0.2}}
\put( -14.34,  -6.12){\circle*{0.2}}
\put( -14.56,  -6.00){\circle*{0.2}}
\put( -14.77,  -5.89){\circle*{0.2}}
\put( -14.98,  -5.79){\circle*{0.2}}
\put( -15.18,  -5.70){\circle*{0.2}}
\put( -15.38,  -5.62){\circle*{0.2}}
\put( -15.58,  -5.56){\circle*{0.2}}
\put( -15.76,  -5.51){\circle*{0.2}}
\put( -15.94,  -5.47){\circle*{0.2}}
\put( -16.11,  -5.45){\circle*{0.2}}
\put( -16.27,  -5.45){\circle*{0.2}}
\put( -16.42,  -5.47){\circle*{0.2}}
\put( -16.56,  -5.50){\circle*{0.2}}
\put( -16.69,  -5.55){\circle*{0.2}}
\put( -16.82,  -5.62){\circle*{0.2}}
\put( -16.93,  -5.71){\circle*{0.2}}
\put( -17.04,  -5.82){\circle*{0.2}}
\put( -17.14,  -5.94){\circle*{0.2}}
\put( -17.23,  -6.08){\circle*{0.2}}
\put( -17.31,  -6.24){\circle*{0.2}}
\put( -17.38,  -6.41){\circle*{0.2}}
\put( -17.45,  -6.60){\circle*{0.2}}
\put( -17.51,  -6.80){\circle*{0.2}}
\put( -17.56,  -7.01){\circle*{0.2}}
\put( -17.61,  -7.24){\circle*{0.2}}
\put( -17.65,  -7.47){\circle*{0.2}}
\put( -17.69,  -7.71){\circle*{0.2}}
\put( -17.72,  -7.97){\circle*{0.2}}
\put( -17.75,  -8.22){\circle*{0.2}}
\put( -17.78,  -8.48){\circle*{0.2}}
\put( -17.81,  -8.75){\circle*{0.2}}
\put( -17.84,  -9.01){\circle*{0.2}}
\put( -17.86,  -9.28){\circle*{0.2}}
\put( -17.89,  -9.54){\circle*{0.2}}
\put( -17.92,  -9.80){\circle*{0.2}}
\put( -17.96, -10.05){\circle*{0.2}}
\put( -17.99, -10.30){\circle*{0.2}}
\put( -18.03, -10.53){\circle*{0.2}}
\put( -18.08, -10.76){\circle*{0.2}}
\put( -18.13, -10.98){\circle*{0.2}}
\put( -18.19, -11.18){\circle*{0.2}}
\put( -18.25, -11.37){\circle*{0.2}}
\put( -18.32, -11.55){\circle*{0.2}}
\put( -18.40, -11.71){\circle*{0.2}}
\put( -18.49, -11.85){\circle*{0.2}}
\put( -18.59, -11.98){\circle*{0.2}}
\put( -18.69, -12.09){\circle*{0.2}}
\put( -18.80, -12.18){\circle*{0.2}}
\put( -18.93, -12.26){\circle*{0.2}}
\put( -19.06, -12.31){\circle*{0.2}}
\put( -19.20, -12.35){\circle*{0.2}}
\put( -19.35, -12.37){\circle*{0.2}}
\put( -19.51, -12.38){\circle*{0.2}}
\put( -19.67, -12.36){\circle*{0.2}}
\put( -19.85, -12.33){\circle*{0.2}}
\put( -20.03, -12.28){\circle*{0.2}}
\put( -20.22, -12.22){\circle*{0.2}}
\put( -20.42, -12.15){\circle*{0.2}}
\put( -20.62, -12.06){\circle*{0.2}}
\put( -20.83, -11.96){\circle*{0.2}}
\put( -21.04, -11.86){\circle*{0.2}}
\put( -21.26, -11.74){\circle*{0.2}}
\put( -21.48, -11.61){\circle*{0.2}}
\put( -21.71, -11.49){\circle*{0.2}}
\put( -21.93, -11.35){\circle*{0.2}}
\put( -22.16, -11.21){\circle*{0.2}}
\put( -22.39, -11.08){\circle*{0.2}}
\put( -22.62, -10.94){\circle*{0.2}}
\put( -22.84, -10.81){\circle*{0.2}}
\put( -23.07, -10.68){\circle*{0.2}}
\put( -23.29, -10.55){\circle*{0.2}}
\put( -23.51, -10.43){\circle*{0.2}}
\put( -23.72, -10.33){\circle*{0.2}}
\put( -23.93, -10.23){\circle*{0.2}}
\put( -24.14, -10.14){\circle*{0.2}}
\put( -24.33, -10.06){\circle*{0.2}}
\put( -24.52, -10.00){\circle*{0.2}}
\put( -24.71, - 9.96){\circle*{0.2}}
\put( -24.88, - 9.92){\circle*{0.2}}
\put( -25.05, - 9.91){\circle*{0.2}}
\put( -25.21, - 9.91){\circle*{0.2}}
\put( -25.36, - 9.93){\circle*{0.2}}
\put( -25.50, - 9.97){\circle*{0.2}}
\put( -25.63, -10.02){\circle*{0.2}}
\put( -25.75, -10.09){\circle*{0.2}}
\put( -25.87, -10.19){\circle*{0.2}}
\put( -25.97, -10.29){\circle*{0.2}}
\put( -26.07, -10.42){\circle*{0.2}}
\put( -26.16, -10.56){\circle*{0.2}}
\put( -26.24, -10.72){\circle*{0.2}}
\put( -26.31, -10.90){\circle*{0.2}}
\put( -26.37, -11.09){\circle*{0.2}}
\put( -26.43, -11.29){\circle*{0.2}}
\put( -26.48, -11.51){\circle*{0.2}}
\put( -26.53, -11.73){\circle*{0.2}}
\put( -26.57, -11.97){\circle*{0.2}}
\put( -26.61, -12.22){\circle*{0.2}}
\put( -26.64, -12.47){\circle*{0.2}}
\put( -26.67, -12.73){\circle*{0.2}}
\put( -26.70, -12.99){\circle*{0.2}}
\put( -26.73, -13.25){\circle*{0.2}}
\put( -26.76, -13.52){\circle*{0.2}}
\put( -26.78, -13.78){\circle*{0.2}}
\put( -26.81, -14.04){\circle*{0.2}}
\put( -26.84, -14.30){\circle*{0.2}}
\put( -26.88, -14.55){\circle*{0.2}}
\put( -26.91, -14.80){\circle*{0.2}}
\put( -26.96, -15.03){\circle*{0.2}}
\put( -27.00, -15.26){\circle*{0.2}}
\put( -27.05, -15.47){\circle*{0.2}}
\put( -27.11, -15.67){\circle*{0.2}}
\put( -27.18, -15.86){\circle*{0.2}}
\put( -27.25, -16.04){\circle*{0.2}}
\put( -27.33, -16.19){\circle*{0.2}}
\put( -27.42, -16.33){\circle*{0.2}}
\put( -27.52, -16.46){\circle*{0.2}}
\put( -27.63, -16.57){\circle*{0.2}}
\put( -27.74, -16.65){\circle*{0.2}}
\put( -27.86, -16.73){\circle*{0.2}}
\put( -28.00, -16.78){\circle*{0.2}}
\put( -28.14, -16.81){\circle*{0.2}}
\put( -28.29, -16.83){\circle*{0.2}}
\put( -28.45, -16.83){\circle*{0.2}}
\put( -28.62, -16.81){\circle*{0.2}}
\put( -28.80, -16.78){\circle*{0.2}}
\put( -28.98, -16.73){\circle*{0.2}}
\put( -29.17, -16.67){\circle*{0.2}}
\put( -29.37, -16.59){\circle*{0.2}}
\put( -29.57, -16.50){\circle*{0.2}}
\put( -29.78, -16.40){\circle*{0.2}}
\put( -30.00, -16.29){\circle*{0.2}}
\put( -30.22, -16.17){\circle*{0.2}}
\put( -30.44, -16.05){\circle*{0.2}}
\put( -30.66, -15.92){\circle*{0.2}}
\put( -30.89, -15.78){\circle*{0.2}}
\put( -31.12, -15.65){\circle*{0.2}}
\put( -31.35, -15.51){\circle*{0.2}}
\put( -31.57, -15.37){\circle*{0.2}}
\put( -31.80, -15.24){\circle*{0.2}}
\put( -32.02, -15.11){\circle*{0.2}}
\put( -32.25, -14.99){\circle*{0.2}}
\put( -32.46, -14.87){\circle*{0.2}}
\put( -32.68, -14.76){\circle*{0.2}}
\put( -32.88, -14.67){\circle*{0.2}}
\put( -33.09, -14.58){\circle*{0.2}}
\put( -33.28, -14.51){\circle*{0.2}}
\put( -33.47, -14.45){\circle*{0.2}}
\put( -33.65, -14.40){\circle*{0.2}}
\put( -33.83, -14.38){\circle*{0.2}}
\put( -33.99, -14.36){\circle*{0.2}}
\put( -34.15, -14.37){\circle*{0.2}}
\put( -34.30, -14.39){\circle*{0.2}}
\put( -34.44, -14.43){\circle*{0.2}}
\put( -34.57, -14.49){\circle*{0.2}}
\put( -34.69, -14.57){\circle*{0.2}}
\put( -34.80, -14.66){\circle*{0.2}}
\put( -34.90, -14.77){\circle*{0.2}}
\put( -35.00, -14.90){\circle*{0.2}}
\put( -35.09, -15.05){\circle*{0.2}}
\put( -35.16, -15.21){\circle*{0.2}}
\put( -35.24, -15.39){\circle*{0.2}}
\put( -35.30, -15.58){\circle*{0.2}}
\put( -35.36, -15.79){\circle*{0.2}}
\put( -35.41, -16.01){\circle*{0.2}}
\put( -35.45, -16.24){\circle*{0.2}}
\put( -35.49, -16.47){\circle*{0.2}}
\put( -35.53, -16.72){\circle*{0.2}}
\put( -35.56, -16.97){\circle*{0.2}}
\put( -35.59, -17.23){\circle*{0.2}}
\put( -35.62, -17.50){\circle*{0.2}}
\put( -35.65, -17.76){\circle*{0.2}}
\put( -35.67, -18.02){\circle*{0.2}}
\put( -35.70, -18.29){\circle*{0.2}}
\put( -35.73, -18.55){\circle*{0.2}}
\put( -35.76, -18.81){\circle*{0.2}}
\put( -35.80, -19.06){\circle*{0.2}}
\put( -35.83, -19.30){\circle*{0.2}}
\put( -35.88, -19.53){\circle*{0.2}}
\put( -35.92, -19.76){\circle*{0.2}}
\put( -35.98, -19.97){\circle*{0.2}}
\put( -36.04, -20.17){\circle*{0.2}}
\put( -36.11, -20.35){\circle*{0.2}}
\put( -36.18, -20.52){\circle*{0.2}}
\put( -36.26, -20.68){\circle*{0.2}}
\put( -36.35, -20.82){\circle*{0.2}}
\put( -36.45, -20.94){\circle*{0.2}}
\put( -36.56, -21.04){\circle*{0.2}}
\put( -36.68, -21.13){\circle*{0.2}}
\put( -36.80, -21.19){\circle*{0.2}}
\put( -36.94, -21.24){\circle*{0.2}}
\put( -37.08, -21.27){\circle*{0.2}}
\put( -37.23, -21.29){\circle*{0.2}}
\put( -37.39, -21.29){\circle*{0.2}}
\put( -37.56, -21.26){\circle*{0.2}}
\put( -37.74, -21.23){\circle*{0.2}}
\put( -37.93, -21.18){\circle*{0.2}}
\put( -38.12, -21.11){\circle*{0.2}}
\put( -38.32, -21.03){\circle*{0.2}}
\put( -38.52, -20.94){\circle*{0.2}}
\put( -38.74, -20.84){\circle*{0.2}}
\put( -38.95, -20.73){\circle*{0.2}}
\put( -39.17, -20.61){\circle*{0.2}}
\put( -39.39, -20.48){\circle*{0.2}}
\put( -39.62, -20.35){\circle*{0.2}}
\put( -39.85, -20.21){\circle*{0.2}}
\put( -40.07, -20.08){\circle*{0.2}}
}

\newsavebox{\brgluon}
\savebox{\brgluon}(0,0)[tl]{
\put(  0.00,  0.00){\circle*{0.2}}
\put(  0.03, -0.26){\circle*{0.2}}
\put(  0.06, -0.53){\circle*{0.2}}
\put(  0.08, -0.79){\circle*{0.2}}
\put(  0.12, -1.04){\circle*{0.2}}
\put(  0.15, -1.29){\circle*{0.2}}
\put(  0.19, -1.53){\circle*{0.2}}
\put(  0.24, -1.76){\circle*{0.2}}
\put(  0.28, -1.98){\circle*{0.2}}
\put(  0.34, -2.19){\circle*{0.2}}
\put(  0.40, -2.39){\circle*{0.2}}
\put(  0.47, -2.57){\circle*{0.2}}
\put(  0.55, -2.74){\circle*{0.2}}
\put(  0.63, -2.89){\circle*{0.2}}
\put(  0.72, -3.02){\circle*{0.2}}
\put(  0.82, -3.14){\circle*{0.2}}
\put(  0.93, -3.24){\circle*{0.2}}
\put(  1.05, -3.32){\circle*{0.2}}
\put(  1.18, -3.38){\circle*{0.2}}
\put(  1.32, -3.43){\circle*{0.2}}
\put(  1.47, -3.45){\circle*{0.2}}
\put(  1.62, -3.46){\circle*{0.2}}
\put(  1.78, -3.45){\circle*{0.2}}
\put(  1.96, -3.43){\circle*{0.2}}
\put(  2.14, -3.39){\circle*{0.2}}
\put(  2.32, -3.33){\circle*{0.2}}
\put(  2.52, -3.27){\circle*{0.2}}
\put(  2.72, -3.18){\circle*{0.2}}
\put(  2.92, -3.09){\circle*{0.2}}
\put(  3.14, -2.98){\circle*{0.2}}
\put(  3.35, -2.87){\circle*{0.2}}
\put(  3.57, -2.75){\circle*{0.2}}
\put(  3.80, -2.62){\circle*{0.2}}
\put(  4.02, -2.49){\circle*{0.2}}
\put(  4.25, -2.35){\circle*{0.2}}
\put(  4.48, -2.22){\circle*{0.2}}
\put(  4.71, -2.08){\circle*{0.2}}
\put(  4.93, -1.94){\circle*{0.2}}
\put(  5.16, -1.81){\circle*{0.2}}
\put(  5.38, -1.68){\circle*{0.2}}
\put(  5.60, -1.56){\circle*{0.2}}
\put(  5.82, -1.45){\circle*{0.2}}
\put(  6.03, -1.35){\circle*{0.2}}
\put(  6.23, -1.26){\circle*{0.2}}
\put(  6.43, -1.18){\circle*{0.2}}
\put(  6.63, -1.11){\circle*{0.2}}
\put(  6.81, -1.06){\circle*{0.2}}
\put(  6.99, -1.02){\circle*{0.2}}
\put(  7.16, -1.00){\circle*{0.2}}
\put(  7.32, -0.99){\circle*{0.2}}
\put(  7.48, -1.01){\circle*{0.2}}
\put(  7.62, -1.04){\circle*{0.2}}
\put(  7.76, -1.09){\circle*{0.2}}
\put(  7.88, -1.15){\circle*{0.2}}
\put(  8.00, -1.24){\circle*{0.2}}
\put(  8.11, -1.34){\circle*{0.2}}
\put(  8.21, -1.46){\circle*{0.2}}
\put(  8.30, -1.60){\circle*{0.2}}
\put(  8.38, -1.75){\circle*{0.2}}
\put(  8.46, -1.92){\circle*{0.2}}
\put(  8.52, -2.10){\circle*{0.2}}
\put(  8.58, -2.30){\circle*{0.2}}
\put(  8.64, -2.51){\circle*{0.2}}
\put(  8.69, -2.74){\circle*{0.2}}
\put(  8.73, -2.97){\circle*{0.2}}
\put(  8.77, -3.21){\circle*{0.2}}
\put(  8.80, -3.46){\circle*{0.2}}
\put(  8.83, -3.72){\circle*{0.2}}
\put(  8.86, -3.98){\circle*{0.2}}
\put(  8.89, -4.24){\circle*{0.2}}
\put(  8.92, -4.51){\circle*{0.2}}
\put(  8.95, -4.77){\circle*{0.2}}
\put(  8.97, -5.03){\circle*{0.2}}
\put(  9.00, -5.29){\circle*{0.2}}
\put(  9.04, -5.55){\circle*{0.2}}
\put(  9.07, -5.79){\circle*{0.2}}
\put(  9.11, -6.03){\circle*{0.2}}
\put(  9.16, -6.26){\circle*{0.2}}
\put(  9.21, -6.48){\circle*{0.2}}
\put(  9.26, -6.69){\circle*{0.2}}
\put(  9.33, -6.88){\circle*{0.2}}
\put(  9.40, -7.06){\circle*{0.2}}
\put(  9.47, -7.22){\circle*{0.2}}
\put(  9.56, -7.37){\circle*{0.2}}
\put(  9.66, -7.50){\circle*{0.2}}
\put(  9.76, -7.61){\circle*{0.2}}
\put(  9.87, -7.71){\circle*{0.2}}
\put(  9.99, -7.79){\circle*{0.2}}
\put( 10.12, -7.85){\circle*{0.2}}
\put( 10.26, -7.89){\circle*{0.2}}
\put( 10.41, -7.91){\circle*{0.2}}
\put( 10.56, -7.92){\circle*{0.2}}
\put( 10.73, -7.91){\circle*{0.2}}
\put( 10.90, -7.88){\circle*{0.2}}
\put( 11.08, -7.84){\circle*{0.2}}
\put( 11.27, -7.78){\circle*{0.2}}
\put( 11.47, -7.71){\circle*{0.2}}
\put( 11.67, -7.62){\circle*{0.2}}
\put( 11.88, -7.53){\circle*{0.2}}
\put( 12.09, -7.42){\circle*{0.2}}
\put( 12.31, -7.30){\circle*{0.2}}
\put( 12.53, -7.18){\circle*{0.2}}
\put( 12.75, -7.05){\circle*{0.2}}
\put( 12.98, -6.92){\circle*{0.2}}
\put( 13.21, -6.78){\circle*{0.2}}
\put( 13.43, -6.65){\circle*{0.2}}
\put( 13.66, -6.51){\circle*{0.2}}
\put( 13.89, -6.38){\circle*{0.2}}
\put( 14.11, -6.24){\circle*{0.2}}
\put( 14.34, -6.12){\circle*{0.2}}
\put( 14.56, -6.00){\circle*{0.2}}
\put( 14.77, -5.89){\circle*{0.2}}
\put( 14.98, -5.79){\circle*{0.2}}
\put( 15.18, -5.70){\circle*{0.2}}
\put( 15.38, -5.62){\circle*{0.2}}
\put( 15.58, -5.56){\circle*{0.2}}
\put( 15.76, -5.51){\circle*{0.2}}
\put( 15.94, -5.47){\circle*{0.2}}
\put( 16.11, -5.45){\circle*{0.2}}
\put( 16.27, -5.45){\circle*{0.2}}
\put( 16.42, -5.47){\circle*{0.2}}
\put( 16.56, -5.50){\circle*{0.2}}
\put( 16.69, -5.55){\circle*{0.2}}
\put( 16.82, -5.62){\circle*{0.2}}
\put( 16.93, -5.71){\circle*{0.2}}
\put( 17.04, -5.82){\circle*{0.2}}
\put( 17.14, -5.94){\circle*{0.2}}
\put( 17.23, -6.08){\circle*{0.2}}
\put( 17.31, -6.24){\circle*{0.2}}
\put( 17.38, -6.41){\circle*{0.2}}
\put( 17.45, -6.60){\circle*{0.2}}
\put( 17.51, -6.80){\circle*{0.2}}
\put( 17.56, -7.01){\circle*{0.2}}
\put( 17.61, -7.24){\circle*{0.2}}
\put( 17.65, -7.47){\circle*{0.2}}
\put( 17.69, -7.71){\circle*{0.2}}
\put( 17.72, -7.97){\circle*{0.2}}
\put( 17.75, -8.22){\circle*{0.2}}
\put( 17.78, -8.48){\circle*{0.2}}
\put( 17.81, -8.75){\circle*{0.2}}
\put( 17.84, -9.01){\circle*{0.2}}
\put( 17.86, -9.28){\circle*{0.2}}
\put( 17.89, -9.54){\circle*{0.2}}
\put( 17.92, -9.80){\circle*{0.2}}
\put( 17.96,-10.05){\circle*{0.2}}
\put( 17.99,-10.30){\circle*{0.2}}
\put( 18.03,-10.53){\circle*{0.2}}
\put( 18.08,-10.76){\circle*{0.2}}
\put( 18.13,-10.98){\circle*{0.2}}
\put( 18.19,-11.18){\circle*{0.2}}
\put( 18.25,-11.37){\circle*{0.2}}
\put( 18.32,-11.55){\circle*{0.2}}
\put( 18.40,-11.71){\circle*{0.2}}
\put( 18.49,-11.85){\circle*{0.2}}
\put( 18.59,-11.98){\circle*{0.2}}
\put( 18.69,-12.09){\circle*{0.2}}
\put( 18.80,-12.18){\circle*{0.2}}
\put( 18.93,-12.26){\circle*{0.2}}
\put( 19.06,-12.31){\circle*{0.2}}
\put( 19.20,-12.35){\circle*{0.2}}
\put( 19.35,-12.37){\circle*{0.2}}
\put( 19.51,-12.38){\circle*{0.2}}
\put( 19.67,-12.36){\circle*{0.2}}
\put( 19.85,-12.33){\circle*{0.2}}
\put( 20.03,-12.28){\circle*{0.2}}
\put( 20.22,-12.22){\circle*{0.2}}
\put( 20.42,-12.15){\circle*{0.2}}
\put( 20.62,-12.06){\circle*{0.2}}
\put( 20.83,-11.96){\circle*{0.2}}
\put( 21.04,-11.86){\circle*{0.2}}
\put( 21.26,-11.74){\circle*{0.2}}
\put( 21.48,-11.61){\circle*{0.2}}
\put( 21.71,-11.49){\circle*{0.2}}
\put( 21.93,-11.35){\circle*{0.2}}
\put( 22.16,-11.21){\circle*{0.2}}
\put( 22.39,-11.08){\circle*{0.2}}
\put( 22.62,-10.94){\circle*{0.2}}
\put( 22.84,-10.81){\circle*{0.2}}
\put( 23.07,-10.68){\circle*{0.2}}
\put( 23.29,-10.55){\circle*{0.2}}
\put( 23.51,-10.43){\circle*{0.2}}
\put( 23.72,-10.33){\circle*{0.2}}
\put( 23.93,-10.23){\circle*{0.2}}
\put( 24.14,-10.14){\circle*{0.2}}
\put( 24.33,-10.06){\circle*{0.2}}
\put( 24.52,-10.00){\circle*{0.2}}
\put( 24.71, -9.96){\circle*{0.2}}
\put( 24.88, -9.92){\circle*{0.2}}
\put( 25.05, -9.91){\circle*{0.2}}
\put( 25.21, -9.91){\circle*{0.2}}
\put( 25.36, -9.93){\circle*{0.2}}
\put( 25.50, -9.97){\circle*{0.2}}
\put( 25.63,-10.02){\circle*{0.2}}
\put( 25.75,-10.09){\circle*{0.2}}
\put( 25.87,-10.19){\circle*{0.2}}
\put( 25.97,-10.29){\circle*{0.2}}
\put( 26.07,-10.42){\circle*{0.2}}
\put( 26.16,-10.56){\circle*{0.2}}
\put( 26.24,-10.72){\circle*{0.2}}
\put( 26.31,-10.90){\circle*{0.2}}
\put( 26.37,-11.09){\circle*{0.2}}
\put( 26.43,-11.29){\circle*{0.2}}
\put( 26.48,-11.51){\circle*{0.2}}
\put( 26.53,-11.73){\circle*{0.2}}
\put( 26.57,-11.97){\circle*{0.2}}
\put( 26.61,-12.22){\circle*{0.2}}
\put( 26.64,-12.47){\circle*{0.2}}
\put( 26.67,-12.73){\circle*{0.2}}
\put( 26.70,-12.99){\circle*{0.2}}
\put( 26.73,-13.25){\circle*{0.2}}
\put( 26.76,-13.52){\circle*{0.2}}
\put( 26.78,-13.78){\circle*{0.2}}
\put( 26.81,-14.04){\circle*{0.2}}
\put( 26.84,-14.30){\circle*{0.2}}
\put( 26.88,-14.55){\circle*{0.2}}
\put( 26.91,-14.80){\circle*{0.2}}
\put( 26.96,-15.03){\circle*{0.2}}
\put( 27.00,-15.26){\circle*{0.2}}
\put( 27.05,-15.47){\circle*{0.2}}
\put( 27.11,-15.67){\circle*{0.2}}
\put( 27.18,-15.86){\circle*{0.2}}
\put( 27.25,-16.04){\circle*{0.2}}
\put( 27.33,-16.19){\circle*{0.2}}
\put( 27.42,-16.33){\circle*{0.2}}
\put( 27.52,-16.46){\circle*{0.2}}
\put( 27.63,-16.57){\circle*{0.2}}
\put( 27.74,-16.65){\circle*{0.2}}
\put( 27.86,-16.73){\circle*{0.2}}
\put( 28.00,-16.78){\circle*{0.2}}
\put( 28.14,-16.81){\circle*{0.2}}
\put( 28.29,-16.83){\circle*{0.2}}
\put( 28.45,-16.83){\circle*{0.2}}
\put( 28.62,-16.81){\circle*{0.2}}
\put( 28.80,-16.78){\circle*{0.2}}
\put( 28.98,-16.73){\circle*{0.2}}
\put( 29.17,-16.67){\circle*{0.2}}
\put( 29.37,-16.59){\circle*{0.2}}
\put( 29.57,-16.50){\circle*{0.2}}
\put( 29.78,-16.40){\circle*{0.2}}
\put( 30.00,-16.29){\circle*{0.2}}
\put( 30.22,-16.17){\circle*{0.2}}
\put( 30.44,-16.05){\circle*{0.2}}
\put( 30.66,-15.92){\circle*{0.2}}
\put( 30.89,-15.78){\circle*{0.2}}
\put( 31.12,-15.65){\circle*{0.2}}
\put( 31.35,-15.51){\circle*{0.2}}
\put( 31.57,-15.37){\circle*{0.2}}
\put( 31.80,-15.24){\circle*{0.2}}
\put( 32.02,-15.11){\circle*{0.2}}
\put( 32.25,-14.99){\circle*{0.2}}
\put( 32.46,-14.87){\circle*{0.2}}
\put( 32.68,-14.76){\circle*{0.2}}
\put( 32.88,-14.67){\circle*{0.2}}
\put( 33.09,-14.58){\circle*{0.2}}
\put( 33.28,-14.51){\circle*{0.2}}
\put( 33.47,-14.45){\circle*{0.2}}
\put( 33.65,-14.40){\circle*{0.2}}
\put( 33.83,-14.38){\circle*{0.2}}
\put( 33.99,-14.36){\circle*{0.2}}
\put( 34.15,-14.37){\circle*{0.2}}
\put( 34.30,-14.39){\circle*{0.2}}
\put( 34.44,-14.43){\circle*{0.2}}
\put( 34.57,-14.49){\circle*{0.2}}
\put( 34.69,-14.57){\circle*{0.2}}
\put( 34.80,-14.66){\circle*{0.2}}
\put( 34.90,-14.77){\circle*{0.2}}
\put( 35.00,-14.90){\circle*{0.2}}
\put( 35.09,-15.05){\circle*{0.2}}
\put( 35.16,-15.21){\circle*{0.2}}
\put( 35.24,-15.39){\circle*{0.2}}
\put( 35.30,-15.58){\circle*{0.2}}
\put( 35.36,-15.79){\circle*{0.2}}
\put( 35.41,-16.01){\circle*{0.2}}
\put( 35.45,-16.24){\circle*{0.2}}
\put( 35.49,-16.47){\circle*{0.2}}
\put( 35.53,-16.72){\circle*{0.2}}
\put( 35.56,-16.97){\circle*{0.2}}
\put( 35.59,-17.23){\circle*{0.2}}
\put( 35.62,-17.50){\circle*{0.2}}
\put( 35.65,-17.76){\circle*{0.2}}
\put( 35.67,-18.02){\circle*{0.2}}
\put( 35.70,-18.29){\circle*{0.2}}
\put( 35.73,-18.55){\circle*{0.2}}
\put( 35.76,-18.81){\circle*{0.2}}
\put( 35.80,-19.06){\circle*{0.2}}
\put( 35.83,-19.30){\circle*{0.2}}
\put( 35.88,-19.53){\circle*{0.2}}
\put( 35.92,-19.76){\circle*{0.2}}
\put( 35.98,-19.97){\circle*{0.2}}
\put( 36.04,-20.17){\circle*{0.2}}
\put( 36.11,-20.35){\circle*{0.2}}
\put( 36.18,-20.52){\circle*{0.2}}
\put( 36.26,-20.68){\circle*{0.2}}
\put( 36.35,-20.82){\circle*{0.2}}
\put( 36.45,-20.94){\circle*{0.2}}
\put( 36.56,-21.04){\circle*{0.2}}
\put( 36.68,-21.13){\circle*{0.2}}
\put( 36.80,-21.19){\circle*{0.2}}
\put( 36.94,-21.24){\circle*{0.2}}
\put( 37.08,-21.27){\circle*{0.2}}
\put( 37.23,-21.29){\circle*{0.2}}
\put( 37.39,-21.29){\circle*{0.2}}
\put( 37.56,-21.26){\circle*{0.2}}
\put( 37.74,-21.23){\circle*{0.2}}
\put( 37.93,-21.18){\circle*{0.2}}
\put( 38.12,-21.11){\circle*{0.2}}
\put( 38.32,-21.03){\circle*{0.2}}
\put( 38.52,-20.94){\circle*{0.2}}
\put( 38.74,-20.84){\circle*{0.2}}
\put( 38.95,-20.73){\circle*{0.2}}
\put( 39.17,-20.61){\circle*{0.2}}
\put( 39.39,-20.48){\circle*{0.2}}
\put( 39.62,-20.35){\circle*{0.2}}
\put( 39.85,-20.21){\circle*{0.2}}
\put( 40.07,-20.08){\circle*{0.2}}
}

\newsavebox{\tlgluon}
\savebox{\tlgluon}(0,0)[br]{
\put(  -0.00,  0.00){\circle*{0.2}}
\put(  -0.03, 0.26){\circle*{0.2}}
\put(  -0.06, 0.53){\circle*{0.2}}
\put(  -0.08, 0.79){\circle*{0.2}}
\put(  -0.12, 1.04){\circle*{0.2}}
\put(  -0.15, 1.29){\circle*{0.2}}
\put(  -0.19, 1.53){\circle*{0.2}}
\put(  -0.24, 1.76){\circle*{0.2}}
\put(  -0.28, 1.98){\circle*{0.2}}
\put(  -0.34, 2.19){\circle*{0.2}}
\put(  -0.40, 2.39){\circle*{0.2}}
\put(  -0.47, 2.57){\circle*{0.2}}
\put(  -0.55, 2.74){\circle*{0.2}}
\put(  -0.63, 2.89){\circle*{0.2}}
\put(  -0.72, 3.02){\circle*{0.2}}
\put(  -0.82, 3.14){\circle*{0.2}}
\put(  -0.93, 3.24){\circle*{0.2}}
\put(  -1.05, 3.32){\circle*{0.2}}
\put(  -1.18, 3.38){\circle*{0.2}}
\put(  -1.32, 3.43){\circle*{0.2}}
\put(  -1.47, 3.45){\circle*{0.2}}
\put(  -1.62, 3.46){\circle*{0.2}}
\put(  -1.78, 3.45){\circle*{0.2}}
\put(  -1.96, 3.43){\circle*{0.2}}
\put(  -2.14, 3.39){\circle*{0.2}}
\put(  -2.32, 3.33){\circle*{0.2}}
\put(  -2.52, 3.27){\circle*{0.2}}
\put(  -2.72, 3.18){\circle*{0.2}}
\put(  -2.92, 3.09){\circle*{0.2}}
\put(  -3.14, 2.98){\circle*{0.2}}
\put(  -3.35, 2.87){\circle*{0.2}}
\put(  -3.57, 2.75){\circle*{0.2}}
\put(  -3.80, 2.62){\circle*{0.2}}
\put(  -4.02, 2.49){\circle*{0.2}}
\put(  -4.25, 2.35){\circle*{0.2}}
\put(  -4.48, 2.22){\circle*{0.2}}
\put(  -4.71, 2.08){\circle*{0.2}}
\put(  -4.93, 1.94){\circle*{0.2}}
\put(  -5.16, 1.81){\circle*{0.2}}
\put(  -5.38, 1.68){\circle*{0.2}}
\put(  -5.60, 1.56){\circle*{0.2}}
\put(  -5.82, 1.45){\circle*{0.2}}
\put(  -6.03, 1.35){\circle*{0.2}}
\put(  -6.23, 1.26){\circle*{0.2}}
\put(  -6.43, 1.18){\circle*{0.2}}
\put(  -6.63, 1.11){\circle*{0.2}}
\put(  -6.81, 1.06){\circle*{0.2}}
\put(  -6.99, 1.02){\circle*{0.2}}
\put(  -7.16, 1.00){\circle*{0.2}}
\put(  -7.32, 0.99){\circle*{0.2}}
\put(  -7.48, 1.01){\circle*{0.2}}
\put(  -7.62, 1.04){\circle*{0.2}}
\put(  -7.76, 1.09){\circle*{0.2}}
\put(  -7.88, 1.15){\circle*{0.2}}
\put(  -8.00, 1.24){\circle*{0.2}}
\put(  -8.11, 1.34){\circle*{0.2}}
\put(  -8.21, 1.46){\circle*{0.2}}
\put(  -8.30, 1.60){\circle*{0.2}}
\put(  -8.38, 1.75){\circle*{0.2}}
\put(  -8.46, 1.92){\circle*{0.2}}
\put(  -8.52, 2.10){\circle*{0.2}}
\put(  -8.58, 2.30){\circle*{0.2}}
\put(  -8.64, 2.51){\circle*{0.2}}
\put(  -8.69, 2.74){\circle*{0.2}}
\put(  -8.73, 2.97){\circle*{0.2}}
\put(  -8.77, 3.21){\circle*{0.2}}
\put(  -8.80, 3.46){\circle*{0.2}}
\put(  -8.83, 3.72){\circle*{0.2}}
\put(  -8.86, 3.98){\circle*{0.2}}
\put(  -8.89, 4.24){\circle*{0.2}}
\put(  -8.92, 4.51){\circle*{0.2}}
\put(  -8.95, 4.77){\circle*{0.2}}
\put(  -8.97, 5.03){\circle*{0.2}}
\put(  -9.00, 5.29){\circle*{0.2}}
\put(  -9.04, 5.55){\circle*{0.2}}
\put(  -9.07, 5.79){\circle*{0.2}}
\put(  -9.11, 6.03){\circle*{0.2}}
\put(  -9.16, 6.26){\circle*{0.2}}
\put(  -9.21, 6.48){\circle*{0.2}}
\put(  -9.26, 6.69){\circle*{0.2}}
\put(  -9.33, 6.88){\circle*{0.2}}
\put(  -9.40, 7.06){\circle*{0.2}}
\put(  -9.47, 7.22){\circle*{0.2}}
\put(  -9.56, 7.37){\circle*{0.2}}
\put(  -9.66, 7.50){\circle*{0.2}}
\put(  -9.76, 7.61){\circle*{0.2}}
\put(  -9.87, 7.71){\circle*{0.2}}
\put(  -9.99, 7.79){\circle*{0.2}}
\put( -10.12, 7.85){\circle*{0.2}}
\put( -10.26, 7.89){\circle*{0.2}}
\put( -10.41, 7.91){\circle*{0.2}}
\put( -10.56, 7.92){\circle*{0.2}}
\put( -10.73, 7.91){\circle*{0.2}}
\put( -10.90, 7.88){\circle*{0.2}}
\put( -11.08, 7.84){\circle*{0.2}}
\put( -11.27, 7.78){\circle*{0.2}}
\put( -11.47, 7.71){\circle*{0.2}}
\put( -11.67, 7.62){\circle*{0.2}}
\put( -11.88, 7.53){\circle*{0.2}}
\put( -12.09, 7.42){\circle*{0.2}}
\put( -12.31, 7.30){\circle*{0.2}}
\put( -12.53, 7.18){\circle*{0.2}}
\put( -12.75, 7.05){\circle*{0.2}}
\put( -12.98, 6.92){\circle*{0.2}}
\put( -13.21, 6.78){\circle*{0.2}}
\put( -13.43, 6.65){\circle*{0.2}}
\put( -13.66, 6.51){\circle*{0.2}}
\put( -13.89, 6.38){\circle*{0.2}}
\put( -14.11, 6.24){\circle*{0.2}}
\put( -14.34, 6.12){\circle*{0.2}}
\put( -14.56, 6.00){\circle*{0.2}}
\put( -14.77, 5.89){\circle*{0.2}}
\put( -14.98, 5.79){\circle*{0.2}}
\put( -15.18, 5.70){\circle*{0.2}}
\put( -15.38, 5.62){\circle*{0.2}}
\put( -15.58, 5.56){\circle*{0.2}}
\put( -15.76, 5.51){\circle*{0.2}}
\put( -15.94, 5.47){\circle*{0.2}}
\put( -16.11, 5.45){\circle*{0.2}}
\put( -16.27, 5.45){\circle*{0.2}}
\put( -16.42, 5.47){\circle*{0.2}}
\put( -16.56, 5.50){\circle*{0.2}}
\put( -16.69, 5.55){\circle*{0.2}}
\put( -16.82, 5.62){\circle*{0.2}}
\put( -16.93, 5.71){\circle*{0.2}}
\put( -17.04, 5.82){\circle*{0.2}}
\put( -17.14, 5.94){\circle*{0.2}}
\put( -17.23, 6.08){\circle*{0.2}}
\put( -17.31, 6.24){\circle*{0.2}}
\put( -17.38, 6.41){\circle*{0.2}}
\put( -17.45, 6.60){\circle*{0.2}}
\put( -17.51, 6.80){\circle*{0.2}}
\put( -17.56, 7.01){\circle*{0.2}}
\put( -17.61, 7.24){\circle*{0.2}}
\put( -17.65, 7.47){\circle*{0.2}}
\put( -17.69, 7.71){\circle*{0.2}}
\put( -17.72, 7.97){\circle*{0.2}}
\put( -17.75, 8.22){\circle*{0.2}}
\put( -17.78, 8.48){\circle*{0.2}}
\put( -17.81, 8.75){\circle*{0.2}}
\put( -17.84, 9.01){\circle*{0.2}}
\put( -17.86, 9.28){\circle*{0.2}}
\put( -17.89, 9.54){\circle*{0.2}}
\put( -17.92, 9.80){\circle*{0.2}}
\put( -17.96,10.05){\circle*{0.2}}
\put( -17.99,10.30){\circle*{0.2}}
\put( -18.03,10.53){\circle*{0.2}}
\put( -18.08,10.76){\circle*{0.2}}
\put( -18.13,10.98){\circle*{0.2}}
\put( -18.19,11.18){\circle*{0.2}}
\put( -18.25,11.37){\circle*{0.2}}
\put( -18.32,11.55){\circle*{0.2}}
\put( -18.40,11.71){\circle*{0.2}}
\put( -18.49,11.85){\circle*{0.2}}
\put( -18.59,11.98){\circle*{0.2}}
\put( -18.69,12.09){\circle*{0.2}}
\put( -18.80,12.18){\circle*{0.2}}
\put( -18.93,12.26){\circle*{0.2}}
\put( -19.06,12.31){\circle*{0.2}}
\put( -19.20,12.35){\circle*{0.2}}
\put( -19.35,12.37){\circle*{0.2}}
\put( -19.51,12.38){\circle*{0.2}}
\put( -19.67,12.36){\circle*{0.2}}
\put( -19.85,12.33){\circle*{0.2}}
\put( -20.03,12.28){\circle*{0.2}}
\put( -20.22,12.22){\circle*{0.2}}
\put( -20.42,12.15){\circle*{0.2}}
\put( -20.62,12.06){\circle*{0.2}}
\put( -20.83,11.96){\circle*{0.2}}
\put( -21.04,11.86){\circle*{0.2}}
\put( -21.26,11.74){\circle*{0.2}}
\put( -21.48,11.61){\circle*{0.2}}
\put( -21.71,11.49){\circle*{0.2}}
\put( -21.93,11.35){\circle*{0.2}}
\put( -22.16,11.21){\circle*{0.2}}
\put( -22.39,11.08){\circle*{0.2}}
\put( -22.62,10.94){\circle*{0.2}}
\put( -22.84,10.81){\circle*{0.2}}
\put( -23.07,10.68){\circle*{0.2}}
\put( -23.29,10.55){\circle*{0.2}}
\put( -23.51,10.43){\circle*{0.2}}
\put( -23.72,10.33){\circle*{0.2}}
\put( -23.93,10.23){\circle*{0.2}}
\put( -24.14,10.14){\circle*{0.2}}
\put( -24.33,10.06){\circle*{0.2}}
\put( -24.52,10.00){\circle*{0.2}}
\put( -24.71, 9.96){\circle*{0.2}}
\put( -24.88, 9.92){\circle*{0.2}}
\put( -25.05, 9.91){\circle*{0.2}}
\put( -25.21, 9.91){\circle*{0.2}}
\put( -25.36, 9.93){\circle*{0.2}}
\put( -25.50, 9.97){\circle*{0.2}}
\put( -25.63,10.02){\circle*{0.2}}
\put( -25.75,10.09){\circle*{0.2}}
\put( -25.87,10.19){\circle*{0.2}}
\put( -25.97,10.29){\circle*{0.2}}
\put( -26.07,10.42){\circle*{0.2}}
\put( -26.16,10.56){\circle*{0.2}}
\put( -26.24,10.72){\circle*{0.2}}
\put( -26.31,10.90){\circle*{0.2}}
\put( -26.37,11.09){\circle*{0.2}}
\put( -26.43,11.29){\circle*{0.2}}
\put( -26.48,11.51){\circle*{0.2}}
\put( -26.53,11.73){\circle*{0.2}}
\put( -26.57,11.97){\circle*{0.2}}
\put( -26.61,12.22){\circle*{0.2}}
\put( -26.64,12.47){\circle*{0.2}}
\put( -26.67,12.73){\circle*{0.2}}
\put( -26.70,12.99){\circle*{0.2}}
\put( -26.73,13.25){\circle*{0.2}}
\put( -26.76,13.52){\circle*{0.2}}
\put( -26.78,13.78){\circle*{0.2}}
\put( -26.81,14.04){\circle*{0.2}}
\put( -26.84,14.30){\circle*{0.2}}
\put( -26.88,14.55){\circle*{0.2}}
\put( -26.91,14.80){\circle*{0.2}}
\put( -26.96,15.03){\circle*{0.2}}
\put( -27.00,15.26){\circle*{0.2}}
\put( -27.05,15.47){\circle*{0.2}}
\put( -27.11,15.67){\circle*{0.2}}
\put( -27.18,15.86){\circle*{0.2}}
\put( -27.25,16.04){\circle*{0.2}}
\put( -27.33,16.19){\circle*{0.2}}
\put( -27.42,16.33){\circle*{0.2}}
\put( -27.52,16.46){\circle*{0.2}}
\put( -27.63,16.57){\circle*{0.2}}
\put( -27.74,16.65){\circle*{0.2}}
\put( -27.86,16.73){\circle*{0.2}}
\put( -28.00,16.78){\circle*{0.2}}
\put( -28.14,16.81){\circle*{0.2}}
\put( -28.29,16.83){\circle*{0.2}}
\put( -28.45,16.83){\circle*{0.2}}
\put( -28.62,16.81){\circle*{0.2}}
\put( -28.80,16.78){\circle*{0.2}}
\put( -28.98,16.73){\circle*{0.2}}
\put( -29.17,16.67){\circle*{0.2}}
\put( -29.37,16.59){\circle*{0.2}}
\put( -29.57,16.50){\circle*{0.2}}
\put( -29.78,16.40){\circle*{0.2}}
\put( -30.00,16.29){\circle*{0.2}}
\put( -30.22,16.17){\circle*{0.2}}
\put( -30.44,16.05){\circle*{0.2}}
\put( -30.66,15.92){\circle*{0.2}}
\put( -30.89,15.78){\circle*{0.2}}
\put( -31.12,15.65){\circle*{0.2}}
\put( -31.35,15.51){\circle*{0.2}}
\put( -31.57,15.37){\circle*{0.2}}
\put( -31.80,15.24){\circle*{0.2}}
\put( -32.02,15.11){\circle*{0.2}}
\put( -32.25,14.99){\circle*{0.2}}
\put( -32.46,14.87){\circle*{0.2}}
\put( -32.68,14.76){\circle*{0.2}}
\put( -32.88,14.67){\circle*{0.2}}
\put( -33.09,14.58){\circle*{0.2}}
\put( -33.28,14.51){\circle*{0.2}}
\put( -33.47,14.45){\circle*{0.2}}
\put( -33.65,14.40){\circle*{0.2}}
\put( -33.83,14.38){\circle*{0.2}}
\put( -33.99,14.36){\circle*{0.2}}
\put( -34.15,14.37){\circle*{0.2}}
\put( -34.30,14.39){\circle*{0.2}}
\put( -34.44,14.43){\circle*{0.2}}
\put( -34.57,14.49){\circle*{0.2}}
\put( -34.69,14.57){\circle*{0.2}}
\put( -34.80,14.66){\circle*{0.2}}
\put( -34.90,14.77){\circle*{0.2}}
\put( -35.00,14.90){\circle*{0.2}}
\put( -35.09,15.05){\circle*{0.2}}
\put( -35.16,15.21){\circle*{0.2}}
\put( -35.24,15.39){\circle*{0.2}}
\put( -35.30,15.58){\circle*{0.2}}
\put( -35.36,15.79){\circle*{0.2}}
\put( -35.41,16.01){\circle*{0.2}}
\put( -35.45,16.24){\circle*{0.2}}
\put( -35.49,16.47){\circle*{0.2}}
\put( -35.53,16.72){\circle*{0.2}}
\put( -35.56,16.97){\circle*{0.2}}
\put( -35.59,17.23){\circle*{0.2}}
\put( -35.62,17.50){\circle*{0.2}}
\put( -35.65,17.76){\circle*{0.2}}
\put( -35.67,18.02){\circle*{0.2}}
\put( -35.70,18.29){\circle*{0.2}}
\put( -35.73,18.55){\circle*{0.2}}
\put( -35.76,18.81){\circle*{0.2}}
\put( -35.80,19.06){\circle*{0.2}}
\put( -35.83,19.30){\circle*{0.2}}
\put( -35.88,19.53){\circle*{0.2}}
\put( -35.92,19.76){\circle*{0.2}}
\put( -35.98,19.97){\circle*{0.2}}
\put( -36.04,20.17){\circle*{0.2}}
\put( -36.11,20.35){\circle*{0.2}}
\put( -36.18,20.52){\circle*{0.2}}
\put( -36.26,20.68){\circle*{0.2}}
\put( -36.35,20.82){\circle*{0.2}}
\put( -36.45,20.94){\circle*{0.2}}
\put( -36.56,21.04){\circle*{0.2}}
\put( -36.68,21.13){\circle*{0.2}}
\put( -36.80,21.19){\circle*{0.2}}
\put( -36.94,21.24){\circle*{0.2}}
\put( -37.08,21.27){\circle*{0.2}}
\put( -37.23,21.29){\circle*{0.2}}
\put( -37.39,21.29){\circle*{0.2}}
\put( -37.56,21.26){\circle*{0.2}}
\put( -37.74,21.23){\circle*{0.2}}
\put( -37.93,21.18){\circle*{0.2}}
\put( -38.12,21.11){\circle*{0.2}}
\put( -38.32,21.03){\circle*{0.2}}
\put( -38.52,20.94){\circle*{0.2}}
\put( -38.74,20.84){\circle*{0.2}}
\put( -38.95,20.73){\circle*{0.2}}
\put( -39.17,20.61){\circle*{0.2}}
\put( -39.39,20.48){\circle*{0.2}}
\put( -39.62,20.35){\circle*{0.2}}
\put( -39.85,20.21){\circle*{0.2}}
\put( -40.07,20.08){\circle*{0.2}}
}

\newsavebox{\topgluon}
\savebox{\topgluon}(0,0)[c]{
\put(  0.00,  2.50){\circle*{0.2}}
\put(  0.13,  2.49){\circle*{0.2}}
\put(  0.25,  2.47){\circle*{0.2}}
\put(  0.38,  2.43){\circle*{0.2}}
\put(  0.51,  2.37){\circle*{0.2}}
\put(  0.64,  2.30){\circle*{0.2}}
\put(  0.76,  2.22){\circle*{0.2}}
\put(  0.89,  2.12){\circle*{0.2}}
\put(  1.02,  2.01){\circle*{0.2}}
\put(  1.15,  1.88){\circle*{0.2}}
\put(  1.27,  1.74){\circle*{0.2}}
\put(  1.40,  1.59){\circle*{0.2}}
\put(  1.53,  1.43){\circle*{0.2}}
\put(  1.66,  1.27){\circle*{0.2}}
\put(  1.78,  1.09){\circle*{0.2}}
\put(  1.91,  0.91){\circle*{0.2}}
\put(  2.04,  0.72){\circle*{0.2}}
\put(  2.17,  0.52){\circle*{0.2}}
\put(  2.29,  0.33){\circle*{0.2}}
\put(  2.42,  0.13){\circle*{0.2}}
\put(  2.55, -0.07){\circle*{0.2}}
\put(  2.68, -0.27){\circle*{0.2}}
\put(  2.80, -0.47){\circle*{0.2}}
\put(  2.93, -0.66){\circle*{0.2}}
\put(  3.06, -0.86){\circle*{0.2}}
\put(  3.18, -1.04){\circle*{0.2}}
\put(  3.31, -1.22){\circle*{0.2}}
\put(  3.44, -1.39){\circle*{0.2}}
\put(  3.57, -1.55){\circle*{0.2}}
\put(  3.69, -1.70){\circle*{0.2}}
\put(  3.82, -1.84){\circle*{0.2}}
\put(  3.95, -1.97){\circle*{0.2}}
\put(  4.08, -2.09){\circle*{0.2}}
\put(  4.20, -2.19){\circle*{0.2}}
\put(  4.33, -2.28){\circle*{0.2}}
\put(  4.46, -2.36){\circle*{0.2}}
\put(  4.59, -2.41){\circle*{0.2}}
\put(  4.71, -2.46){\circle*{0.2}}
\put(  4.84, -2.49){\circle*{0.2}}
\put(  4.97, -2.50){\circle*{0.2}}
\put(  5.10, -2.50){\circle*{0.2}}
\put(  5.22, -2.48){\circle*{0.2}}
\put(  5.35, -2.44){\circle*{0.2}}
\put(  5.48, -2.39){\circle*{0.2}}
\put(  5.61, -2.32){\circle*{0.2}}
\put(  5.73, -2.24){\circle*{0.2}}
\put(  5.86, -2.15){\circle*{0.2}}
\put(  5.99, -2.04){\circle*{0.2}}
\put(  6.11, -1.91){\circle*{0.2}}
\put(  6.24, -1.78){\circle*{0.2}}
\put(  6.37, -1.63){\circle*{0.2}}
\put(  6.50, -1.48){\circle*{0.2}}
\put(  6.62, -1.31){\circle*{0.2}}
\put(  6.75, -1.14){\circle*{0.2}}
\put(  6.88, -0.96){\circle*{0.2}}
\put(  7.01, -0.77){\circle*{0.2}}
\put(  7.13, -0.58){\circle*{0.2}}
\put(  7.26, -0.38){\circle*{0.2}}
\put(  7.39, -0.18){\circle*{0.2}}
\put(  7.52,  0.02){\circle*{0.2}}
\put(  7.64,  0.22){\circle*{0.2}}
\put(  7.77,  0.42){\circle*{0.2}}
\put(  7.90,  0.61){\circle*{0.2}}
\put(  8.03,  0.80){\circle*{0.2}}
\put(  8.15,  0.99){\circle*{0.2}}
\put(  8.28,  1.17){\circle*{0.2}}
\put(  8.41,  1.34){\circle*{0.2}}
\put(  8.54,  1.51){\circle*{0.2}}
\put(  8.66,  1.66){\circle*{0.2}}
\put(  8.79,  1.81){\circle*{0.2}}
\put(  8.92,  1.94){\circle*{0.2}}
\put(  9.04,  2.06){\circle*{0.2}}
\put(  9.17,  2.17){\circle*{0.2}}
\put(  9.30,  2.26){\circle*{0.2}}
\put(  9.43,  2.34){\circle*{0.2}}
\put(  9.55,  2.40){\circle*{0.2}}
\put(  9.68,  2.45){\circle*{0.2}}
\put(  9.81,  2.48){\circle*{0.2}}
\put(  9.94,  2.50){\circle*{0.2}}
\put( 10.06,  2.50){\circle*{0.2}}
\put( 10.19,  2.48){\circle*{0.2}}
\put( 10.32,  2.45){\circle*{0.2}}
\put( 10.45,  2.40){\circle*{0.2}}
\put( 10.57,  2.34){\circle*{0.2}}
\put( 10.70,  2.27){\circle*{0.2}}
\put( 10.83,  2.17){\circle*{0.2}}
\put( 10.96,  2.07){\circle*{0.2}}
\put( 11.08,  1.95){\circle*{0.2}}
\put( 11.21,  1.82){\circle*{0.2}}
\put( 11.34,  1.67){\circle*{0.2}}
\put( 11.46,  1.52){\circle*{0.2}}
\put( 11.59,  1.36){\circle*{0.2}}
\put( 11.72,  1.19){\circle*{0.2}}
\put( 11.85,  1.01){\circle*{0.2}}
\put( 11.97,  0.82){\circle*{0.2}}
\put( 12.10,  0.63){\circle*{0.2}}
\put( 12.23,  0.43){\circle*{0.2}}
\put( 12.36,  0.23){\circle*{0.2}}
\put( 12.48,  0.03){\circle*{0.2}}
\put( 12.61, -0.16){\circle*{0.2}}
\put( 12.74, -0.36){\circle*{0.2}}
\put( 12.87, -0.56){\circle*{0.2}}
\put( 12.99, -0.75){\circle*{0.2}}
\put( 13.12, -0.94){\circle*{0.2}}
\put( 13.25, -1.12){\circle*{0.2}}
\put( 13.38, -1.30){\circle*{0.2}}
\put( 13.50, -1.46){\circle*{0.2}}
\put( 13.63, -1.62){\circle*{0.2}}
\put( 13.76, -1.77){\circle*{0.2}}
\put( 13.89, -1.90){\circle*{0.2}}
\put( 14.01, -2.03){\circle*{0.2}}
\put( 14.14, -2.14){\circle*{0.2}}
\put( 14.27, -2.23){\circle*{0.2}}
\put( 14.39, -2.32){\circle*{0.2}}
\put( 14.52, -2.38){\circle*{0.2}}
\put( 14.65, -2.44){\circle*{0.2}}
\put( 14.78, -2.47){\circle*{0.2}}
\put( 14.90, -2.49){\circle*{0.2}}
\put( 15.03, -2.50){\circle*{0.2}}
\put( 15.16, -2.49){\circle*{0.2}}
\put( 15.29, -2.46){\circle*{0.2}}
\put( 15.41, -2.42){\circle*{0.2}}
\put( 15.54, -2.36){\circle*{0.2}}
\put( 15.67, -2.29){\circle*{0.2}}
\put( 15.80, -2.20){\circle*{0.2}}
\put( 15.92, -2.10){\circle*{0.2}}
\put( 16.05, -1.98){\circle*{0.2}}
\put( 16.18, -1.85){\circle*{0.2}}
\put( 16.31, -1.71){\circle*{0.2}}
\put( 16.43, -1.56){\circle*{0.2}}
\put( 16.56, -1.40){\circle*{0.2}}
\put( 16.69, -1.23){\circle*{0.2}}
\put( 16.82, -1.05){\circle*{0.2}}
\put( 16.94, -0.87){\circle*{0.2}}
\put( 17.07, -0.68){\circle*{0.2}}
\put( 17.20, -0.49){\circle*{0.2}}
\put( 17.32, -0.29){\circle*{0.2}}
\put( 17.45, -0.09){\circle*{0.2}}
\put( 17.58,  0.11){\circle*{0.2}}
\put( 17.71,  0.31){\circle*{0.2}}
\put( 17.83,  0.51){\circle*{0.2}}
\put( 17.96,  0.70){\circle*{0.2}}
\put( 18.09,  0.89){\circle*{0.2}}
\put( 18.22,  1.07){\circle*{0.2}}
\put( 18.34,  1.25){\circle*{0.2}}
\put( 18.47,  1.42){\circle*{0.2}}
\put( 18.60,  1.58){\circle*{0.2}}
\put( 18.73,  1.73){\circle*{0.2}}
\put( 18.85,  1.87){\circle*{0.2}}
\put( 18.98,  2.00){\circle*{0.2}}
\put( 19.11,  2.11){\circle*{0.2}}
\put( 19.24,  2.21){\circle*{0.2}}
\put( 19.36,  2.30){\circle*{0.2}}
\put( 19.49,  2.37){\circle*{0.2}}
\put( 19.62,  2.42){\circle*{0.2}}
\put( 19.75,  2.47){\circle*{0.2}}
\put( 19.87,  2.49){\circle*{0.2}}
}

\newsavebox{\bottomgluon}
\savebox{\bottomgluon}(0,0)[c]{
\put(  0.00, -2.50){\circle*{0.2}}
\put(  0.13, -2.49){\circle*{0.2}}
\put(  0.25, -2.47){\circle*{0.2}}
\put(  0.38, -2.43){\circle*{0.2}}
\put(  0.51, -2.37){\circle*{0.2}}
\put(  0.64, -2.30){\circle*{0.2}}
\put(  0.76, -2.22){\circle*{0.2}}
\put(  0.89, -2.12){\circle*{0.2}}
\put(  1.02, -2.01){\circle*{0.2}}
\put(  1.15, -1.88){\circle*{0.2}}
\put(  1.27, -1.74){\circle*{0.2}}
\put(  1.40, -1.59){\circle*{0.2}}
\put(  1.53, -1.43){\circle*{0.2}}
\put(  1.66, -1.27){\circle*{0.2}}
\put(  1.78, -1.09){\circle*{0.2}}
\put(  1.91, -0.91){\circle*{0.2}}
\put(  2.04, -0.72){\circle*{0.2}}
\put(  2.17, -0.52){\circle*{0.2}}
\put(  2.29, -0.33){\circle*{0.2}}
\put(  2.42, -0.13){\circle*{0.2}}
\put(  2.55,  0.07){\circle*{0.2}}
\put(  2.68,  0.27){\circle*{0.2}}
\put(  2.80,  0.47){\circle*{0.2}}
\put(  2.93,  0.66){\circle*{0.2}}
\put(  3.06,  0.86){\circle*{0.2}}
\put(  3.18,  1.04){\circle*{0.2}}
\put(  3.31,  1.22){\circle*{0.2}}
\put(  3.44,  1.39){\circle*{0.2}}
\put(  3.57,  1.55){\circle*{0.2}}
\put(  3.69,  1.70){\circle*{0.2}}
\put(  3.82,  1.84){\circle*{0.2}}
\put(  3.95,  1.97){\circle*{0.2}}
\put(  4.08,  2.09){\circle*{0.2}}
\put(  4.20,  2.19){\circle*{0.2}}
\put(  4.33,  2.28){\circle*{0.2}}
\put(  4.46,  2.36){\circle*{0.2}}
\put(  4.59,  2.41){\circle*{0.2}}
\put(  4.71,  2.46){\circle*{0.2}}
\put(  4.84,  2.49){\circle*{0.2}}
\put(  4.97,  2.50){\circle*{0.2}}
\put(  5.10,  2.50){\circle*{0.2}}
\put(  5.22,  2.48){\circle*{0.2}}
\put(  5.35,  2.44){\circle*{0.2}}
\put(  5.48,  2.39){\circle*{0.2}}
\put(  5.61,  2.32){\circle*{0.2}}
\put(  5.73,  2.24){\circle*{0.2}}
\put(  5.86,  2.15){\circle*{0.2}}
\put(  5.99,  2.04){\circle*{0.2}}
\put(  6.11,  1.91){\circle*{0.2}}
\put(  6.24,  1.78){\circle*{0.2}}
\put(  6.37,  1.63){\circle*{0.2}}
\put(  6.50,  1.48){\circle*{0.2}}
\put(  6.62,  1.31){\circle*{0.2}}
\put(  6.75,  1.14){\circle*{0.2}}
\put(  6.88,  0.96){\circle*{0.2}}
\put(  7.01,  0.77){\circle*{0.2}}
\put(  7.13,  0.58){\circle*{0.2}}
\put(  7.26,  0.38){\circle*{0.2}}
\put(  7.39,  0.18){\circle*{0.2}}
\put(  7.52, -0.02){\circle*{0.2}}
\put(  7.64, -0.22){\circle*{0.2}}
\put(  7.77, -0.42){\circle*{0.2}}
\put(  7.90, -0.61){\circle*{0.2}}
\put(  8.03, -0.80){\circle*{0.2}}
\put(  8.15, -0.99){\circle*{0.2}}
\put(  8.28, -1.17){\circle*{0.2}}
\put(  8.41, -1.34){\circle*{0.2}}
\put(  8.54, -1.51){\circle*{0.2}}
\put(  8.66, -1.66){\circle*{0.2}}
\put(  8.79, -1.81){\circle*{0.2}}
\put(  8.92, -1.94){\circle*{0.2}}
\put(  9.04, -2.06){\circle*{0.2}}
\put(  9.17, -2.17){\circle*{0.2}}
\put(  9.30, -2.26){\circle*{0.2}}
\put(  9.43, -2.34){\circle*{0.2}}
\put(  9.55, -2.40){\circle*{0.2}}
\put(  9.68, -2.45){\circle*{0.2}}
\put(  9.81, -2.48){\circle*{0.2}}
\put(  9.94, -2.50){\circle*{0.2}}
\put( 10.06, -2.50){\circle*{0.2}}
\put( 10.19, -2.48){\circle*{0.2}}
\put( 10.32, -2.45){\circle*{0.2}}
\put( 10.45, -2.40){\circle*{0.2}}
\put( 10.57, -2.34){\circle*{0.2}}
\put( 10.70, -2.27){\circle*{0.2}}
\put( 10.83, -2.17){\circle*{0.2}}
\put( 10.96, -2.07){\circle*{0.2}}
\put( 11.08, -1.95){\circle*{0.2}}
\put( 11.21, -1.82){\circle*{0.2}}
\put( 11.34, -1.67){\circle*{0.2}}
\put( 11.46, -1.52){\circle*{0.2}}
\put( 11.59, -1.36){\circle*{0.2}}
\put( 11.72, -1.19){\circle*{0.2}}
\put( 11.85, -1.01){\circle*{0.2}}
\put( 11.97, -0.82){\circle*{0.2}}
\put( 12.10, -0.63){\circle*{0.2}}
\put( 12.23, -0.43){\circle*{0.2}}
\put( 12.36, -0.23){\circle*{0.2}}
\put( 12.48, -0.03){\circle*{0.2}}
\put( 12.61,  0.16){\circle*{0.2}}
\put( 12.74,  0.36){\circle*{0.2}}
\put( 12.87,  0.56){\circle*{0.2}}
\put( 12.99,  0.75){\circle*{0.2}}
\put( 13.12,  0.94){\circle*{0.2}}
\put( 13.25,  1.12){\circle*{0.2}}
\put( 13.38,  1.30){\circle*{0.2}}
\put( 13.50,  1.46){\circle*{0.2}}
\put( 13.63,  1.62){\circle*{0.2}}
\put( 13.76,  1.77){\circle*{0.2}}
\put( 13.89,  1.90){\circle*{0.2}}
\put( 14.01,  2.03){\circle*{0.2}}
\put( 14.14,  2.14){\circle*{0.2}}
\put( 14.27,  2.23){\circle*{0.2}}
\put( 14.39,  2.32){\circle*{0.2}}
\put( 14.52,  2.38){\circle*{0.2}}
\put( 14.65,  2.44){\circle*{0.2}}
\put( 14.78,  2.47){\circle*{0.2}}
\put( 14.90,  2.49){\circle*{0.2}}
\put( 15.03,  2.50){\circle*{0.2}}
\put( 15.16,  2.49){\circle*{0.2}}
\put( 15.29,  2.46){\circle*{0.2}}
\put( 15.41,  2.42){\circle*{0.2}}
\put( 15.54,  2.36){\circle*{0.2}}
\put( 15.67,  2.29){\circle*{0.2}}
\put( 15.80,  2.20){\circle*{0.2}}
\put( 15.92,  2.10){\circle*{0.2}}
\put( 16.05,  1.98){\circle*{0.2}}
\put( 16.18,  1.85){\circle*{0.2}}
\put( 16.31,  1.71){\circle*{0.2}}
\put( 16.43,  1.56){\circle*{0.2}}
\put( 16.56,  1.40){\circle*{0.2}}
\put( 16.69,  1.23){\circle*{0.2}}
\put( 16.82,  1.05){\circle*{0.2}}
\put( 16.94,  0.87){\circle*{0.2}}
\put( 17.07,  0.68){\circle*{0.2}}
\put( 17.20,  0.49){\circle*{0.2}}
\put( 17.32,  0.29){\circle*{0.2}}
\put( 17.45,  0.09){\circle*{0.2}}
\put( 17.58, -0.11){\circle*{0.2}}
\put( 17.71, -0.31){\circle*{0.2}}
\put( 17.83, -0.51){\circle*{0.2}}
\put( 17.96, -0.70){\circle*{0.2}}
\put( 18.09, -0.89){\circle*{0.2}}
\put( 18.22, -1.07){\circle*{0.2}}
\put( 18.34, -1.25){\circle*{0.2}}
\put( 18.47, -1.42){\circle*{0.2}}
\put( 18.60, -1.58){\circle*{0.2}}
\put( 18.73, -1.73){\circle*{0.2}}
\put( 18.85, -1.87){\circle*{0.2}}
\put( 18.98, -2.00){\circle*{0.2}}
\put( 19.11, -2.11){\circle*{0.2}}
\put( 19.24, -2.21){\circle*{0.2}}
\put( 19.36, -2.30){\circle*{0.2}}
\put( 19.49, -2.37){\circle*{0.2}}
\put( 19.62, -2.42){\circle*{0.2}}
\put( 19.75, -2.47){\circle*{0.2}}
\put( 19.87, -2.49){\circle*{0.2}}
}

\newcommand{\be}{\begin{equation}}
\newcommand{\ee}{\end{equation}}
\newcommand{\bea}{\begin{eqnarray}}
\newcommand{\eea}{\end{eqnarray}}

\newcommand{\NPB}[3]{Nucl.\ Phys.\ {\bf B{#1}} (19{#2}) {#3}}
\newcommand{\PRD}[3]{Phys.\ Rev.\ {\bf D{#1}} (19{#2}) {#3}}
\newcommand{\PLB}[3]{Phys.\ Lett.\ {\bf B{#1}} (19{#2}) {#3}}

\begin{document}

\begin{titlepage}

\begin{center}

{\Large Centre de Physique Th\'eorique\footnote{Unit\'e Propre de
Recherche 7061} - CNRS - Luminy, Case 907}

{\Large F-13288 Marseille Cedex 9 - France}

\vspace{3 cm}

{\large {\bf EFFECTIVE CHARGES IN NON-ABELIAN
GAUGE THEORIES}}\footnote{
To appear in the proceedings of the Ringberg Workshop on
``Perspectives for Electroweak Interactions in $e^{+}e^{-}$ Collisions'',
hosted by the Max Planck Institut f\"ur Physik at Ringberg Castle,
Munich, February 5-8, 1995, editor B.\ Kniehl.}

\vspace{0.3 cm}

{\bf N.J. Watson}\footnote{email: watson@cptsu4.univ-mrs.fr}

\vspace{2.5 cm}

{\bf Abstract}

\end{center}

The question of the definition of effective charges for
non-abelian gauge theories is discussed, focusing in particular on
both the pinch technique and background field method approaches.
It is argued that there does exist a unique generalization of the
QED concept of an effective charge to non-abelian theories,
and that this generalization is given by the pinch technique.
The discussion is set in the wider (and controversial) context of the
definition of gauge-independent Green function-like quantities
in general in such theories.

\vspace{2 cm}


\bigskip

\noindent Number of figures: 4

\bigskip

\noindent May 1995

\noindent CPT-95/P.3198

\bigskip

\noindent anonymous ftp or gopher: cpt.univ-mrs.fr

\end{titlepage}

\centerline{\normalsize\bf EFFECTIVE CHARGES IN NON-ABELIAN
GAUGE THEORIES}
\baselineskip=16pt

\vspace*{0.6cm}
\centerline{\footnotesize N.J. WATSON}
\baselineskip=13pt
\centerline{\footnotesize\it Centre de Physique Th\'eorique, CNRS Luminy,}
\baselineskip=12pt
\centerline{\footnotesize\it F-13288 Marseille Cedex 9, France.}
\centerline{\footnotesize E-mail: watson@cptsu4.univ-mrs.fr}

\vspace*{0.9cm}
\abstracts{The question of the definition of effective charges for
non-abelian gauge theories is discussed, focusing in particular on
both the pinch technique and background field method approaches.
It is argued that there does exist a unique generalization of the
QED concept of an effective charge to non-abelian theories,
and that this generalization is given by the pinch technique.
The discussion is set in the wider (and controversial) context of the
definition of gauge-independent Green function-like quantities
in general in such theories.}

\normalsize\baselineskip=15pt
\setcounter{footnote}{0}
\renewcommand{\thefootnote}{\alph{footnote}}
\section{Introduction}
With the advent of the LEP machine at CERN, the study of radiative
corrections to the tree level predictions of the electroweak sector of
the Standard Model (SM) has become of enormous importance. Because these
corrections are formulated in the framework of perturbation theory, in
which one necessarily has to break the gauge invariance of the
classical lagrangian, issues of gauge dependence naturally arise.
While physical observables are known from general proofs to be
independent of the particular gauge fixing procedure used, the
intermediate steps of any perturbative calculation involve Green
functions which in general are gauge-dependent. Although this
gauge dependence necessarily cancels in the end, the gauge dependence
of the individual Green functions may be very strong, even introducing
spurious singularities. Indeed, in some gauges the behaviour of the
individual Green functions may be such as to obscure completely
important characteristics of the theory. Perhaps the most
spectacular example of this is the standard statement that the
electroweak sector is non-renormalizable in the unitary gauge.

As long as one is only interested in calculating S-matrix elements in
a given theory, this gauge dependence of the Green functions does not
represent a problem. But in attempting to parameterize and detect
deviations from the SM predictions resulting from the
effects of ``new physics'', particularly at accuracies beyond tree
level, the fact that the basic building blocks of perturbation theory
are gauge-dependent causes difficulties. For example, in attempting to
parameterize the deviations from the SM tree level predictions for
the electroweak triple gauge
vertices soon to be measured at LEP, the conventional one-loop
proper three-point
functions calculated in the SM for off-shell particles are not only
gauge-dependent, but in the Feynman gauge
involve contributions which are both infrared divergent
and badly behaved at high energies.
This is before any attempt has been made to include the effects of
possible non-SM physics.

Over the last few years there have been various
suggestions\cite{kennedy1}$^-$\cite{kuroda} as to how
to reorganize in a more physical and intuitive way the Green functions
for one-loop radiative corrections occurring in non-abelian gauge
theories. The basic idea is to rearrange parts of the contributions
from the various one-loop $n$-point functions to form sets of
gauge-independent self-energy-like, vertex-like and box-like
functions. The most systematic of these schemes is
the pinch technique (PT), introduced originally by
Cornwall\cite{cornwall1}$^-$\cite{cornwall4} in the context
of QCD, and since extensively developed and applied by Papavassiliou and
collaborators\cite{papav1}$^-$\cite{degrassi2}.
The PT provides a well-defined algorithm
for the rearrangement of one-loop corrections to tree-level processes,
with the resulting functions, in addition to being gauge-independent,
displaying many theoretically desirable properties.
In particular, the PT one-loop functions satisfy the same set of Ward
identities as the corresponding tree level Green functions.
Furthermore, it has been shown by Degrassi and Sirlin\cite{degrassi1}
that the PT algorithm in fact
corresponds to a systematic use of current algebra, thus demonstrating
explicitly the PT's basis in the underlying gauge symmetry of the theory.

An alternative approach to the calculation of radiative corrections
is the background field method (BFM)\cite{dewitt1}$^-$\cite{abbott2}.
By splitting the gauge fields into
background and quantum components and then choosing a gauge fixing for
the quantum fields such that explicit gauge invariance of the
background fields is retained, the BFM
provides a very appealing framework in which to carry out the
calculation of radiative corrections. This has recently been
demonstrated explicitly with the application of the BFM to the
electroweak sector of the SM\cite{denner1}$^-$\cite{denner2}.
It has been shown that the PT functions are obtained directly
in the BFM as the background field Green functions
for the particular choice of the Feynman quantum gauge
$\xi_{q} = 1$\cite{rafael}$^-$\cite{hashimoto},
while all of the desirable properties
of the PT functions are obtained in the BFM for {\em any} choice of the
quantum gauge parameter $\xi_{q}$\cite{denner1}$^-$\cite{denner2}.
This includes the background field
Green functions satisfying to all orders the PT tree level-like Ward
identities, a direct result of the exact background gauge invariance
of the BFM effective action.
It was argued\cite{denner1}$^-$\cite{denner2}
that the PT is
therefore not distinguished on physical grounds, but rather
represents just one of an infinity of choices to obtain well-behaved
Green function-like quantities, this choice being parameterized in the
BFM by the quantum gauge fixing parameter $\xi_{q}$.

If this conclusion is correct, then clearly it makes (or rather leaves)
difficult the analysis and interpretation of the triple gauge vertex
measurements at LEP: beyond tree level, what one regards as the vertex
function must be a matter of convention.
However, it will be argued here that the PT precisely {\it is}
distinguished on physical grounds from the general BFM approach.
The discussion centres on the specific question, does there
exist a unique, unambiguous way to extend the concept of an
effective charge from QED to non-abelian gauge theories?
It will therefore concentrate on the gauge boson two-point function.
This question is of interest not just from a phenomenological point of
view but also, because the effective charge sums an infinite
Dyson series of radiative corrections
and so goes beyond perturbation theory,
is central to renormalon approaches to QCD.
After reviewing the problem, a discussion is given of the gauge boson
two-point functions obtained in both the PT and BFM approaches.
It is then argued that there does indeed exist a natural
generalization
of the idea of an effective charge to non-abelian theories, this
generalization being given by the PT.
For simplicity, only the case of an unbroken non-abelian
theory---SU($N$) QCD---is considered.

\section{The Gauge Boson Two-Point Function}

In the classical lagrangian for QED,
\begin{equation}
{\cal L}_{\rm cl}
=
-{\textstyle \frac{1}{4}}F_{\mu\nu}F^{\mu\nu}
+ \overline{\psi}[i\gamma^{\mu}(\partial_{\mu} - ieA_{\mu}) - m]\psi
\end{equation}
there is only a single interaction vertex, viz.\ that of the gauge boson
(photon) with a fermion-antifermion (electron-positron) pair.
The photon self-energy due to this interaction
is transverse and gauge-independent to all orders in perturbation theory:
\begin{eqnarray}
& & \nonumber \\
\hspace*{5cm}
&=&
(q^{2}g_{\mu\nu} - q_{\mu}q_{\nu})i\Pi(q^{2}). \\
& & \nonumber
\end{eqnarray}
\noindent
\begin{picture}(-8,0)(-20,483)

\thicklines

\put( 70,540){\makebox(0,0)[r]{$A_{\mu}$}}
\put( 80,540){\usebox{\gluon}}
\put( 90,540){\usebox{\rgluonarrow}}
\put(100,540){\usebox{\gluon}}
\put(100,525){\makebox(0,0)[c]{$q$}}
\put(140,540){\circle{39}}
\multiput(130,550)( 0.5, 0.5){18}{\circle*{0.2}}
\multiput(130,550)(-0.5,-0.5){18}{\circle*{0.2}}
\multiput(135,545)( 0.5, 0.5){26}{\circle*{0.2}}
\multiput(135,545)(-0.5,-0.5){25}{\circle*{0.2}}
\multiput(140,540)( 0.5, 0.5){27}{\circle*{0.2}}
\multiput(140,540)(-0.5,-0.5){27}{\circle*{0.2}}
\multiput(145,535)( 0.5, 0.5){25}{\circle*{0.2}}
\multiput(145,535)(-0.5,-0.5){25}{\circle*{0.2}}
\multiput(150,530)( 0.5, 0.5){18}{\circle*{0.2}}
\multiput(150,530)(-0.5,-0.5){18}{\circle*{0.2}}
\put(160,540){\usebox{\gluon}}
\put(170,540){\usebox{\rgluonarrow}}
\put(180,525){\makebox(0,0)[c]{$q$}}
\put(180,540){\usebox{\gluon}}
\put(210,540){\makebox(0,0)[l]{$A_{\nu}$}}

\thinlines

\end{picture}
After renormalization, this 1PI
photon self-energy may be summed in a Dyson series to give, at a
given order of perturbation theory, the renormalized dressed photon
propagator
\begin{equation}\label{qedprop}
i\Delta_{R\mu\nu}(q)
=
\frac{i}{q^{2} + i\epsilon}\Biggl\{
\biggl( - g_{\mu\nu} + \frac{q_{\mu}q_{\nu}}{q^{2}}\biggr)d_{R}(q^{2})
- \xi\frac{q_{\mu}q_{\nu}}{q^{2}}
\Biggr\}
\end{equation}
where $\xi$ is the gauge parameter in the class of conventional
$R_{\xi}$ gauges. This photon propagator then naturally defines a
gauge-independent effective charge for the abelian theory:
\begin{equation}\label{qedeffcharge}
e_{R}^{2}d_{R}(q^{2})
=
\frac{e_{R}^{2}}{1 - \Pi_{R}(q^{2})}
=
e_{\rm eff}^{2}(q^{2})
\end{equation}
where $e_{R}$ is the renormalized coupling constant.
At $q^{2} = 0$ (the Thomson limit),
this effective charge matches on to the fine structure
constant ($e_{\rm eff}^{2}(0)/4\pi = \alpha = 1/137.03\ldots$).
Furthermore, in addition to being gauge-independent,
as a result of the famous QED relation $Z_{1} = Z_{2}$, the QED
effective charge is also renormalization scale-independent.
At asymptotic $q^{2}$ values, it therefore obeys a homogeneous
Callan-Symanzik equation involving the QED $\beta$ function.

In the classical lagrangian for QCD with $n_{f}$ flavours of fermion,
\begin{equation}\label{qcd}
{\cal L}_{\rm cl}
=
-{\textstyle \frac{1}{4}}F_{\mu\nu}^{a}F^{a\mu\nu}
+ \sum_{f=1}^{n_{f}}
\overline{\psi}^{^{(f)}}
[i\gamma^{\mu}(\partial_{\mu} - igA_{\mu}^{a}T^{a})
- m_{f}]\psi^{^{(f)}}
\end{equation}
in addition to the interaction of the gauge bosons (gluons) with the
fermions (quarks) similar to that of QED,
the gauge bosons also couple directly to one
another in triple and quadruple gauge boson vertices. Although as in
QED a Ward (more correctly Slavnov-Taylor)
identity guarantees that the gauge boson self-energy
is transverse, as a result of the gauge boson self-interactions it is
gauge-dependent:
\begin{eqnarray}
& & \nonumber \\
\hspace*{5cm}
&=&
(q^{2}g_{\mu\nu} - q_{\mu}q_{\nu})\delta^{ab}i\Pi(\xi,q^{2}). \\
& & \nonumber
\end{eqnarray}
\noindent
\begin{picture}(-8,0)(-20,483)

\thicklines

\put( 70,540){\makebox(0,0)[r]{$A_{\mu}^{a}$}}
\put( 80,540){\usebox{\gluon}}
\put( 90,540){\usebox{\rgluonarrow}}
\put(100,540){\usebox{\gluon}}
\put(100,525){\makebox(0,0)[c]{$q$}}
\put(140,540){\circle{39}}
\multiput(130,550)( 0.5, 0.5){18}{\circle*{0.2}}
\multiput(130,550)(-0.5,-0.5){18}{\circle*{0.2}}
\multiput(135,545)( 0.5, 0.5){25}{\circle*{0.2}}
\multiput(135,545)(-0.5,-0.5){25}{\circle*{0.2}}
\multiput(140,540)( 0.5, 0.5){27}{\circle*{0.2}}
\multiput(140,540)(-0.5,-0.5){27}{\circle*{0.2}}
\multiput(145,535)( 0.5, 0.5){25}{\circle*{0.2}}
\multiput(145,535)(-0.5,-0.5){25}{\circle*{0.2}}
\multiput(150,530)( 0.5, 0.5){18}{\circle*{0.2}}
\multiput(150,530)(-0.5,-0.5){18}{\circle*{0.2}}
\put(160,540){\usebox{\gluon}}
\put(170,540){\usebox{\rgluonarrow}}
\put(180,525){\makebox(0,0)[c]{$q$}}
\put(180,540){\usebox{\gluon}}
\put(210,540){\makebox(0,0)[l]{$A_{\nu}^{b}$}}

\thinlines

\end{picture}
Furthermore, while it is possible after renormalization
to sum this self-energy in a Dyson
series to give a radiatively-corrected gauge field propagator, the
quantity $g_{\rm eff}(\xi,q^{2})$ defined by
analogy with the QED effective charge Eq.\ (\ref{qedeffcharge})
does not at all have the high energy
behaviour expected from the QCD $\beta$ function. The simple
QED correspondence between the gauge boson two-point function and an
effective charge for the theory is therefore lost.

\section{The Pinch Technique}
The pinch technique (PT) is based on the
observation\cite{cornwall1}$^-$\cite{cornwall4} that in a
non-abelian gauge theory, one-loop diagrams which appear to give only
vertex or box corrections to tree level processes in fact implicitly
contain propagator-like components. It is important to
emphasize immediately that this statement is not simply to do with the
kinematics of a given process, but has a precise mathematical
expression in terms of the tree level Feynman rules of the theory.

In order to illustrate this, consider the four fermion scattering process
$\psi_{i}^{^{(f)}}\psi_{i'}^{^{(f')}}
\rightarrow \psi_{j}^{^{(f)}}\psi_{j'}^{^{(f')}}$ in
SU($N$) QCD, Eq.\ (\ref{qcd}). The
complete set of one-loop diagrams for this process is shown in
Fig.\ 1. The contribution of the diagram in Fig.\ 1(a) involving the
conventional gauge boson two-point function is given by
\begin{equation}\label{3.1}
{\rm Fig.\,\,1(a)}
=
\Bigl(\overline{u}_{j'}ig\gamma^{\alpha}T_{j'i'}^{a}u_{i'}\Bigr)
\,\frac{-i}{q^{2}}\,
iq^{2}\Pi(\xi,q^{2})
\,\frac{-i}{q^{2}}\,
\Bigl(\overline{u}_{j}ig\gamma_{\alpha}T_{ji}^{a}u_{i}\Bigr)
\end{equation}
(the particular indices $i,i',j,j'$ are not summed).
The effect of the PT algorithm is to extract the propagator-like
components of the remaining diagrams in Fig.\ 1,
{\it defined as those parts of the diagrams which have exactly the form
of Eq.\ (\ref{3.1}), i.e.\ a function of $q^{2}$ between two tree level
vector-fermion-fermion vertices}. These are the pinch parts of the
diagrams, shown in Figs.\ 1(c) and (e). Adding these pinch parts of
the vertex and box diagrams to the diagram Fig.\ 1(a) involving the
conventional gauge boson two-point function gives the full one-loop
propagator-like contribution to the four fermion process,
illustrated schematically in Fig.\ 1(h), and defines the PT
gauge boson self-energy $\hat{\Pi}(q^{2})$:
\begin{equation}\label{3.2}
{\rm Fig.\,\,1(h)}
=
\Bigl(\overline{u}_{j'}ig\gamma^{\alpha}T_{j'i'}^{a}u_{i'}\Bigr)
\,\frac{-i}{q^{2}}\,
iq^{2}\hat{\Pi}(q^{2})
\,\frac{-i}{q^{2}}\,
\Bigl(\overline{u}_{j}ig\gamma_{\alpha}T_{ji}^{a}u_{i}\Bigr).
\end{equation}

\pagebreak

\noindent
\begin{picture}(570,685)(-20,25)

\put(-20,75){\makebox(0,0)[l]{\small
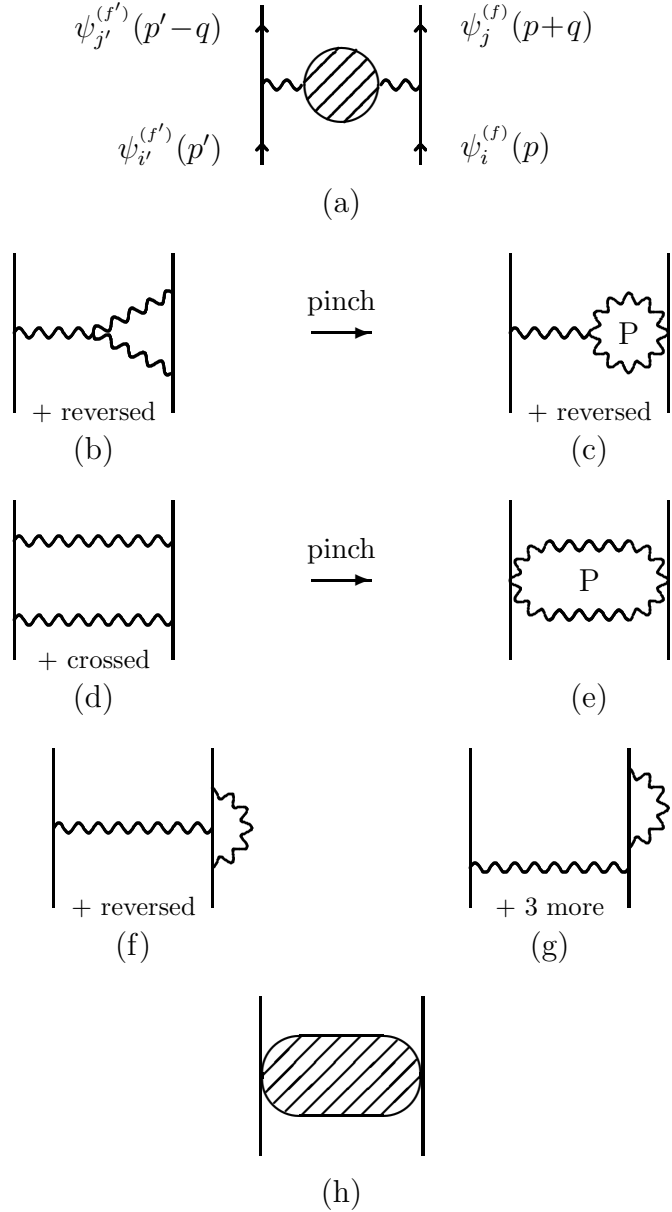
Fig.\ 1. The set of one-loop diagrams for the four fermion scattering
process and their }}
\put(-20,55){\makebox(0,0)[l]{\small
pinch parts, denoted by ``P''. The pinch technique self-energy is
shown in (h).}}

\thicklines


\put(205,635){\makebox(0,0)[r]{$\psi_{i'}^{^{(f')}}(p')$}}
\put(205,695){\makebox(0,0)[r]{$\psi_{j'}^{^{(f')}}(p'\!-\!q)$}}
\put(225,625){\line(0,1){80}}
\multiput(225,695)(-0.1,-0.1){28}{\circle*{0.2}}
\multiput(225,695)( 0.1,-0.1){28}{\circle*{0.2}}
\multiput(225,635)(-0.1,-0.1){28}{\circle*{0.2}}
\multiput(225,635)( 0.1,-0.1){28}{\circle*{0.2}}
\put(225,665){\usebox{\gluon}}
\put(265,665){\circle{39}}
\multiput(255,675)( 0.5, 0.5){18}{\circle*{0.2}}
\multiput(255,675)(-0.5,-0.5){18}{\circle*{0.2}}
\multiput(260,670)( 0.5, 0.5){26}{\circle*{0.2}}
\multiput(260,670)(-0.5,-0.5){25}{\circle*{0.2}}
\multiput(265,665)( 0.5, 0.5){27}{\circle*{0.2}}
\multiput(265,665)(-0.5,-0.5){27}{\circle*{0.2}}
\multiput(270,660)( 0.5, 0.5){25}{\circle*{0.2}}
\multiput(270,660)(-0.5,-0.5){25}{\circle*{0.2}}
\multiput(275,655)( 0.5, 0.5){18}{\circle*{0.2}}
\multiput(275,655)(-0.5,-0.5){18}{\circle*{0.2}}
\put(285,665){\usebox{\gluon}}
\put(305,625){\line(0,1){80}}
\multiput(305,695)(-0.1,-0.1){28}{\circle*{0.2}}
\multiput(305,695)( 0.1,-0.1){28}{\circle*{0.2}}
\multiput(305,635)(-0.1,-0.1){28}{\circle*{0.2}}
\multiput(305,635)( 0.1,-0.1){28}{\circle*{0.2}}
\put(325,635){\makebox(0,0)[l]{$\psi_{i}^{^{(f)}}(p)$}}
\put(325,695){\makebox(0,0)[l]{$\psi_{j}^{^{(f)}}(p\!+\!q)$}}
\put(265,605){\makebox(0,0)[c]{(a)}}


\put(100,500){\line(0,1){80}}
\put(100,540){\usebox{\gluon}}
\put(120,540){\usebox{\gluon}}
\put(140,540){\usebox{\trgluonone}}
\put(140,540){\usebox{\trgluontwo}}
\put(140,540){\usebox{\brgluon}}
\put(180,500){\line(0,1){80}}
\put(140,500){\makebox(0,0)[c]{\footnotesize + reversed}}
\put(140,480){\makebox(0,0)[c]{(b)}}

\put(265,555){\makebox(0,0)[c]{\small pinch}}
\put(250,540){\vector(1, 0){30}}

\put(350,500){\line(0,1){80}}
\put(350,540){\usebox{\gluon}}
\put(370,540){\usebox{\gluon}}
\put(410,540){\makebox(0,0)[c]{P}}
\put(410,540){\usebox{\trgluonloop}}
\put(410,540){\usebox{\brgluonloop}}
\put(410,540){\usebox{\blgluonloop}}
\put(410,540){\usebox{\tlgluonloop}}
\put(430,500){\line(0,1){80}}
\put(390,500){\makebox(0,0)[c]{\footnotesize + reversed}}
\put(390,480){\makebox(0,0)[c]{(c)}}


\put(100,375){\line(0,1){80}}
\put(100,395){\usebox{\gluon}}
\put(120,395){\usebox{\gluon}}
\put(140,395){\usebox{\gluon}}
\put(160,395){\usebox{\gluon}}
\put(100,435){\usebox{\gluon}}
\put(120,435){\usebox{\gluon}}
\put(140,435){\usebox{\gluon}}
\put(160,435){\usebox{\gluon}}
\put(180,375){\line(0,1){80}}
\put(140,375){\makebox(0,0)[c]{\footnotesize + crossed}}
\put(140,355){\makebox(0,0)[c]{(d)}}

\put(265,430){\makebox(0,0)[c]{\small pinch}}
\put(250,415){\vector(1, 0){30}}

\put(350,375){\line(0,1){80}}
\put(370,415){\usebox{\blgluonloop}}
\put(370,415){\usebox{\tlgluonloop}}
\put(390,415){\makebox(0,0)[c]{P}}
\put(370,397.5){\usebox{\bottomgluon}}
\put(390,397.5){\usebox{\bottomgluon}}
\put(370,432.5){\usebox{\topgluon}}
\put(390,432.5){\usebox{\topgluon}}
\put(410,415){\usebox{\trgluonloop}}
\put(410,415){\usebox{\brgluonloop}}
\put(430,375){\line(0,1){80}}
\put(390,355){\makebox(0,0)[c]{(e)}}


\put(120,250){\line(0,1){80}}
\put(120,290){\usebox{\gluon}}
\put(140,290){\usebox{\gluon}}
\put(160,290){\usebox{\gluon}}
\put(180,290){\usebox{\gluon}}
\put(200,250){\line(0,1){80}}
\put(200,290){\usebox{\trgluonloop}}
\put(200,290){\usebox{\brgluonloop}}
\put(160,250){\makebox(0,0)[c]{\footnotesize + reversed}}
\put(160,230){\makebox(0,0)[c]{(f)}}

\put(330,250){\line(0,1){80}}
\put(330,270){\usebox{\gluon}}
\put(350,270){\usebox{\gluon}}
\put(370,270){\usebox{\gluon}}
\put(390,270){\usebox{\gluon}}
\put(410,250){\line(0,1){80}}
\put(410,300){\usebox{\trgluonloop}}
\put(410,300){\usebox{\brgluonloop}}
\put(370,250){\makebox(0,0)[c]{\footnotesize + 3 more}}
\put(370,230){\makebox(0,0)[c]{(g)}}


\put(224,125){\line(0,1){80}}
\put(265,165){\oval(80,40)}
\put(275,185){\line(-1,-1){38}}
\put(265,185){\line(-1,-1){34}}
\put(255,185){\line(-1,-1){28}}
\put(245,185){\line(-1,-1){20}}
\put(245,145){\line(1,1){40}}
\put(255,145){\line(1,1){38}}
\put(265,145){\line(1,1){34}}
\put(275,145){\line(1,1){28}}
\put(285,145){\line(1,1){20}}
\put(306,125){\line(0,1){80}}

\put(265,105){\makebox(0,0)[c]{(h)}}

\end{picture}

Having defined $\hat{\Pi}(q^{2})$ in this way, one can now
argue that it must be
gauge-independent: up to a trivial dependence on the external spinors,
the component Eq.\ (\ref{3.2}) of the S-matrix element for
the four fermion process depends
only on the $t$-channel momentum transfer $q^{2}$, and not on the
$s$-channel momentum transfer $(p+p')^{2}$
or the external fermions' masses. It must therefore
be individually gauge-independent, as can be
verified by explicit calculation.

The identification of the pinch parts is made at the level of the
Feynman integrals for the diagrams using the tree level
vector-fermion-fermion vertex Ward identity
\begin{equation}\label{fermionwid}
q^{\alpha}\gamma_{\alpha}
=
S^{-1}(p+q) - S^{-1}(p)
\end{equation}
where $S^{-1}(p) = p\!\!/ -m$ is the inverse fermion propagator,
$q$ is the four-momentum of the
incoming gauge boson and $p$ and $p+q$ are the four-momenta of the
fermions. In general, such factors of four-momentum arise both from
the longitudinal components of the gauge field propagators and from
triple gauge vertices. The latter may be decomposed as
\begin{equation}
g\Gamma_{\alpha\beta\gamma}^{abc}
=
gf^{abc}\Bigl(
\Gamma_{\alpha\beta\gamma}^{F} + \Gamma_{\alpha\beta\gamma}^{P}
\Bigr)
\end{equation}
where
\begin{eqnarray}
\Gamma_{\alpha\beta\gamma}^{F}(q,k,-k-q)
&=&
(2k+q)_{\alpha}g_{\beta\gamma}
-2q_{\beta}g_{\gamma\alpha}
+2q_{\gamma}g_{\alpha\beta} \\
\Gamma_{\alpha\beta\gamma}^{P}(q,k,-k-q)
&=&
-k_{\beta}g_{\gamma\alpha}
-(k+q)_{\gamma}g_{\alpha\beta}.
\end{eqnarray}
The part $\Gamma_{\alpha\beta\gamma}^{F}$ gives no pinch
contribution and obeys a simple QED-like Ward identity involving the
difference of two inverse gauge field propagators in the Feynman gauge,
$q^{\alpha}\Gamma_{\alpha\beta\gamma}^{F}(q,k,-k-q)
=[(k+q)^{2} - k^{2}]g_{\beta\gamma}$.
The part $\Gamma_{\alpha\beta\gamma}^{P}$ gives two pinch
contributions, one from $k_{\beta}$, the other from $(k+q)_{\gamma}$.
Use of the Ward identity Eq.\ (\ref{fermionwid}) for {\it all}
such factors of longitudinal
four-momentum appearing in the denominator of the
integral enables one to isolate
the components of the diagrams in which the
fermion propagators associated with the external lines are exactly
cancelled.

Given that the PT gauge boson self-energy
$\hat{\Pi}(q^{2})$ is gauge-independent, it
may be calculated using the most convenient choice of
gauge. Evidently, this is the
Feynman gauge $\xi = 1$, since then the gauge field propagators
$iD_{\mu\nu}^{ab}$ are proportional to $g_{\mu\nu}$, leaving only the
triple gauge vertex as a source of longitudinal four-momentum factors.
The box diagrams Fig.\ 1(d)
therefore have zero pinch part in this gauge, and the
pinch contribution to $\hat{\Pi}(q^{2})$ is given entirely by the
diagrams in Fig.\ 1(c).

The vertex in Fig.\ 1(b) is shown in more detail in Fig.\ 2.
Using the Ward identity Eq.\ (\ref{fermionwid}), the vertex diagram
in Fig.\ 2 can be written
\begin{eqnarray}\label{fig2a}
{\rm Fig.\ 2}
&=&
\mu^{2\epsilon}\int\frac{d^{n}k}{(2\pi)^{n}}\frac{1}{k^{2}(k+q)^{2}}
\Biggl\{
-gf^{abc}\Gamma_{\alpha\beta\gamma}^{F}(q,k,-k\!-\!q)
\,ig\gamma^{\gamma}T^{c}\,iS(p\!-\!k)\,ig\gamma^{\beta}T^{b}
\nonumber \\
& &
\hspace*{10mm}+ \textstyle{\frac{1}{2}}ig^{2}N\Bigl(
ig\gamma_{\alpha}T^{a}\,S(p-k)\,S^{-1}(p)
+ S^{-1}(p+q)\,S(p-k)\,ig\gamma_{\alpha}T^{a} \Bigr)
\Biggl\}
\nonumber \\ \label{fig2}
& &
- ig^{2}NA(q)\,ig\gamma_{\alpha}T^{a}
\end{eqnarray}
where $A(q)$ is defined by
\begin{equation}\label{A}
A(q)
=
\mu^{2\epsilon}\int\frac{d^{n}k}{(2\pi)^{n}}\frac{1}{k^{2}(k+q)^{2}}
\end{equation}
(we use always dimensional regularization in $n = 4-2\epsilon$
dimensions with 't Hooft
mass \hfill scale \hfill $\mu$).
When \hfill contracted \hfill into \hfill the \hfill spinors \hfill
for \hfill the \hfill on-shell \hfill external \hfill fermions,

\pagebreak

\noindent
\begin{picture}(570,260)(-20,45)

\put(-20,75){\makebox(0,0)[l]{\small
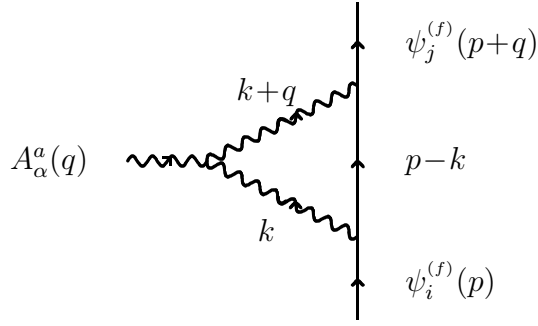
Fig.\ 2. The one-loop vector-fermion-fermion vertex contributing
a pinch part.}}

\thicklines

\put(165,210){\makebox(0,0)[r]{$A_{\alpha}^{a}(q)$}}
\put(185,210){\usebox{\gluon}}
\put(205,210){\usebox{\gluon}}
\put(195,210){\usebox{\rgluonarrow}}
\put(225,210){\usebox{\trgluonone}}
\put(225,210){\usebox{\trgluontwo}}
\put(252,224){\usebox{\trgluonarrow}}
\put(261,228){\usebox{\trgluonone}}
\put(261,228){\usebox{\trgluontwo}}
\put(225,210){\usebox{\brgluon}}
\put(252,197){\usebox{\brgluonarrow}}
\put(261,192){\usebox{\brgluon}}
\put(255,175){\makebox(0,0)[c]{$k$}}
\put(255,245){\makebox(0,0)[c]{$k\!+\!q$}}
\put(301,130){\line(0,1){160}}
\multiput(301,150)(-0.1,-0.1){28}{\circle*{0.2}}
\multiput(301,150)( 0.1,-0.1){28}{\circle*{0.2}}
\multiput(301,210)(-0.1,-0.1){28}{\circle*{0.2}}
\multiput(301,210)( 0.1,-0.1){28}{\circle*{0.2}}
\multiput(301,270)(-0.1,-0.1){28}{\circle*{0.2}}
\multiput(301,270)( 0.1,-0.1){28}{\circle*{0.2}}
\put(325,150){\makebox(0,0)[l]{$\psi_{i}^{^{(f)}}(p)$}}
\put(325,210){\makebox(0,0)[l]{$p\!-\!k$}}
\put(325,270){\makebox(0,0)[l]{$\psi_{j}^{^{(f)}}(p\!+\!q)$}}

\end{picture}

\noindent
the terms in the second line of Eq.\ (\ref{fig2}) vanish identically.
The term in the third line proportional to the tree level
vector-fermion-fermion vertex $ig\gamma_{\alpha}T^{a}$
is the pinch part of the diagram. The
remaining terms in the first line of Eq.\ (\ref{fig2}) are part of the
PT one-loop gauge-independent vector-fermion-fermion vertex, the other
part being given by the ``abelian-like'' vertex involved in Fig.\ 1(f).

Given that the box diagrams give zero pinch contribution in this
gauge, the PT gauge-independent
gauge boson self-energy is obtained just by
extracting the pinch part of the vertex in Fig.\ 2 together with the
identical contribution from the reversed diagram, and adding them to
the conventional self-energy:
\begin{equation}
\hat{\Pi}(q^{2})
=
\Pi(\xi\! =\! 1,q^{2}) - 2ig^{2}NA(q).
\end{equation}

The PT function $\hat{d}_{R}(q^{2})$ is defined via the Dyson
summation of the renormalized PT self-energy $\hat{\Pi}_{R}(q^{2})$:
\begin{equation}
\hat{d}_{R}(q^{2})
=
1 + \hat{\Pi}_{R}(q^{2}) + \hat{\Pi}_{R}^{2}(q^{2})
+ \hat{\Pi}_{R}^{3}(q^{2}) + \ldots
=
\Bigl(1 - \hat{\Pi}_{R}(q^{2})\Bigr)^{-1}.
\end{equation}
Using this function, it is then possible
to define the renormalized dressed gauge boson propagator
\begin{equation}\label{ptprop}
i\hat{\Delta}_{R\mu\nu}^{ab}(q)
=
\frac{i\delta^{ab}}{q^{2} + i\epsilon}\Biggl\{
\biggl( - g_{\mu\nu} + \frac{q_{\mu}q_{\nu}}{q^{2}}\biggr)
\hat{d}_{R}(q^{2})
- \xi\frac{q_{\mu}q_{\nu}}{q^{2}}
\Biggr\}
\end{equation}
and, by analogy with Eq.\ (\ref{qedeffcharge}), the corresponding
PT effective charge for QCD
\begin{equation}
g_{R}^{2}\hat{d}_{R}(q^{2})
=
\frac{g_{R}^{2}}{1 - \hat{\Pi}_{R}(q^{2})}
=
\hat{g}_{\rm eff}^{2}(q^{2}).
\end{equation}
At asymptotic values of $q^{2}$, this PT effective charge
satisfies a homogeneous Callan-Symanzik equation involving the
QCD $\beta$ function.

Thus, by extracting the parts of the one-loop
vertex and box diagrams for the four fermion scattering process
which are proportional to the {\it tree level} vector-fermion-fermion
vertices, and then combining these with the conventional one-loop gauge
boson self-energy, the PT self-energy $\hat{\Pi}(q^{2})$ accounts
for {\it all} of the one-loop corrections to the single
gauge boson mediating the tree level interaction between two fermion
lines.

\section{The Background Field Method}

The background field method (BFM) provides a way of calculating the
quantum corrections to tree level processes without losing
explicit gauge invariance of the
classical fields\cite{dewitt1}$^-$\cite{abbott2}. This is achieved by
first making a shift of the gauge field variable in the generating
functional
\begin{equation}\label{4.1}
A_{\mu}^{a}
\,\,\rightarrow\,\,
\tilde{A}_{\mu}^{a} +  A_{\mu}^{a}
\end{equation}
where $\tilde{A}_{\mu}^{a}$ is the background field and
$A_{\mu}^{a}$ is the quantum field, the latter being the integration
variable. The conventional gauge fixing term ${\cal L}_{\rm gf}$
is then replaced by two
gauge fixing terms, one for the quantum fields, ${\cal L}_{\rm qgf}$,
and one for the background fields, ${\cal L}_{\rm bgf}$:
\begin{equation}
-\frac{1}{2\xi}(\partial_{\mu}A^{a\mu})^{2}
\,\,\rightarrow\,\,
-\frac{1}{2\xi_{q}}(\tilde{D}_{\mu}^{ab}A^{b\mu})^{2}
-\frac{1}{2\xi_{b}}(\partial_{\mu}\tilde{A}^{a\mu})^{2}
\end{equation}
where $\tilde{D}_{\mu}$ is the background covariant derivative and an
$R_{\xi}$-like gauge has been chosen for the background fields. The
gauge parameters $\xi_{q}$ and $\xi_{b}$ are completely independent of
one another. The ghost term is constructed from the variation of
$\tilde{D}_{\mu}^{ab}A^{b\mu}$ under a quantum gauge transformation
\begin{equation}
\delta_{q}\tilde{A}_{\mu}^{a}
=
0,
\hspace{3cm}
\delta_{q}A_{\mu}^{a}
=
D_{\mu}^{ab}\omega^{b}.
\end{equation}
The terms
${\cal L}_{\rm cl} + {\cal L}_{\rm qgf} + {\cal L}_{\rm gh}$
involved in the path integral over the quantum fields therefore
remain exactly invariant under a background gauge transformation
\begin{equation}
\delta_{b}\tilde{A}_{\mu}^{a}
=
\tilde{D}_{\mu}^{ab}\omega^{b},
\hspace{3cm}
\delta_{b}A_{\mu}^{a}
=
gf^{abc}A_{\mu}^{b}\omega^{c}.
\end{equation}
Only the term ${\cal L}_{\rm bgf}$, which is not involved in the path
integral, breaks this exact background gauge invariance.
It is important to note the distinction between background gauge
invariance and quantum gauge independence: the first does not imply
the second.

With a source term $J_{\mu}^{a}A^{a\mu}$ for the quantum fields added to
the lagrangian, the BFM
generating functional $\tilde{Z}$ is a functional of the background
fields and the sources. The generating functional
$\tilde{\Gamma}[\tilde{A}_{\mu}^{a}, \langle A_{\mu}^{a} \rangle]$
for 1PI background field Green functions is given by the Legendre
transform of the effective action
$\tilde{W}[\tilde{A}_{\mu}^{a}, J_{\mu}^{a}] =
-i\log\tilde{Z}[\tilde{A}_{\mu}^{a}, J_{\mu}^{a}]$ with respect to the
sources $J_{\mu}^{a}$.
Demanding that the shift Eq.\ (\ref{4.1}) be made such that
the quantum fields have zero vacuum expectation value,
$\tilde{\Gamma}[\tilde{A}_{\mu}^{a}, 0]$ is exactly invariant under
ordinary gauge transformations of the background fields. The
background fields are then the
classical fields which do not appear inside loops (since the path
integral is only over the quantum fields), while the quantum fields do
not appear as external particles (since
$\langle A_{\mu}^{a} \rangle = 0$).

As a result of the exact invariance of $\tilde{\Gamma}$ under
background gauge transformations, $\delta_{b}\tilde{\Gamma} = 0$,
the background gauge boson self-energy is transverse:
\begin{eqnarray}
& & \nonumber \\
\hspace*{5cm}
&=&
(q^{2}g_{\mu\nu} - q_{\mu}q_{\nu})\delta^{ab}i
\tilde{\Pi}(\xi_{q},q^{2}). \\
& & \nonumber
\end{eqnarray}
\noindent
\begin{picture}(-8,0)(-20,476)

\thicklines

\put( 70,540){\makebox(0,0)[r]{$\tilde{A}_{\mu}^{a}$}}
\put( 80,540){\usebox{\gluon}}
\put( 90,540){\usebox{\rgluonarrow}}
\put(100,540){\usebox{\gluon}}
\put(100,525){\makebox(0,0)[c]{$q$}}
\put(140,540){\circle{39}}
\multiput(130,550)( 0.5, 0.5){18}{\circle*{0.2}}
\multiput(130,550)(-0.5,-0.5){18}{\circle*{0.2}}
\multiput(135,545)( 0.5, 0.5){25}{\circle*{0.2}}
\multiput(135,545)(-0.5,-0.5){25}{\circle*{0.2}}
\multiput(140,540)( 0.5, 0.5){27}{\circle*{0.2}}
\multiput(140,540)(-0.5,-0.5){27}{\circle*{0.2}}
\multiput(145,535)( 0.5, 0.5){25}{\circle*{0.2}}
\multiput(145,535)(-0.5,-0.5){25}{\circle*{0.2}}
\multiput(150,530)( 0.5, 0.5){18}{\circle*{0.2}}
\multiput(150,530)(-0.5,-0.5){18}{\circle*{0.2}}
\put(160,540){\usebox{\gluon}}
\put(170,540){\usebox{\rgluonarrow}}
\put(180,525){\makebox(0,0)[c]{$q$}}
\put(180,540){\usebox{\gluon}}
\put(210,540){\makebox(0,0)[l]{$\tilde{A}_{\nu}^{b}$}}

\thinlines

\end{picture}
However, as indicated explicitly, $\tilde{\Pi}$ retains a dependence
on the quantum gauge fixing parameter $\xi_{q}$. Summing
the renormalized form of this
self-energy in a Dyson series gives, at a given order in
perturbation theory, the dressed, renormalized
background gauge field propagator
\begin{equation}
i\tilde{\Delta}_{R\mu\nu}^{ab}(q)
=
\frac{i\delta^{ab}}{q^{2} + i\epsilon}\Biggl\{
\biggl( - g_{\mu\nu} + \frac{q_{\mu}q_{\nu}}{q^{2}}\biggr)
\tilde{d}_{R}(\xi_{q},q^{2})
- \xi_{b}\frac{q_{\mu}q_{\nu}}{q^{2}}
\Biggr\}.
\end{equation}
It is then possible to define a BFM effective charge
by analogy with Eq.\ (\ref{qedeffcharge})
\begin{equation}
g_{R}^{2}\tilde{d}_{R}(\xi_{q},q^{2})
=
\frac{g_{R}^{2}}{1 - \tilde{\Pi}_{R}(\xi_{q},q^{2})}
=
\tilde{g}_{\rm eff}^{2}(\xi_{q},q^{2}).
\end{equation}
Although the one-loop
background field Green functions are in general quantum
gauge-dependent, their divergent parts, and hence the renormalization
counterterms, are $\xi_{q}$-independent
(Kallosh's Theorem\cite{kallosh}).
Also, just as in QED, $\tilde{g}_{\rm eff}(\xi_{q},q^{2})$ is
scale-independent as a result of the BFM Ward identites.
The BFM one-loop
effective charge therefore has the high energy behaviour
expected from the QCD $\beta$ function for all values of $\xi_{q}$.

As a result of the explicitly-retained background gauge invariance of
the BFM generating functionals, the background field Green functions
in general obey to all orders and for all values of $\xi_{q}$
the same set of Ward identities as the
corresponding tree level functions. Indeed, explicit background gauge
invariance is the fundamental reason for the BFM Green functions
displaying for all $\xi_{q}$
the theoretically desirable properties possessed by PT functions.
Furthermore, the one-loop
PT functions exactly coincide with the background
field Green functions calculated in the particular quantum gauge
$\xi_{q} = 1$, both in QCD\cite{hashimoto}
and the electroweak sector of the
SM\cite{denner1}$^-$\cite{rafael}.
Thus, for the particular case of the one-loop two-point function
\begin{equation}
\tilde{\Pi}(\xi_{q} = 1,q^{2})
=
\hat{\Pi}(q^{2})
\,\,\Rightarrow\,\,
\tilde{g}_{\rm eff}^{2}(\xi_{q}\!=\!1,q^{2})
=
\hat{g}_{\rm eff}^{2}(q^{2}).
\end{equation}
On the basis of these observations, it has been
argued\cite{denner1}$^-$\cite{denner2}
that there
is nothing unique about the PT functions, the PT being
just one of an infinite choice of prescriptions to obtain well-behaved
Green functions.
In particular, this implies that, away from the asymptotic region
governed by the $\beta$ function, there is no unique way to define
an effective charge in a non-abelian gauge theory.

\section{The ``Effective'' Gauge Boson Two-Point Function}

We will now argue against this conclusion. This involves
comparing again the
cases of QED and QCD, and introducing the idea of the ``effective''
gauge boson two-point function.

The interaction term in the QED classical lagrangian is given by
\begin{equation}\label{qedli}
{\cal L}_{\rm cl}^{\rm int}
=
eJ_{\mu}A^{\mu}
\end{equation}
where $J_{\mu}$ is the electromagnetic current. At tree level, the
interaction between currents at spatial points $x_{1}$ and $x_{2}$
is mediated by a single photon and is given by
\begin{eqnarray}
& & \nonumber \\
\label{qedtreeamp}
\hspace*{4cm}
&=&
ieJ^{\mu}(x_{1})\,iD_{\mu\nu}(x_{1} - x_{2})\,ieJ^{\nu}(x_{2})
\\
& & \nonumber
\end{eqnarray}
\noindent
\begin{picture}(-8,0)(0,476)

\thicklines

\put( 70,540){\makebox(0,0)[r]{$x_{1}$}}
\put( 80,510){\line(0,1){60}}
\put( 80,540){\usebox{\gluon}}
\put(100,540){\usebox{\gluon}}
\put(120,540){\usebox{\gluon}}
\put(140,540){\usebox{\gluon}}
\put(160,540){\usebox{\gluon}}
\put(180,540){\usebox{\gluon}}
\put(130,540){\usebox{\rgluonarrow}}
\put(140,525){\makebox(0,0)[c]{$q$}}
\put(200,510){\line(0,1){60}}
\put(210,540){\makebox(0,0)[l]{$x_{2}$}}

\thinlines

\end{picture}
where the Feynman diagram in Eq.\ (\ref{qedtreeamp})
is in position space and
$iD_{\mu\nu}(x_{1} - x_{2})$ is the
Fourier transform of the tree level photon propagator
$iD_{\mu\nu}(q)$. Beyond tree level in perturbation theory,
the renormalized
interaction between the two currents at $x_{1}$ and $x_{2}$ is
given by
\begin{eqnarray}
& & \nonumber \\
\label{qeddressedamp}
\hspace*{5cm}
&=&
ie_{R}J^{\mu}(x_{1})\,i\Delta_{R\mu\nu}(x_{1} - x_{2})\,
ie_{R}J^{\nu}(x_{2})
\\
& & \nonumber
\end{eqnarray}
\noindent
\begin{picture}(-8,0)(0,476)

\thicklines

\put( 70,540){\makebox(0,0)[r]{$x_{1}$}}
\put( 80,510){\line(0,1){60}}
\put( 80,540){\usebox{\gluon}}
\put( 90,540){\usebox{\rgluonarrow}}
\put(100,540){\usebox{\gluon}}
\put(100,525){\makebox(0,0)[c]{$q$}}
\put(140,540){\circle{39}}
\multiput(130,550)( 0.5, 0.5){18}{\circle*{0.2}}
\multiput(130,550)(-0.5,-0.5){18}{\circle*{0.2}}
\multiput(135,545)( 0.5, 0.5){25}{\circle*{0.2}}
\multiput(135,545)(-0.5,-0.5){25}{\circle*{0.2}}
\multiput(140,540)( 0.5, 0.5){27}{\circle*{0.2}}
\multiput(140,540)(-0.5,-0.5){27}{\circle*{0.2}}
\multiput(145,535)( 0.5, 0.5){25}{\circle*{0.2}}
\multiput(145,535)(-0.5,-0.5){25}{\circle*{0.2}}
\multiput(150,530)( 0.5, 0.5){18}{\circle*{0.2}}
\multiput(150,530)(-0.5,-0.5){18}{\circle*{0.2}}
\multiput(130,530)( 0.5,-0.5){18}{\circle*{0.2}}
\multiput(130,530)(-0.5, 0.5){18}{\circle*{0.2}}
\multiput(135,535)( 0.5,-0.5){25}{\circle*{0.2}}
\multiput(135,535)(-0.5, 0.5){25}{\circle*{0.2}}
\multiput(140,540)( 0.5,-0.5){27}{\circle*{0.2}}
\multiput(140,540)(-0.5, 0.5){27}{\circle*{0.2}}
\multiput(145,545)( 0.5,-0.5){25}{\circle*{0.2}}
\multiput(145,545)(-0.5, 0.5){25}{\circle*{0.2}}
\multiput(150,550)( 0.5,-0.5){18}{\circle*{0.2}}
\multiput(150,550)(-0.5, 0.5){18}{\circle*{0.2}}
\put(160,540){\usebox{\gluon}}
\put(170,540){\usebox{\rgluonarrow}}
\put(180,525){\makebox(0,0)[c]{$q$}}
\put(180,540){\usebox{\gluon}}
\put(200,510){\line(0,1){60}}
\put(210,540){\makebox(0,0)[l]{$x_{2}$}}

\thinlines

\end{picture}
where $i\Delta_{R\mu\nu}(x_{1} - x_{2})$ is the Fourier transform of
the renormalized dressed photon propagator Eq.\ (\ref{qedprop}),
involving the Dyson sum of the 1PI photon self-energy $\Pi(q^{2})$.
Thus, in QED the photon propagator can just as well be defined
in terms of the two-point interaction between physical
currents $J_{\mu}(x_{1})$, $J_{\nu}(x_{2})$
appearing in
${\cal L}_{\rm cl}^{\rm int}$ as from the conventional
two-point Green function
$\langle 0|T(A_{\mu}(x_{1})A_{\nu}(x_{2}))|0\rangle$.
Precisely because of this,
the radiative corrections to the tree level propagator $iD_{\mu\nu}$
included in the dressed propagator $i\Delta_{R\mu\nu}$
can be fully accounted for essentially
just by appropriately changing the coupling of the
theory appearing in ${\cal L}_{\rm cl}^{\rm int}$ Eq.\ (\ref{qedli}),
i.e.\  by making the replacement
$e\rightarrow e_{\rm eff}(q^{2})$, and then
using the tree level propagator $iD_{\mu\nu}$.

In QCD, the interaction part of the classical lagrangian
may be written
\begin{equation}\label{qcdli}
{\cal L}_{\rm cl}^{\rm int}
=
g\Bigl(J_{\mu}^{a} + V_{\mu}^{a} + gT_{\mu}^{a}\Bigr)A^{a\mu}
\end{equation}
where
\begin{eqnarray}
J_{\mu}^{a}
&=&
{\textstyle \sum_{f=1}^{n_{f}}}J_{\mu}^{^{(f)}a} =
{\textstyle \sum_{f=1}^{n_{f}}}
\overline{\psi}^{^{(f)}}\gamma_{\mu}T^{a}\psi^{^{(f)}} \\
V_{\mu}^{a}
&=&
{\textstyle -\frac{1}{3}}f^{abc}\Bigl(
A_{\nu}^{b}(\partial_{\mu}A_{\nu}^{c})
+
(\partial_{\nu} A_{\mu}^{b})A_{\nu}^{c}
+
A_{\nu}^{b}A_{\mu}^{c}\partial_{\nu} \Bigr)
\\
T_{\mu}^{a}
&=&
{\textstyle -\frac{1}{4}}f^{rab}f^{rcd}A_{\nu}^{b}A_{\mu}^{c}A_{\nu}^{d}
\end{eqnarray}
and the derivative has been symmetrized in $V_{\mu}^{a}A^{a\mu}$.
The tree level interaction between two fermionic currents
at points $x_{1}$ and $x_{2}$ is given by
\begin{eqnarray}
& & \nonumber \\
\label{qcdtreeamp}
\hspace*{4cm}
&=&
igJ^{^{(f)}a\mu}(x_{1})
\,iD_{\mu\nu}^{ab}(x_{1} - x_{2})\,igJ^{^{(f')}b\nu}(x_{2})
\\
& & \nonumber
\end{eqnarray}
\noindent
\begin{picture}(-8,0)(0,476)

\thicklines

\put( 70,540){\makebox(0,0)[r]{$x_{1}$}}
\put( 80,510){\line(0,1){60}}
\put( 80,540){\usebox{\gluon}}
\put(100,540){\usebox{\gluon}}
\put(120,540){\usebox{\gluon}}
\put(140,540){\usebox{\gluon}}
\put(160,540){\usebox{\gluon}}
\put(180,540){\usebox{\gluon}}
\put(130,540){\usebox{\rgluonarrow}}
\put(140,525){\makebox(0,0)[c]{$q$}}
\put(200,510){\line(0,1){60}}
\put(210,540){\makebox(0,0)[l]{$x_{2}$}}

\thinlines

\end{picture}
where $iD_{\mu\nu}^{ab}(x_{1} - x_{2})$ is the Fourier transform of
the tree level gauge boson propagator $iD_{\mu\nu}^{ab}(q)$.
Beyond tree level in perturbation theory, the renormalized
interaction between the two currents at $x_{1}$ and $x_{2}$ may
be written
\begin{eqnarray}
& & \nonumber \\
\label{qcddressedamp}
\hspace*{5cm}
&=&
ig_{R}J^{^{(f)}a\mu}(x_{1})\,i\Delta_{R\mu\nu}^{ab}(x_{1} - x_{2})\,
ig_{R}J^{^{(f')}b\nu}(x_{2}).
\\
& & \nonumber
\end{eqnarray}
\noindent
\begin{picture}(-8,0)(0,476)

\thicklines

\put( 70,540){\makebox(0,0)[r]{$x_{1}$}}
\put( 79,510){\line(0,1){60}}
\put(140,540){\oval(120,40)}

\put(130,560){\line(-1,-1){38}}
\put(120,560){\line(-1,-1){34}}
\put(110,560){\line(-1,-1){28}}
\put(100,560){\line(-1,-1){20}}

\put(100,520){\line(1,1){40}}
\put(110,520){\line(1,1){40}}
\put(120,520){\line(1,1){40}}
\put(130,520){\line(1,1){40}}
\put(140,520){\line(1,1){40}}

\put(150,520){\line(1,1){38}}
\put(160,520){\line(1,1){34}}
\put(170,520){\line(1,1){28}}
\put(180,520){\line(1,1){20}}

\put(130,520){\line(-1,1){38}}
\put(120,520){\line(-1,1){34}}
\put(110,520){\line(-1,1){28}}
\put(100,520){\line(-1,1){20}}

\put(100,560){\line(1,-1){40}}
\put(110,560){\line(1,-1){40}}
\put(120,560){\line(1,-1){40}}
\put(130,560){\line(1,-1){40}}
\put(140,560){\line(1,-1){40}}

\put(150,560){\line(1,-1){38}}
\put(160,560){\line(1,-1){34}}
\put(170,560){\line(1,-1){28}}
\put(180,560){\line(1,-1){20}}

\put(201,510){\line(0,1){60}}
\put(210,540){\makebox(0,0)[l]{$x_{2}$}}

\thinlines

\end{picture}
The quantity
$i\Delta_{R\mu\nu}^{ab}(x_{1} - x_{2})$, represented by the hatched
blob in Eq.\ (\ref{qcddressedamp}),
is by definition the QCD analogue of the QED propagator
$i\Delta_{R\mu\nu}(x_{1} - x_{2})$
in Eq.\ (\ref{qeddressedamp}): it is the function which accounts
fully for the two-point interaction between fermionic currents
$J_{\mu}^{^{(f)}a}(x_{1})$, $J_{\nu}^{^{(f')}b}(x_{2})$
appearing in ${\cal L}_{\rm cl}^{\rm int}$.
By construction, $i\Delta_{R\mu\nu}^{ab}(x_{1} - x_{2})$
is the propagator the effects of which can be fully accounted for
essentially just by appropriately changing the coupling appearing in
${\cal L}_{\rm cl}^{\rm int}$ Eq.\ (\ref{qcdli}).
But at the one-loop level, $i\Delta_{R\mu\nu}^{ab}(x_{1} - x_{2})$
is exactly the Fourier transform of the PT gauge boson
propagator Eq.\ (\ref{ptprop}), involving the Dyson summation of
the one-loop PT self-energy $\hat{\Pi}(q^{2})$
defined in Sec.\ 3.
At the one-loop level, in exact analogy with QED,
the radiative corrections to the tree level interaction Eq.\
(\ref{qcdtreeamp}) included in the dressed amplitude Eq.\
(\ref{qcddressedamp}) may therefore be fully accounted
for essentially just by making the replacement
$g\rightarrow \hat{g}_{\rm eff}(q^{2})$ at the vertices in the tree level
amplitude Eq.\ (\ref{qcdtreeamp}).

Thus, by defining the gauge boson propagator in terms of the
physical fermionic
currents $J_{\mu}^{^{(f)}a}(x_{1})$, $J_{\nu}^{^{(f')}b}(x_{2})$
which appear in ${\cal L}_{\rm cl}^{\rm int}$ and
between which the gauge boson mediates the interaction,
rather than in terms of
the conventional two-point Green function
$\langle 0|T(A_{\mu}^{a}(x_{1})A_{\nu}^{b}(x_{2})|0\rangle$,
the natural and unambiguous extension to QCD of the QED concept of an
effective charge is obtained. The gauge boson propagator defined in
this way may be termed the ``effective'' gauge boson
two-point function. It is
given precisely by PT propagator Eq.\ (\ref{ptprop}).

However, in the non-abelian theory there also occur in the interaction
part of the
classical \hfill lagrangian \hfill
Eq.\ (\ref{qcdli}) \hfill
the \hfill
triple \hfill
and \hfill
quadruple \hfill
gauge \hfill
vertices---the \hfill  original

\pagebreak

\noindent
\begin{picture}(570,435)(-20,275)

\put(-20,325){\makebox(0,0)[l]{\small
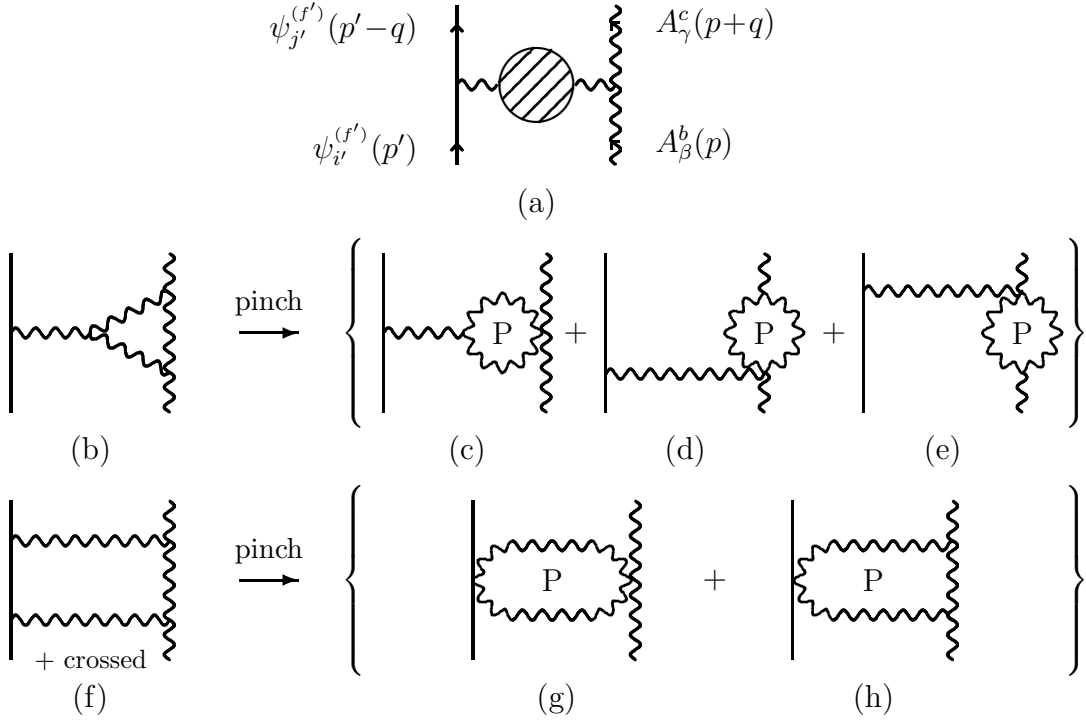
Fig.\ 3. The set of one-loop
diagrams for the fermion-gluon scattering process (corrections }}
\put(-20,305){\makebox(0,0)[l]{\small
involving the fermion line are not shown) and their pinch parts,
denoted by ``P''.}}

\thicklines


\put(205,635){\makebox(0,0)[r]{$\psi_{i'}^{^{(f')}}(p')$}}
\put(205,695){\makebox(0,0)[r]{$\psi_{j'}^{^{(f')}}(p'\!-\!q)$}}
\put(225,625){\line(0,1){80}}
\multiput(225,695)(-0.1,-0.1){28}{\circle*{0.2}}
\multiput(225,695)( 0.1,-0.1){28}{\circle*{0.2}}
\multiput(225,635)(-0.1,-0.1){28}{\circle*{0.2}}
\multiput(225,635)( 0.1,-0.1){28}{\circle*{0.2}}
\put(225,665){\usebox{\gluon}}
\put(265,665){\circle{39}}
\multiput(255,675)( 0.5, 0.5){18}{\circle*{0.2}}
\multiput(255,675)(-0.5,-0.5){18}{\circle*{0.2}}
\multiput(260,670)( 0.5, 0.5){26}{\circle*{0.2}}
\multiput(260,670)(-0.5,-0.5){25}{\circle*{0.2}}
\multiput(265,665)( 0.5, 0.5){27}{\circle*{0.2}}
\multiput(265,665)(-0.5,-0.5){27}{\circle*{0.2}}
\multiput(270,660)( 0.5, 0.5){25}{\circle*{0.2}}
\multiput(270,660)(-0.5,-0.5){25}{\circle*{0.2}}
\multiput(275,655)( 0.5, 0.5){18}{\circle*{0.2}}
\multiput(275,655)(-0.5,-0.5){18}{\circle*{0.2}}
\put(285,665){\usebox{\gluon}}
\put(305,625){\usebox{\vgluon}}
\put(305,625){\usebox{\ugluonarrow}}
\put(305,645){\usebox{\vgluon}}
\put(305,665){\usebox{\vgluon}}
\put(305,685){\usebox{\vgluon}}
\put(305,685){\usebox{\ugluonarrow}}
\put(325,635){\makebox(0,0)[l]{$A_{\beta}^{b}(p)$}}
\put(325,695){\makebox(0,0)[l]{$A_{\gamma}^{c}(p\!+\!q)$}}
\put(265,605){\makebox(0,0)[c]{(a)}}


\put(  0,500){\line(0,1){80}}
\put(  0,540){\usebox{\gluon}}
\put( 20,540){\usebox{\gluon}}
\put( 40,540){\usebox{\trgluonone}}
\put( 40,540){\usebox{\trgluontwo}}
\put( 40,540){\usebox{\brgluon}}
\put( 80,500){\usebox{\vgluon}}
\put( 80,520){\usebox{\vgluon}}
\put( 80,540){\usebox{\vgluon}}
\put( 80,560){\usebox{\vgluon}}
\put( 40,480){\makebox(0,0)[c]{(b)}}

\put(130,555){\makebox(0,0)[c]{\small pinch}}
\put(115,540){\vector(1, 0){30}}
\put(180,540){\makebox(0,0)[c]{$
\left\{ \begin{array}{c} \\ \\ \\ \\ \\ \end{array}\right.
$}}

\put(188,500){\line(0,1){80}}
\put(188,540){\usebox{\gluon}}
\put(208,540){\usebox{\gluon}}
\put(248,540){\makebox(0,0)[c]{P}}
\put(248,540){\usebox{\trgluonloop}}
\put(248,540){\usebox{\brgluonloop}}
\put(248,540){\usebox{\blgluonloop}}
\put(248,540){\usebox{\tlgluonloop}}
\put(270,500){\usebox{\vgluon}}
\put(270,520){\usebox{\vgluon}}
\put(270,540){\usebox{\vgluon}}
\put(270,560){\usebox{\vgluon}}
\put(230,480){\makebox(0,0)[c]{(c)}}

\put(285,540){\makebox(0,0)[c]{$+$}}

\put(300,500){\line(0,1){80}}
\put(300,519){\usebox{\gluon}}
\put(320,519){\usebox{\gluon}}
\put(340,519){\usebox{\gluon}}
\put(360,519){\usebox{\gluon}}
\put(380,500){\usebox{\vgluon}}
\put(380,540){\makebox(0,0)[c]{P}}
\put(380,540){\usebox{\trgluonloop}}
\put(380,540){\usebox{\brgluonloop}}
\put(380,540){\usebox{\blgluonloop}}
\put(380,540){\usebox{\tlgluonloop}}
\put(380,560){\usebox{\vgluon}}
\put(340,480){\makebox(0,0)[c]{(d)}}

\put(415,540){\makebox(0,0)[c]{$+$}}

\put(430,500){\line(0,1){80}}
\put(430,561){\usebox{\gluon}}
\put(450,561){\usebox{\gluon}}
\put(470,561){\usebox{\gluon}}
\put(490,561){\usebox{\gluon}}
\put(510,500){\usebox{\vgluon}}
\put(510,540){\makebox(0,0)[c]{P}}
\put(510,540){\usebox{\trgluonloop}}
\put(510,540){\usebox{\brgluonloop}}
\put(510,540){\usebox{\blgluonloop}}
\put(510,540){\usebox{\tlgluonloop}}
\put(510,560){\usebox{\vgluon}}
\put(470,480){\makebox(0,0)[c]{(e)}}

\put(530,540){\makebox(0,0)[c]{$
\left. \begin{array}{c} \\ \\ \\ \\ \\ \end{array}\right\}
$}}


\put(  0,375){\line(0,1){80}}
\put(  0,395){\usebox{\gluon}}
\put( 20,395){\usebox{\gluon}}
\put( 40,395){\usebox{\gluon}}
\put( 60,395){\usebox{\gluon}}
\put(  0,435){\usebox{\gluon}}
\put( 20,435){\usebox{\gluon}}
\put( 40,435){\usebox{\gluon}}
\put( 60,435){\usebox{\gluon}}
\put( 80,375){\usebox{\vgluon}}
\put( 80,395){\usebox{\vgluon}}
\put( 80,415){\usebox{\vgluon}}
\put( 80,435){\usebox{\vgluon}}
\put( 40,375){\makebox(0,0)[c]{\footnotesize + crossed}}
\put( 40,355){\makebox(0,0)[c]{(f)}}

\put(130,430){\makebox(0,0)[c]{\small pinch}}
\put(115,415){\vector(1, 0){30}}
\put(180,415){\makebox(0,0)[c]{$
\left\{ \begin{array}{c} \\ \\ \\ \\ \\ \end{array}\right.
$}}

\put(233,375){\line(0,1){80}}
\put(253,415){\usebox{\blgluonloop}}
\put(253,415){\usebox{\tlgluonloop}}
\put(273,415){\makebox(0,0)[c]{P}}
\put(253,397.5){\usebox{\bottomgluon}}
\put(273,397.5){\usebox{\bottomgluon}}
\put(253,432.5){\usebox{\topgluon}}
\put(273,432.5){\usebox{\topgluon}}
\put(293,415){\usebox{\trgluonloop}}
\put(293,415){\usebox{\brgluonloop}}
\put(315,375){\usebox{\vgluon}}
\put(315,395){\usebox{\vgluon}}
\put(315,415){\usebox{\vgluon}}
\put(315,435){\usebox{\vgluon}}
\put(275,355){\makebox(0,0)[c]{(g)}}

\put(355,415){\makebox(0,0)[c]{$+$}}

\put(394,375){\line(0,1){80}}
\put(415,415){\usebox{\blgluonloop}}
\put(415,415){\usebox{\tlgluonloop}}
\put(435,415){\makebox(0,0)[c]{P}}
\put(415,397.5){\usebox{\bottomgluon}}
\put(435,397.5){\usebox{\bottomgluon}}
\put(455,397.5){\usebox{\bottomgluon}}
\put(415,432.5){\usebox{\topgluon}}
\put(435,432.5){\usebox{\topgluon}}
\put(455,432.5){\usebox{\topgluon}}
\put(475,375){\usebox{\vgluon}}
\put(475,395){\usebox{\vgluon}}
\put(475,415){\usebox{\vgluon}}
\put(475,435){\usebox{\vgluon}}
\put(435,355){\makebox(0,0)[c]{(h)}}

\put(530,415){\makebox(0,0)[c]{$
\left. \begin{array}{c} \\ \\ \\ \\ \\ \end{array}\right\}
$}}

\end{picture}

\noindent
source of the difficulties
described in Sec.\ 2. If the concept of
an effective charge for the theory is to be valid, then it must
account for the radiative corrections to the two-point interaction
between {\it any} pair among
the terms $J_{\mu}^{^{(f)}a}$, $V_{\mu}^{a}$, $gT_{\mu}^{a}$ in
${\cal L}_{\rm cl}^{\rm int}$.
In other words, the PT self-energy must be universal.

To illustrate that this is indeed the case,
consider the scattering process
$A_{\beta}^{b}\psi_{i'}^{^{(f')}}
\rightarrow A_{\gamma}^{c}\psi_{j'}^{^{(f')}}$.
The set of one-loop diagrams for this process, except for
those involving corrections to the fermion line, are shown in
Fig.\ 3, together with the associated pinch diagrams.
In an exactly similar way to the four fermion case in Sec.\ 3, the PT
self-energy, shown in Fig.\ 3(o), is defined as the function of
$q^{2}$ appearing between the tree level vertices
$ig\gamma_{\alpha}T^{a}$ and
$g\Gamma_{\alpha\beta\gamma}^{abc}(q,p,-p\!-\!q)$:
\begin{equation}
{\rm Fig.\,\,3(o)}
=
\Bigl(\overline{u}_{j'}ig\gamma^{\alpha}T_{j'i'}^{a}u_{i'}\Bigr)
\,\frac{-i}{q^{2}}\,
iq^{2}\hat{\Pi}(q^{2})
\,\frac{-i}{q^{2}}\,
\Bigl(g\Gamma_{\alpha\beta\gamma}^{abc}(q,p,-p\!-\!q)
\epsilon^{\beta}(p)\epsilon^{*\gamma}(p\!+\!q)\Bigr)
\end{equation}
where the $\epsilon$'s are the external
gauge boson polarization vectors.

In the diagrams shown in Fig.\ 3,
the pinch parts are identified using the tree-level Ward identity
\begin{equation}\label{gaugewid}
q_{1}^{\alpha}\Gamma_{\alpha\beta\gamma}(q_{1},q_{2},q_{3})
=
P_{\beta\sigma}(q_{3})D_{\sigma\gamma}^{-1}(q_{3}) -
P_{\beta\sigma}(q_{2})D_{\sigma\gamma}^{-1}(q_{2})
\end{equation}
where $P_{\mu\nu}(q)= g_{\mu\nu} - q_{\mu}q_{\nu}/q^{2}$
is the transverse projection operator and $iD_{\mu\nu}(q)$ is the
tree level gauge boson propagator. We work again in the Feynman gauge
$\xi = 1$.

\pagebreak

\noindent
\begin{picture}(570,415)(-20,-80)

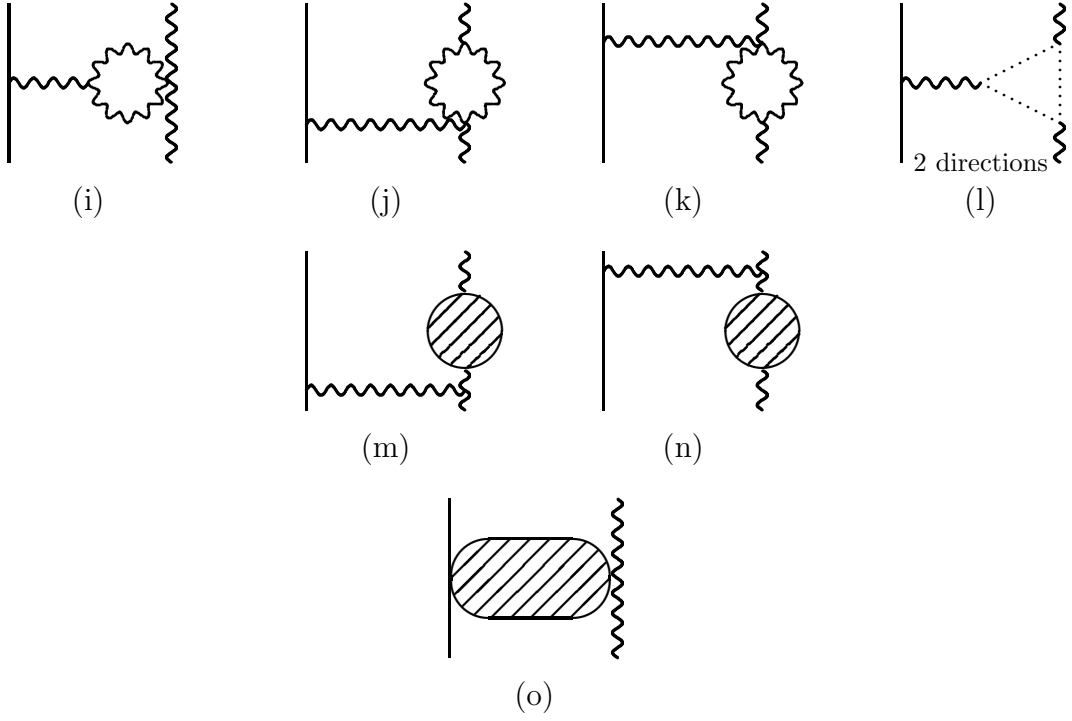
\put(-20,-50){\makebox(0,0)[l]{\small Fig.\ 3 continued. The pinch
technique self-energy is shown in (o).}}

\thicklines


\put(  0,250){\line(0,1){80}}
\put(  0,290){\usebox{\gluon}}
\put( 20,290){\usebox{\gluon}}
\put( 60,290){\usebox{\trgluonloop}}
\put( 60,290){\usebox{\brgluonloop}}
\put( 60,290){\usebox{\blgluonloop}}
\put( 60,290){\usebox{\tlgluonloop}}
\put( 82,250){\usebox{\vgluon}}
\put( 82,270){\usebox{\vgluon}}
\put( 82,290){\usebox{\vgluon}}
\put( 82,310){\usebox{\vgluon}}
\put( 40,230){\makebox(0,0)[c]{(i)}}

\put(150,250){\line(0,1){80}}
\put(150,269){\usebox{\gluon}}
\put(170,269){\usebox{\gluon}}
\put(190,269){\usebox{\gluon}}
\put(210,269){\usebox{\gluon}}
\put(230,250){\usebox{\vgluon}}
\put(230,290){\usebox{\trgluonloop}}
\put(230,290){\usebox{\brgluonloop}}
\put(230,290){\usebox{\blgluonloop}}
\put(230,290){\usebox{\tlgluonloop}}
\put(230,310){\usebox{\vgluon}}
\put(190,230){\makebox(0,0)[c]{(j)}}

\put(300,250){\line(0,1){80}}
\put(300,311){\usebox{\gluon}}
\put(320,311){\usebox{\gluon}}
\put(340,311){\usebox{\gluon}}
\put(360,311){\usebox{\gluon}}
\put(380,250){\usebox{\vgluon}}
\put(380,290){\usebox{\trgluonloop}}
\put(380,290){\usebox{\brgluonloop}}
\put(380,290){\usebox{\blgluonloop}}
\put(380,290){\usebox{\tlgluonloop}}
\put(380,310){\usebox{\vgluon}}
\put(340,230){\makebox(0,0)[c]{(k)}}

\put(450,250){\line(0,1){80}}
\put(450,290){\usebox{\gluon}}
\put(470,290){\usebox{\gluon}}
\multiput(490,290)(4, 2){11}{\circle*{0.4}}
\multiput(490,290)(4,-2){11}{\circle*{0.4}}
\multiput(530,270)(0,4.47){9}{\circle*{0.4}}
\put(530,250){\usebox{\vgluon}}
\put(530,310){\usebox{\vgluon}}
\put(490,250){\makebox(0,0)[c]{\footnotesize 2 directions}}
\put(490,230){\makebox(0,0)[c]{(l)}}


\put(150,125){\line(0,1){80}}
\put(150,135){\usebox{\gluon}}
\put(170,135){\usebox{\gluon}}
\put(190,135){\usebox{\gluon}}
\put(210,135){\usebox{\gluon}}
\put(230,125){\usebox{\vgluon}}
\put(230,165){\circle{39}}
\multiput(220,175)( 0.5, 0.5){18}{\circle*{0.2}}
\multiput(220,175)(-0.5,-0.5){18}{\circle*{0.2}}
\multiput(225,170)( 0.5, 0.5){26}{\circle*{0.2}}
\multiput(225,170)(-0.5,-0.5){25}{\circle*{0.2}}
\multiput(230,165)( 0.5, 0.5){27}{\circle*{0.2}}
\multiput(230,165)(-0.5,-0.5){27}{\circle*{0.2}}
\multiput(235,160)( 0.5, 0.5){25}{\circle*{0.2}}
\multiput(235,160)(-0.5,-0.5){25}{\circle*{0.2}}
\multiput(240,155)( 0.5, 0.5){18}{\circle*{0.2}}
\multiput(240,155)(-0.5,-0.5){18}{\circle*{0.2}}
\put(230,185){\usebox{\vgluon}}
\put(190,105){\makebox(0,0)[c]{(m)}}

\put(300,125){\line(0,1){80}}
\put(300,195){\usebox{\gluon}}
\put(320,195){\usebox{\gluon}}
\put(340,195){\usebox{\gluon}}
\put(360,195){\usebox{\gluon}}
\put(380,125){\usebox{\vgluon}}
\put(380,165){\circle{39}}
\multiput(370,175)( 0.5, 0.5){18}{\circle*{0.2}}
\multiput(370,175)(-0.5,-0.5){18}{\circle*{0.2}}
\multiput(375,170)( 0.5, 0.5){26}{\circle*{0.2}}
\multiput(375,170)(-0.5,-0.5){25}{\circle*{0.2}}
\multiput(380,165)( 0.5, 0.5){27}{\circle*{0.2}}
\multiput(380,165)(-0.5,-0.5){27}{\circle*{0.2}}
\multiput(385,160)( 0.5, 0.5){25}{\circle*{0.2}}
\multiput(385,160)(-0.5,-0.5){25}{\circle*{0.2}}
\multiput(390,155)( 0.5, 0.5){18}{\circle*{0.2}}
\multiput(390,155)(-0.5,-0.5){18}{\circle*{0.2}}
\put(380,185){\usebox{\vgluon}}
\put(340,105){\makebox(0,0)[c]{(n)}}


\put(222,  0){\line(0,1){80}}
\put(263,40){\oval(80,40)}
\put(273,60){\line(-1,-1){38}}
\put(263,60){\line(-1,-1){34}}
\put(253,60){\line(-1,-1){28}}
\put(243,60){\line(-1,-1){20}}
\put(243,20){\line(1,1){40}}
\put(253,20){\line(1,1){38}}
\put(263,20){\line(1,1){34}}
\put(273,20){\line(1,1){28}}
\put(283,20){\line(1,1){20}}
\put(307,  0){\usebox{\vgluon}}
\put(307, 20){\usebox{\vgluon}}
\put(307, 40){\usebox{\vgluon}}
\put(307, 60){\usebox{\vgluon}}
\put(265,-20){\makebox(0,0)[c]{(o)}}

\end{picture}

The vertex part of the diagram Fig.\ 3(b) is shown in more detail in
Fig.\ 4. Using the Ward identity Eq.\ (\ref{gaugewid}), it may be
written
\begin{eqnarray}
{\rm Fig.\ 4}
&=&
-\frac{ig^{3}Nf^{abc}}{2}
\mu^{2\epsilon}\int\frac{d^{n}k}{(2\pi)^{n}}
\frac{1}{k_{1}^{2}k_{2}^{2}k_{3}^{2}}\times
\nonumber \\
& &
\biggl\{
\Gamma_{\alpha\tau\sigma}^{F}(q_{1},k_{3},-k_{2})
\Gamma_{\beta\rho\tau}^{F}   (q_{2},k_{1},-k_{3})
\Gamma_{\gamma\sigma\rho}^{F}(q_{3},k_{2},-k_{1})
\nonumber \\
& &
-2(k_{2} + k_{3})_{\alpha}(k_{3} + k_{1})_{\beta}(k_{1} + k_{2})_{\gamma}
+ k_{2\alpha}k_{3\beta}k_{1\gamma} + k_{3\alpha}k_{1\beta}k_{2\gamma}
\nonumber \\
& &
+q_{1}^{2}P_{\alpha\rho}(q_{1})B_{\rho\beta\gamma}(k_{1},q_{2},q_{3})
+q_{2}^{2}P_{\beta\rho}(q_{2})B_{\rho\gamma\alpha}(k_{2},q_{3},q_{1})
\nonumber \\
& &
+q_{3}^{2}P_{\gamma\rho}(q_{3})B_{\rho\alpha\beta}(k_{3},q_{1},q_{2})
\biggl\}
\nonumber \\
& &
- ig^{2}NA(q_{1})\Bigl(
g\Gamma_{\alpha\beta\gamma}^{abc}(q_{1},q_{2},q_{3})
+\textstyle{\frac{7}{4}}gf^{abc}(q_{1\beta}g_{\gamma\alpha}
                               - q_{1\gamma}g_{\beta\alpha}) \Bigl)
\nonumber \\
& &
- ig^{2}NA(q_{2})\Bigl(
g\Gamma_{\alpha\beta\gamma}^{abc}(q_{1},q_{2},q_{3})
+\textstyle{\frac{7}{4}}gf^{abc}(q_{2\gamma}g_{\alpha\beta}
                               - q_{2\alpha}g_{\beta\gamma}) \Bigl)
\nonumber \\
\label{fig4a}
& &
- ig^{2}NA(q_{3})\Bigl(
g\Gamma_{\alpha\beta\gamma}^{abc}(q_{1},q_{2},q_{3})
+ \textstyle{\frac{7}{4}}gf^{abc}(q_{3\alpha}g_{\beta\gamma}
                                - q_{3\beta}g_{\gamma\alpha}) \Bigl)
\end{eqnarray}
where $A$ is given in Eq.\ (\ref{A}) and $B$ is given by
\begin{eqnarray}
B_{\rho\beta\gamma}(k_{1},q_{2},q_{3})
&=&
-\Gamma_{\rho\beta\gamma}(q_{1},q_{2},q_{3})
-\Gamma_{\rho\beta\gamma}( 0,-k_{1},k_{1}) \nonumber \\
& &
-{\textstyle \frac{1}{2}}(2k_{1} + q_{2} - q_{3})_{\rho}g_{\beta\gamma}
-(2k_{1} + q_{2})_{\beta}g_{\gamma\rho}
-(2k_{1} - q_{3})_{\gamma}g_{\beta\rho}.
\end{eqnarray}

\pagebreak

\noindent
\begin{picture}(570,260)(-20,45)

\put(-20,75){\makebox(0,0)[l]{\small
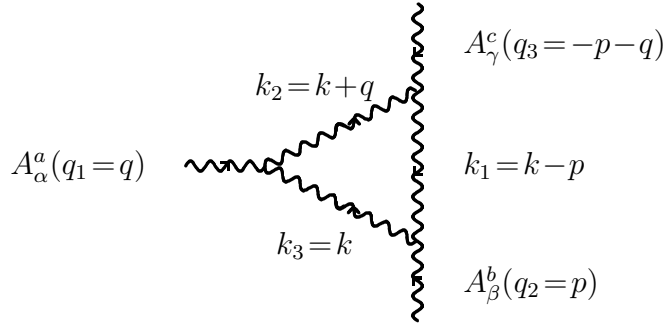
Fig.\ 4. The one-loop triple gauge vertex contributing pinch parts.}}

\thicklines

\put(165,210){\makebox(0,0)[r]{$A_{\alpha}^{a}(q_{1}\!=\! q)$}}
\put(185,210){\usebox{\gluon}}
\put(205,210){\usebox{\gluon}}
\put(195,210){\usebox{\rgluonarrow}}
\put(225,210){\usebox{\trgluonone}}
\put(225,210){\usebox{\trgluontwo}}
\put(252,224){\usebox{\trgluonarrow}}
\put(261,228){\usebox{\trgluonone}}
\put(261,228){\usebox{\trgluontwo}}
\put(225,210){\usebox{\brgluon}}
\put(252,197){\usebox{\brgluonarrow}}
\put(261,192){\usebox{\brgluon}}
\put(302,132){\usebox{\vgluon}}
\put(302,152){\usebox{\vgluon}}
\put(302,142){\usebox{\ugluonarrow}}
\put(302,172){\usebox{\vgluon}}
\put(302,192){\usebox{\vgluon}}
\put(312,207.5){\usebox{\lgluonarrow}}
\put(302,212){\usebox{\vgluon}}
\put(302,232){\usebox{\vgluon}}
\put(302,252){\usebox{\vgluon}}
\put(312,267.5){\usebox{\lgluonarrow}}
\put(302,272){\usebox{\vgluon}}
\put(250,170){\makebox(0,0)[c]{$k_{3}\! =\! k$}}
\put(250,250){\makebox(0,0)[c]{$k_{2}\! =\! k\!+\!q$}}
\put(325,150){\makebox(0,0)[l]{$A_{\beta}^{b}(q_{2}\! =\! p)$}}
\put(325,210){\makebox(0,0)[l]{$k_{1}\! =\! k\!-\!p$}}
\put(325,270){\makebox(0,0)[l]{$A_{\gamma}^{c}(q_{3}\!=\! -p\!-\!q)$}}

\end{picture}

The three terms proportional to
$g\Gamma_{\alpha\beta\gamma}^{abc}(q_{1},q_{2},q_{3})$
in Eq.\ (\ref{fig4a}) are the three pinch parts of the
diagram, shown in the overall amplitude
in Figs.\ 3(c)-(e). We see immediately that each of these
pinch terms has {\it exactly} the same dependence $-ig^{2}NA(q_{i})$
on the corresponding momentum $q_{i}$ as the pinch
term proportional to $ig\gamma_{\alpha}T^{a}$ in the one-loop
vector-fermion-fermion vertex Eq.\ (\ref{fig2a}) ($q_{1}\equiv q$).
Thus the pinch parts of the one-loop triple gauge vertex
in Fig.\ 4 make
exactly the required contribution, not only to the conventional
self-energy appearing in the propagator Fig.\ 3(a) but also to that
appearing in each of the external leg corrections Figs.\ 3(m) and (n),
to give the PT self-energy $\hat{\Pi}$
(recall that there is a symmetry factor $\frac{1}{2}$ associated with
the latter diagrams).
Of the three terms proportional to the transverse projection operators
$P$, those depending on $q_{2}\equiv p$ and
$q_{3}\equiv -p-q$ vanish for the on-shell
external gauge bosons. The remaining one depending on $q_{1} \equiv q$ is
{\it exactly} cancelled in the overall amplitude by the vertex-like
pinch part Fig.\ 3(h) of box diagrams.
The remaining terms in Eq.\ (\ref{fig4a}) are part of the
PT one-loop gauge-independent triple gauge vertex, the other parts
being given by the vertices in diagrams Figs.\ 3(i)-(k)
involving the quadruple gauge vertex and in Fig.\ 3(l) involving the
ghosts.

It is important to note that, although the loop integral for the
diagram in Fig.\ 3(i) depends only on
the four-momentum transfer $q$, it does {\it not}
contribute to the PT self-energy  since it is not proportional to the
tree level triple gauge vertex $g\Gamma_{\alpha\beta\gamma}^{abc}$:
\begin{equation}
{\rm Vertex\,\, of\,\, Fig.\ 3(i)}
=
-ig^{2}NA(q_{1})\,\textstyle{\frac{9}{4}} gf^{abc}
(q_{1\beta}g_{\gamma\alpha}
                       - q_{1\gamma}g_{\beta\alpha}).
\end{equation}
Similarly, the ghost diagrams
contribute only to the PT one-loop vertex.
Of the remaining diagrams in Fig.\ 3, the propagator-like
pinch part Fig.\ 3(g) of the box diagrams vanishes in the Feynman
gauge, just as in the four fermion scattering process.

Thus, we see that exactly the same PT self-energy is obtained from the
fermion-gluon scattering process as from fermion-fermion process. It
has recently been shown\cite{watson}
how this result generalizes to all the combinations of fields
(including scalars) to which a gauge boson boson couples at tree level.

%

\section{Conclusions}

I have tried to argue here how the concept of an effective charge can
be naturally extended from QED, where it arises automatically, to
non-abelian gauge theories. The argument starts from the simple
fact that, in both abelian and non-abelian theories, the
charge $g$ appearing in the interaction term $gJ\cdot A$
of the classical lagrangian defines the
strength of the tree level interaction between two fermionic
currents $J$ at points $x_{1}$ and $x_{2}$
due to the exchange of a single
gauge boson $A$. The role of the effective charge
$g_{\rm eff}$ is then simply to
account for the change in strength of this two-point interaction due to
radiative corrections. In QED, this can be calculated in perturbation
theory by considering just the conventional photon two-point function.
But in a non-abelian theory, there are contributions to the
interaction between currents at two points which are not included in
the conventional non-abelian gauge boson two-point function. This is
precisely the observation upon which the pinch technique
of Cornwall and Papavassiliou is based;
it is made particularly transparent in the current algebra
formulation of the pinch technique due to Degrassi and Sirlin.
Thus, to calculate the change in strength of the interaction
between currents at two points due to radiative corrections,
rather than considering the conventional gauge boson two-point
function, one needs to consider the PT ``effective'' gauge boson
two-point function. This then gives the effective charge for the
non-abelian theory.

However, in a non-abelian theory there are further sets of fields
in the interaction part of the classical lagrangian
which interact with one another at tree level via the exchange of
a single gauge boson with strength governed by $g$. If the concept of
the PT ``effective'' gauge boson two-point function is to be valid,
then it must also account for the radiative corrections between
any pair of these combinations of fields at two points.
The natural way in which this happens has been demonstrated for
the case of the triple gauge vertex here and in general
elsewhere\cite{watson}.

Returning to the general discussion of the definition of
gauge-independent Green function-like quantities, the lesson to be
learned from the background field method is that it is easy to define
sets of well-behaved Green functions. Although
the BFM Green functions themselves are certainly not gauge-independent,
it is precisely {\it because} they are dependent on an arbitrary
parameter ($\xi_{q}$) and yet
still possess all of the other properties one desires
that they demonstrate that some further field-theoretic
criterion is required to define gauge-independent Green functions
unambiguously. This I believe will turn out to be
supplied by the concept of
the ``effective'' $n$-point function obtained in the pinch technique
and described above for $n=2$.

\section{Acknowledgements}

I wish to thank J. Papavassiliou, A. Sirlin, G. Degrassi,
A. Denner and G. Weiglein for many very stimulating discussions.
It is a pleasure to thank Bernd Kniehl for organising such a
lively workshop.

This work was supported by EC grant ERB4001GT933989.

%

\end{document}